\documentclass[journal,twoside]{IEEEtran}
\usepackage{amsmath}
\usepackage{amssymb}
\usepackage{graphicx}
\usepackage{scalefnt}
\usepackage[font=footnotesize]{caption}
\usepackage[font=footnotesize,labelformat=simple]{subcaption}

\usepackage[nopostdot,nonumberlist,acronym]{glossaries}
\setglossarysection{chapter}
\setacronymstyle{long-short}
\usepackage{url}
\usepackage{multirow}
\usepackage{tabularx}
\usepackage{algorithmic}
\usepackage[ruled]{algorithm2e}
\usepackage{longtable}
\usepackage{xcolor}
\usepackage[subtle]{savetrees}

\newcommand{\quot}[1]{``#1''}
\newcommand*\nestedgls[1]{%
	\protect\ifglsused{#1}{%
		\glsentryshort{#1}%
	}{%
		\glsentrylong{#1}%
	}%
}
\newacronym{NR}{NR}{new radio}
\newacronym{3GPP}{3GPP}{3rd generation partnership project}
\newacronym{eMBB}{eMBB}{enhanced mobile broadband}
\newacronym{uRLLC}{uRLLC}{ultra-reliable and low-latency communications}
\newacronym{mMTC}{mMTC}{massive \nestedgls{MTC}}
\newacronym{VOD}{VOD}{video on demand}
\newacronym{LTI}{LTI}{linear time invariant}
\newacronym{mmWave}{mmWave}{millimetre wave}
\newacronym{LAA}{LAA}{licence assisted access}
\newacronym{IoT}{IoT}{internet-of-things}
\newacronym{D2D}{D2D}{device-to-device}
\newacronym{UMTS}{UMTS}{Universal Mobile Telecommunications System}
\newacronym{WLAN}{WLAN}{Wireless Local Area Network}
\newacronym{AP}{AP}{access point}
\newacronym{LiFi}{LiFi}{light-fidelity}
\newacronym{FET}{FET}{field-effect transistor}
\newacronym{PED}{PED}{portable electronic device}
\newacronym{MCRT}{MCRT}{Monte Carlo ray-tracing}
\newacronym{DL}{DL}{downlink}
\newacronym{UL}{UL}{uplink}
\newacronym{ICI}{ICI}{inter-cell interference}
\newacronym{LED}{LED}{light emitting diode}
\newacronym{PD}{PD}{photo-diode}
\newacronym{IFC}{IFC}{in-flight connectivity}
\newacronym{OIFC}{OIFC}{optical \nestedgls{IFC}}
\newacronym{IFE}{IFE}{in-flight entertainment}
\newacronym{IM/DD}{IM/DD}{intensity-modulation and direct-detection}
\newacronym{IR}{IR}{infra-red}
\newacronym{VL}{VL}{visible light}
\newacronym{OW}{OW}{optical wireless}
\newacronym{OWC}{OWC}{optical wireless communications}
\newacronym{CAD}{CAD}{computer-aided-design}
\newacronym{VLC}{VLC}{visible light communications}
\newacronym{SIR}{SIR}{signal-to-interference-ratio}
\newacronym{SNR}{SNR}{signal-to-noise-ratio}
\newacronym{SINR}{SINR}{signal-to-interference-plus-noise-ratio}
\newacronym{QoS}{QoS}{quality-of-service}
\newacronym{RF}{RF}{radio frequency}
\newacronym{EM}{EM}{electromagnetic}
\newacronym{OOK}{OOK}{on-off keying}
\newacronym{PL}{PL}{path loss}
\newacronym{WDM}{WDM}{wavelength-division-multiplexing}
\newacronym{RGB}{RGB}{red-green-blue}
\newacronym{LoS}{LoS}{line-of-sight}
\newacronym{NLoS}{NLoS}{non-\nestedgls{LoS}}
\newacronym{VoM}{VoM}{volume of mobility}
\newacronym{FoV}{FoV}{field-of-view}
\newacronym{SM}{SM}{spatial modulation}
\newacronym{GSM}{GSM}{generalised \nestedgls{SM}}
\newacronym{OSM}{OSM}{optical \nestedgls{SM}}
\newacronym{SISO}{SISO}{single-input-single-output}
\newacronym{SIMO}{SIMO}{single-input-multiple-output}
\newacronym{MISO}{MISO}{multiple-input-single-output}
\newacronym{MIMO}{MIMO}{multiple-input-multiple-output}
\newacronym{OMIMO}{OMIMO}{optical \nestedgls{MIMO}}
\newacronym{VR}{VR}{virtual reality}
\newacronym{AR}{AR}{augmented reality}
\newacronym{MTC}{MTC}{machine type communications}
\newacronym{CAGR}{CAGR}{compound annual growth rate}
\newacronym{InM}{InM}{intensity modulation}
\newacronym{DD}{DD}{direct detection}
\newacronym{TX}{TX}{transmitter}
\newacronym{RX}{RX}{receiver}
\newacronym{PAM}{PAM}{pulse amplitude modulation}
\newacronym{QAM}{QAM}{quadrature amplitude modulation}
\newacronym{MPAM}{$M$-PAM}{$M$-ary pulse amplitude modulation}
\newacronym{PAPR}{PAPR}{peak to average power ratio}
\newacronym{NRZ-OOK}{NRZ-OOK}{non-return-zero \nestedgls{OOK}}
\newacronym{SSK}{SSK}{space shift keying}
\newacronym{A/D}{A/D}{analog-to-digital}
\newacronym{D/A}{D/A}{digital-to-analog}
\newacronym{O/E}{O/E}{optical-to-electrical}
\newacronym{E/O}{E/O}{electrical-to-optical}
\newacronym{DC}{DC}{direct-current}
\newacronym{SER}{SER}{symbol error ratio}
\newacronym{BER}{BER}{bit error ratio}
\newacronym{OFDM}{OFDM}{orthogonal frequency division multiplexing}
\newacronym{OFDMA}{OFDMA}{orthogonal frequency division multiple access}
\newacronym{AWGN}{AWGN}{additive white Gaussian noise}
\newacronym{ML}{ML}{maximum-likelihood}
\newacronym{ME-OSM}{ME-OSM}{minimum error OSM}
\newacronym{CCI}{CCI}{co-channel interference}
\newacronym{CCD}{CCD}{charge-coupled device}
\newacronym{CIR}{CIR}{channel impulse response}
\newacronym{CFR}{CFR}{channel frequency response}
\newacronym{IFOWC}{IFOWC}{in-flight \nestedgls{OWC}}
\newacronym{LD}{LD}{laser diode}
\newacronym{FR}{FR}{frequency reuse}
\newacronym{WR}{WR}{wavelength reuse}
\newacronym{W-LED}{W-LED}{white \nestedgls{LED}}
\newacronym{PB}{PB}{phosphorescence-based}
\newacronym{RR}{RR}{retro-reflector}
\newacronym{LTE}{LTE}{long term evolution}
\newacronym{Wi-Fi}{Wi-Fi}{wireless fidelity}
\newacronym{5G}{5G}{fifth generation}
\newacronym{FAA}{FAA}{Federal Aviation Administration}
\newacronym{FCC}{FCC}{Federal Communications Comission}
\newacronym{ATENAA}{ATENAA}{Advanced Technologies for Networking in Avionic Applications}
\newacronym{RMS}{RMS}{root-mean-squared}
\newacronym{PSU}{PSU}{passenger service unit}
\newacronym{SIC}{SIC}{successive interference cancellation}
\newacronym{SRT}{SRT}{sequential ray tracing}
\newacronym{NSRT}{NSRT}{non-\nestedgls{SRT}}
\newacronym{RDB}{RDB}{ray database}
\newacronym{CCT}{CCT}{correlated colour temperature}
\newacronym{SMD}{SMD}{surface-mounted device}
\newacronym{USGS}{USGS}{United States Geological Survey}
\newacronym{FSO}{FSO}{free space optical communication}
\newacronym{w.r.t.}{w.r.t.}{with respect to}
\newacronym{HPBW}{HPBW}{half power beam width}
\newacronym{PSD}{PSD}{power spectral density}
\newacronym{DSP}{DSP}{digital signal processing}
\newacronym{UE}{UE}{user equipment}
\newacronym{BRDF}{BRDF}{bidirectional reflectance distribution function}
\newacronym{BSDF}{BSDF}{bidirectional scatter distribution function}
\newacronym{PCM}{PCM}{pulse code modulation}
\newacronym{PPM}{PPM}{pulse position modulation}
\newacronym{DPPM}{DPPM}{differential \nestedgls{PPM}}
\newacronym{U-PAM}{U-PAM}{unipolar \nestedgls{PAM}}
\newacronym{ISI}{ISI}{inter-symbol interference}
\newacronym{DCO-OFDM}{DCO-OFDM}{\nestedgls{DC} biased optical \nestedgls{OFDM}}
\newacronym{ACO-OFDM}{ACO-OFDM}{asymetrically clipped optical \nestedgls{OFDM}}
\newacronym{NDC-OFDM}{NDC-OFDM}{non-DC biased \nestedgls{OFDM}}
\newacronym{PAM-DMT}{PAM-DMT}{pulse amplitude modulated discrete multitone}
\newacronym{U-OFDM}{U-OFDM}{unipolar \nestedgls{OFDM}}
\newacronym{eU-OFDM}{eU-OFDM}{enhanced unipolar \nestedgls{OFDM}}
\newacronym{FFT}{FFT}{fast Fourier transform}
\newacronym{IFFT}{IFFT}{inverse \nestedgls{FFT}}
\newacronym{DAC}{DAC}{digital-to-analog converter}
\newacronym{ADC}{ADC}{analog-to-digital converter}
\newacronym{EOC}{EOC}{electrical-to-optical converter}
\newacronym{OEC}{OEC}{optical-to-electrical converter}
\newacronym{PDF}{PDF}{probability density function}
\newacronym{CP}{CP}{cyclic prefix}
\newacronym{IM}{IM}{index modulation}
\newacronym{SIM}{SIM}{subcarrier \nestedgls{IM}}
\newacronym{DCO-OFDM-SIM}{DCO-OFDM-SIM}{\nestedgls{DCO-OFDM} with \nestedgls{SIM}}
\newacronym{ACO-OFDM-SIM}{ACO-OFDM-SIM}{\nestedgls{ACO-OFDM} with \nestedgls{SIM}}
\newacronym{THP}{THP}{Tomlinson-Harashima Precoder}
\newacronym{LRU}{LRU}{line-replaceable unit}
\newacronym{ACI}{ACI}{adjacent channel interference}
\newacronym{TDD}{TDD}{time division duplex}
\newacronym{MMSE}{MMSE}{minimum mean square error}
\newacronym{CDMA}{CDMA}{code division multiple access}
\newacronym{FWHM}{FWHM}{full width at half maximum}
\newacronym{PM}{PM}{permutation modulation}
\newacronym{PC}{PC}{permutation combinatory}
\newacronym{SS}{SS}{spread spectrum}
\newacronym{PSK}{PSK}{phase shift keying}
\newacronym{BPSK}{BPSK}{binary \nestedgls{PSK}}
\newacronym{QPSK}{QPSK}{quadrature \nestedgls{PSK}}
\newacronym{ICaI}{ICaI}{inter carrier interference}
\newacronym{SE-SIM}{SE-SIM}{spectrally efficient \nestedgls{SIM}}
\newacronym{OFDM-SIM}{OFDM-SIM}{\nestedgls{OFDM} with \nestedgls{SIM}}
\newacronym{APM}{APM}{amplitude and phase modulation}
\newacronym{QSM}{QSM}{quadrature \nestedgls{SM}}
\newacronym{LLR}{LLR}{log-likelihood ratio}
\newacronym{DM}{DM}{dual mode}
\newacronym{DSM}{DSM}{differential \nestedgls{SM}}
\newacronym{DSIM}{DSIM}{differential \nestedgls{SIM}}
\newacronym{CSI}{CSI}{channel state information}
\newacronym{CPU}{CPU}{central process unit}
\newacronym{VoD}{VoD}{video on demand}
\newacronym{EMI}{EMI}{\nestedgls{EM} interference}
\newacronym{FRP}{FRP}{fiber reinforced plastic}
\newacronym{SMX}{SMX}{spatial multiplexing}
\newacronym{JT}{JT}{joint transmission}
\newacronym{SP}{SP}{single point}
\newacronym{FFR}{FFR}{fractional frequency reuse}
\newacronym{CoMP}{CoMP}{coordinated multi-point}
\newacronym{BYOD}{BYOD}{bring your own device}
\newacronym{3D}{3D}{three dimensional}
\newacronym{ICeI}{ICeI}{inter-cell-interference}
\newacronym{FSK}{FSK}{frequency shift keying}
\newacronym{CBE-SM}{CBE-SM}{complex-bipolar encoded \nestedgls{SM}}
\newacronym{AH}{AH}{antenna hopping}
\newacronym{PMF}{PMF}{probability mass function}
\newacronym{AWG}{AWG}{arbitrary waveform generator}
\newacronym{FBE-SM}{FBE-SM}{fractional bit encoded \nestedgls{SM}}
\newacronym{STSK}{STSK}{space-time shift keying}
\newacronym{ESM}{ESM}{enhanced \nestedgls{SM}}
\newacronym{TRX-SM}{TRX-SM}{transceiver \nestedgls{SM}}
\newacronym{ADR}{ADR}{angle diversity receiver}
\newacronym{FLIM}{FLIM}{flexible \nestedgls{LED} index modulation}
\newacronym{GSSK}{GSSK}{generalized \nestedgls{SSK}}
\newacronym{GPSSK}{GPSSK}{generalized pulse position modulated \nestedgls{SSK}}
\newacronym{MA-GSM}{MA-GSM}{multiple active \nestedgls{GSM}}
\newacronym{PEP}{PEP}{pairwise error probability}
\newacronym{APEP}{APEP}{average pairwise error probability}
\newacronym{ABEP}{ABEP}{average \nestedgls{BEP}}
\newacronym{ABER}{ABER}{average \nestedgls{BER}}
\newacronym{BEP}{BEP}{bit error probability}
\newacronym{FEC}{FEC}{forward error correction}
\newacronym{TSM}{TSM}{transmit \nestedgls{SM}}
\newacronym{R-SM}{R-SM}{receive \nestedgls{SM}}
\newacronym{iid}{i.i.d.}{independent identically distributed}
\newacronym{CSIT}{CSIT}{\nestedgls{CSI} at the transmitter}
\newacronym{SC}{SC}{single carrier}
\newacronym{MC}{MC}{multi carrier}
\newacronym{TCM-SM}{TCM-SM}{trellis coded modulation \nestedgls{SM}}
\newacronym{D-SM}{D-SM}{differential \nestedgls{SM}}
\newacronym{DSTSK}{DSTSK}{differential space-time shift keying}
\newacronym{PSM}{PSM}{precoding aided \nestedgls{SM}}
\newacronym{PLS}{PLS}{physical layer security}
\newacronym{F-GSSK}{F-GSSK}{flexible \nestedgls{GSSK}}
\newacronym{Q-SM}{Q-SM}{quadrature \nestedgls{SM}}
\newacronym{E-SM}{E-SM}{enhanced \nestedgls{SM}}
\newacronym{V-SM}{V-SM}{virtual \nestedgls{SM}}
\newacronym{TR-SM}{TR-SM}{transceive \nestedgls{SM}}
\newacronym{SVD}{SVD}{singular value decomposition}
\newacronym{CSIR}{CSIR}{channel state information at the \nestedgls{RX}}
\newacronym{JSM}{JSM}{joint transmitter-receiver \nestedgls{SM}}
\newacronym{GLIM}{GLIM}{generalized \nestedgls{LED} index modulation}
\newacronym{e-GLIM}{e-GLIM}{extended \nestedgls{GLIM}}
\newacronym{DFT}{DFT}{discrete Fourier transform}
\newacronym{IDFT}{IDFT}{inverse \nestedgls{DFT}}
\newacronym{ZF}{ZF}{zero-forcing}
\newacronym{MAP}{MAP}{maximum-a-posteriori-probability}
\newacronym{TD-SM}{TD-SM}{time domain \nestedgls{SM}}
\newacronym{FD-SM}{FD-SM}{frequency domain \nestedgls{SM}}
\newacronym{RC}{RC}{repetition coding}
\newacronym{STBC}{STBC}{space-time block coding}
\newacronym{P/S}{P/S}{space-time block coding}
\newacronym{CN}{CN}{condition number}
\newacronym{eTD-SM}{eTD-SM}{enhanced \nestedgls{TD-SM}}
\newacronym{FDE}{FDE}{frequency domain equalization}
\newacronym{IChI}{IChI}{inter channel interference}
\newacronym{MAC}{MAC}{media access control}
\newacronym{MCS}{MCS}{modulation and coding scheme}
\newacronym{AMC}{AMC}{adaptive modulation and coding}
\newacronym{CQI}{CQI}{channel quality indicator}
\newacronym{RB}{RB}{resource block}
\newacronym{PDCCH}{PDCCH}{physical downlink control channel}
\newacronym{TDMA}{TDMA}{time division multiple access}
\newacronym{SMA}{SMA}{SubMiniature version A}
\newacronym{SPCD}{SPCD}{spectrum}
\newacronym{modem}{modem}{modulator-demodulator}
\newacronym{PLC}{PLC}{powerline communications}
\newacronym{PoE}{PoE}{power over Ethernet}
\newacronym{IATA}{IATA}{The International Air Transport Association}
\newacronym{LSE}{LSE}{The London School of Economics and Political Science}
\newacronym{bpcu}{bpcu}{bits per channel use}
\newacronym{MSE}{MSE}{mean squared error}

\begin{document}
%
\title{Channel Modelling and Error Performance Investigation for Reading Lights Based In-flight LiFi}
\author{Anil~Yesilkaya,~\IEEEmembership{Member,~IEEE,}
	and~Harald~Haas,~\IEEEmembership{Fellow,~IEEE}
\thanks{Manuscript received July xx, 2021; revised December xx, 2021.}
\thanks{Authors are with LiFi Research and Development Center (LRDC),
	Department of Electronic and Electrical Engineering, University of Strathclyde, Glasgow G1 1RD, UK. This research has been supported in part by Zodiac Inflight Innovations (TriaGnoSys GmbH), EPSRC under Established Career Fellowship Grant EP/R007101/1, Wolfson Foundation and European Commission’s
	Horizon 2020 research and innovation program under grant agreement 871428, 5G-CLARITY project.}}

\markboth{Transactions on Vehicular Technology,~Vol.~xx, No.~x, July~2021}%
{Yesilkaya \MakeLowercase{\textit{et al.}}: Channel Modelling and Error Performance Investigation for Reading Lights Based In-flight LiFi}
%
\IEEEoverridecommandlockouts
\IEEEpubid{\makebox[\columnwidth]{\copyright~2022 IEEE. Digital Object Identifier 10.1109/TVT.2022.3148796 \hfill} \hspace{\columnsep}\makebox[\columnwidth]{ }}

\maketitle

\begin{abstract}
The new generation of communication technologies are constantly being pushed to meet a diverse range of user requirements such as high data rate, low power consumption, very low latency, very high reliability and broad availability. To address all these demands, \gls{5G} radio access technologies have been extended into a wide range of new services. However, there are still only a limited number of applications for \gls{RF} based wireless communications inside aircraft cabins that comply with the \gls{5G} vision. Potential interference and safety issues in on-board wireless communications pose significant deployment challenges. By transforming each reading light into an optical wireless \gls{AP}, \gls{LiFi}, could provide seamless on-board connectivity in dense cabin environments without \gls{RF} interference. Furthermore, the utilization of available reading lights allows for a relatively simple, cost-effective deployment with the high energy and spectral efficiency. To successfully implement the aeronautical cabin \gls{LiFi} applications, comprehensive optical channel characterization is required. In this paper, we propose a novel \gls{MCRT} channel modelling technique to capture the details of in-flight \gls{LiFi} links. Accordingly, a realistic channel simulator, which takes the cabin models, interior elements and measurement based optical source, receiver, surface material characteristics into account is developed. The effect of the operation wavelength, cabin model accuracy and user terminal mobility on the optical channel conditions is also investigated. As a final step, the on-board \gls{DCO-OFDM} performance is evaluated by using obtained in-flight \gls{LiFi} channels. Numerical results show that the location of a mobile terminal and accurate aircraft cabin modelling yield as much as $12$ and $2$ dB performance difference, respectively.
\end{abstract}

\begin{IEEEkeywords}
In-flight communications (IFC), LiFi, Monte-Carlo Ray Tracing (MCRT), channel modelling, DC-biased optical OFDM (DCO-OFDM), bit-error-ratio (BER)
\end{IEEEkeywords}

\IEEEpeerreviewmaketitle

\glsresetall

\section{Introduction}
The rapid paradigm shift from the voice oriented circuit-switched domain to the data centric packet-switched network has emerged as a major driver of innovation in wireless communication technologies. The user demand for high speed video streaming and online gaming services, \gls{AR}/\gls{VR} applications, wearable devices and \gls{IoT} create a data greedy ecosystem. Recent forecasts by Cisco showed that monthly data traffic will reach more than $77$ exabytes by 2022, with $93\%$ of this emerging due to smart phones and tablets \cite{cisco_2017_2022}.  However, industrial automation, autonomous cars, remote controlling and tactile internet require strict reliability and low latency features instead of high data rates. Furthermore, \gls{MTC}, which is based on sensor networks, requires low data rate, power consumption and high user density to build smart entities. In order to cope with these diverse and ever increasing demands, services in \gls{5G} \gls{NR} networks are categorized into three main categories; \gls{eMBB}, \gls{uRLLC} and \gls{mMTC}.

As a natural extension of the digital revolution and \gls{5G} \gls{NR} targets, aircraft infotainment and broadband internet access became one of the most important revenue sources for the airline industry. Accordingly, the market potential forecasts show that the revenue for in-flight broadband enabled passenger services for the years 2028 and 2035 are expected to be more than 36 and 63 billion US dollars, respectively \cite{skyhigh_CH1}. Although the majority of airline companies are already providing various on-board connectivity solutions, passengers are still obligated to keep their cellular connections off during flight. Nonetheless, the responses taken from passenger surveys show that $92\%$ of passengers are interested in using their own devices for on-board infotainment \cite{inmarsat}. Extreme user density, approximately $3.03~\text{passengers}/\text{m}^2$ for a narrow-body aircraft, and potential interference issues in small confined areas such as aircraft cabins, pose significant challenges. Specifically, safety concerns and related regulations in the aviation industry prevents the wide adoption of on-board \gls{RF} cellular networks.

One of the key enablers of \gls{5G} \gls{NR} to meet the user demand is the utilization of the higher frequency portion (e.g. Sub-6 and millimetre wave) of the spectrum along with the existing bands to provide additional capacity. To comply with 5G \gls{NR} innovations, \gls{LiFi}, which is a bi-directional and seamless optical wireless broadband networking solution, could overcome the challenges mentioned above. The non-interfering nature of the light with \gls{RF} and the unregulated spectrum, which is 2600 times larger than the entire \gls{RF} band \cite{wjzzlh1701,hycvppai2001}, are the main advantages of \gls{LiFi} to realize small cells and high frequency reuse. Moreover, the deployment, energy, cost efficiency and enhanced security aspects of \gls{LiFi} could also be harvested via the utilization of the pre-existing in-built reading lights in the cabin as broadband in-flight wireless \glspl{AP}. To achieve mentioned goals, accurate channel modelling plays an important role in designing highly efficient \gls{LiFi} networks. Unlike \gls{RF} channels with slow, fast fading and shadowing effects, indoor \gls{LiFi} applications have relatively more deterministic channel characteristics due to the intrinsic nature of \gls{IM/DD} transmission \cite{kb9701}. However, due to the high geometric and spectral dependency of the on-board optical channels, each specific application requires it's own modelling procedure. Furthermore, both the analytical \gls{LoS} and recursive channel modelling approaches in the literature do not always yield accurate results, as some of the optical characteristics are omitted for calculation simplicity. To capture details of the \gls{IFOWC} channels and provide deeper time and frequency-domain analyses, a comprehensive in-flight channel modelling approach is proposed in this paper. Accordingly, both \gls{IR} and \gls{VL} spectrum channels are characterized by using realistic source, receiver, cabin, seating and coating material models within a \gls{MCRT} simulation environment. 

The contributions of this work can be summarised as follows:
\begin{itemize}
	\item A novel optical channel modelling toolkit for broadband in-flight \gls{LiFi} is proposed. Accordingly, realistic geometry of the aircraft cabin and the auxiliary parts; seats and overhead luggage compartment are designed first. Then, geometrical, spatial, angular and spectral profiles of the front-end optics and surface materials are modelled and inputted into the toolkit. The \gls{MCRT} based simulations are conducted both in the \gls{IR} and \gls{VL} bands for the realistic and simplified aircraft cabin structures and for various \gls{UE} locations.
	
	\item The time dispersive in-flight \gls{LiFi} channels are characterized by calculating the \gls{CIR}, \gls{CFR}, \gls{DC} gain, mean delay, \gls{RMS} delay spread and flatness factor parameters. Moreover, the bandwidth properties of the channels are also devised by frequency domain analyses of the \gls{MCRT} based channel data.
	
	\item Lastly, a practical on-board \gls{LiFi} system performance under both the multipath optical channel and non-linear transmit \gls{LED} clipping effects are presented. The analytical \gls{BER} expression for \gls{DCO-OFDM} under the mentioned channel impairments are obtained and compared with the computer simulation results. The importance of the accuracy of the aircraft cabin modelling, transmission wavelength and the location of the \gls{UE} on a practical system performance are also investigated.
\end{itemize}

The reminder of the paper is organized as follows. In Section II, a comprehensive \gls{IFOWC} channel modelling literature review is provided. In Section III, the details of the proposed \gls{MCRT} based \gls{LiFi} channel modelling environment is presented. In Section IV, the mathematical background of the channel simulations and characterization along with the time and frequency domain analyses are provided and discussed. In Section V, the analytical expression and computer simulations based \gls{BER} performance curves of \gls{DCO-OFDM} obtained under the frequency selectivity and non-linear \gls{LED} clipping noise are also presented. Finally, the conclusions are drawn in Section VI.

\emph{Notation}: Throughout the paper, vectors are in bold lowercase letters. The $x$, $y$ and $z$ axes elements of the vector $\mathbf{a}=[a_x,~a_y,~~a_z]$ are given by $a_x$, $a_y$ and $a_z$, respectively. The Dirac delta, inverse tangent and limit functions are expressed by $\delta(\cdot)$, $\arctan(\cdot)$ and $\lim\{\cdot\}$, respectively. The magnitude of a complex number $z$, logarithm of a number to base $b$, statistical expectation, circular convolution, element-wise multiplication and ceiling operations are also denoted by $\lvert z \rvert$, $\log_b\left( \cdot \right)$, $\text{E}\{ \cdot \}$, $\circledast$, $\circ$ and $\lceil \cdot \rceil$, respectively. The real-valued normal distribution with mean, $\mu$, and variance, $\sigma^2$, is given by $\mathcal{N}(\mu,\sigma^2)$. The Q-function is defined by $Q(x)=\frac{1}{\sqrt{2\pi}}\int_{x}^{\infty}e^{-u^2/2}\text{d}u$.

\section{Multi-bounce IFOWC Channel Modelling Literature}
In this section, a literature review for multi-bounce \gls{IFOWC} channel modelling techniques, which includes \gls{LoS} and \gls{NLoS} components, will be presented. The \gls{PL} and \gls{RMS} delay spread characteristics of the multi-bounce optical channel for ceiling (wash) light-based \gls{IR} links is investigated in \cite{wc0801}. Accordingly, the \gls{MCRT} analysis inside a simplified Boeing 777 cabin model is presented. Similarly, in \cite{bfzs0801}, an \gls{IFOWC} system with \gls{IR}-\gls{LD} based ceiling lights is considered where the data is envisaged to be delivered to seat backs within a cellular deployment, which covers 5 seats. Moreover, multipath \gls{CIR} is obtained by \gls{MCRT} for a simplified cabin model and results are validated by mock-up cabin measurements. Another ceiling light based channel characterization for the \gls{IR} band \gls{IFOWC} is proposed in \cite{dmhcob0901} and \cite{dmhcob0902}. Accordingly, the multi-bounce \gls{MCRT} channel model for a simplified cabin model is utilized to estimate the parameters of \gls{PL} and shadowing. Moreover, the cell edge and centre \gls{SIR} maps are also obtained. Note that the omni-directional \gls{TX}-\gls{RX} pair is designed in \cite{dmhcob0901,dmhcob0902} to show that the \gls{RF} \gls{PL} model (Friis' formula) with shadowing (slow fading) effects can also be modelled in incoherent \gls{OWC}. Furthermore, the effect of \gls{FR} schemes on the \gls{SIR} distribution is also investigated in \cite{dmhcob0902}. The ceiling mounted passive retro-reflector based communication between \glspl{PSU} via \gls{IR} directed links is proposed in \cite{yilass1101}. However, only a single bounce, $\kappa\in \{0,1\}$, \gls{CIR} with a simplified cabin model is calculated by using the method in \cite{kkc9501}. Note that the parameter $\kappa$ will be used to refer the light rays after the $\kappa^\text{th}$ order reflections, where $0 \leq \kappa \leq \kappa_\text{max}$. The parameter $\kappa_\text{max}$ is the maximum number of reflections considered within the system. Furthermore, $0^\text{th}$ bounce corresponds to the \gls{LoS} component of the channel. In \cite{qgrrj1301}, an \gls{MCRT} based \gls{IFOWC} channel modelling technique, which is able to encounter up to $10$-bounces, is used to obtain \gls{SIR} and \gls{SNR} maps. Accordingly, \gls{VL} and \gls{IR} band sources in \gls{UL} and \gls{DL} directions, respectively, are employed. However, a very primitive cabin and aircraft interior model is presented in  \cite{qgrrj1301}, where the details of the optical channels are also neglected. In the same work, a hardware implementation of the \gls{IFOWC} network and user side adapter prototypes are also presented. Another ray tracing based \gls{VL} band \gls{IFOWC} \gls{CIR} modelling method for a realistic Boeing 737-900 cabin and seating scenario is given by \cite{kavehrad15}. In the paper, both the \gls{BER} spatial distributions and respective outage probability values are also obtained. The \gls{SINR} maps for a simplified Boeing 737 cabin and seat models is presented for both phosphorescence and red-green-blue based white \glspl{LED} in \cite{tc1601}. The modified \gls{MCRT} method is used in order to trace a million rays up to $5$-bounce while Lambertian and Phong radiation patterns are assumed for the \glspl{LED}. Similarly, in \cite{tc1701}, $5$-bounce, $0 \leq \kappa \leq 5$, ray tracing simulations for a simplified Boeing 727 cabin model and seating is obtained to simulate \gls{UL} \gls{IR} band \gls{IFOWC}. The effect of passenger movements on the channel and system performance is also investigated, where it has been reported that $10$ Mbit/s can be supported by \gls{UL} \gls{IFOWC}. A modified \gls{MCRT} method to obtain \gls{VL} band reading light based \gls{DL} \gls{IFOWC} \gls{CIR} is presented with a realistic Boeing 737 cabin model and seating in \cite{tmcs1801}. Specifically, various sources with different beam angles and solar radiation coming from the windows are taken into consideration in the \gls{SINR} calculations. The same approach to obtain the \gls{CIR} is also adopted for a simplified Airbus A320-200 cabin and seat models in \cite{tc1901}. The investigation of an \gls{IR} band \gls{IFOWC} inside the cockpit, which aims to provide bidirectional connectivity between the ceiling unit and the pilot headset, is given by \cite{dssvcmab1901}. Moreover, a Lambertian bidirectional reflectance distribution function, source semi-angles, realistic pilot body models and the effect of users head movement are investigated in \cite{dssvcmab1901}. The same approach is enhanced in \cite{dsh1201} to capture the effect that both body and head movements have on the \gls{IR} band \gls{CIR}.
\section{In-flight LiFi Channel Modelling Methodology}
The signal propagation in the \gls{IR} and \gls{VL} spectrum is highly dependent on geometry and the surface structure of the considered environment/scenario due to the weak penetration characteristics of \gls{EM} waves. Thus, it is very important to ensure that geometry and optical characteristics of both the interior and exterior elements of an aeronautical cabin environment are captured accurately as possible for comprehensive channel modelling. However, it is emphasized by the previous section that the existing channel modelling approaches are unable to provide realistic \gls{TX}, \gls{RX}, cabin geometry/interior, coating material and reflection/refraction characteristics as a whole. Furthermore, deeper analyses on important channel parameters; frequency response, coherence time/bandwidth, flatness and mobility effects are also omitted. Therefore, in this section, we propose a \gls{NSRT} based in-flight \gls{LiFi} channel modelling approach, which is able to encompass a complete set of \gls{TX}, \gls{RX} and environment parameters. Accordingly, the shortcomings of the \gls{MCRT} based \gls{CIR} modelling technique, which is given in \cite{mu1501,mup1501,mu2001,802.11bb_refchannel}, are resolved by the proposed method. Hence, a complete optical channel modelling capability in realistic aircraft environments are supported in the proposed technique by including the \gls{RX} spectral, angular and spatial characterization.

\begin{figure}[!t]
	\centering
	\includegraphics[width=1.0\columnwidth]{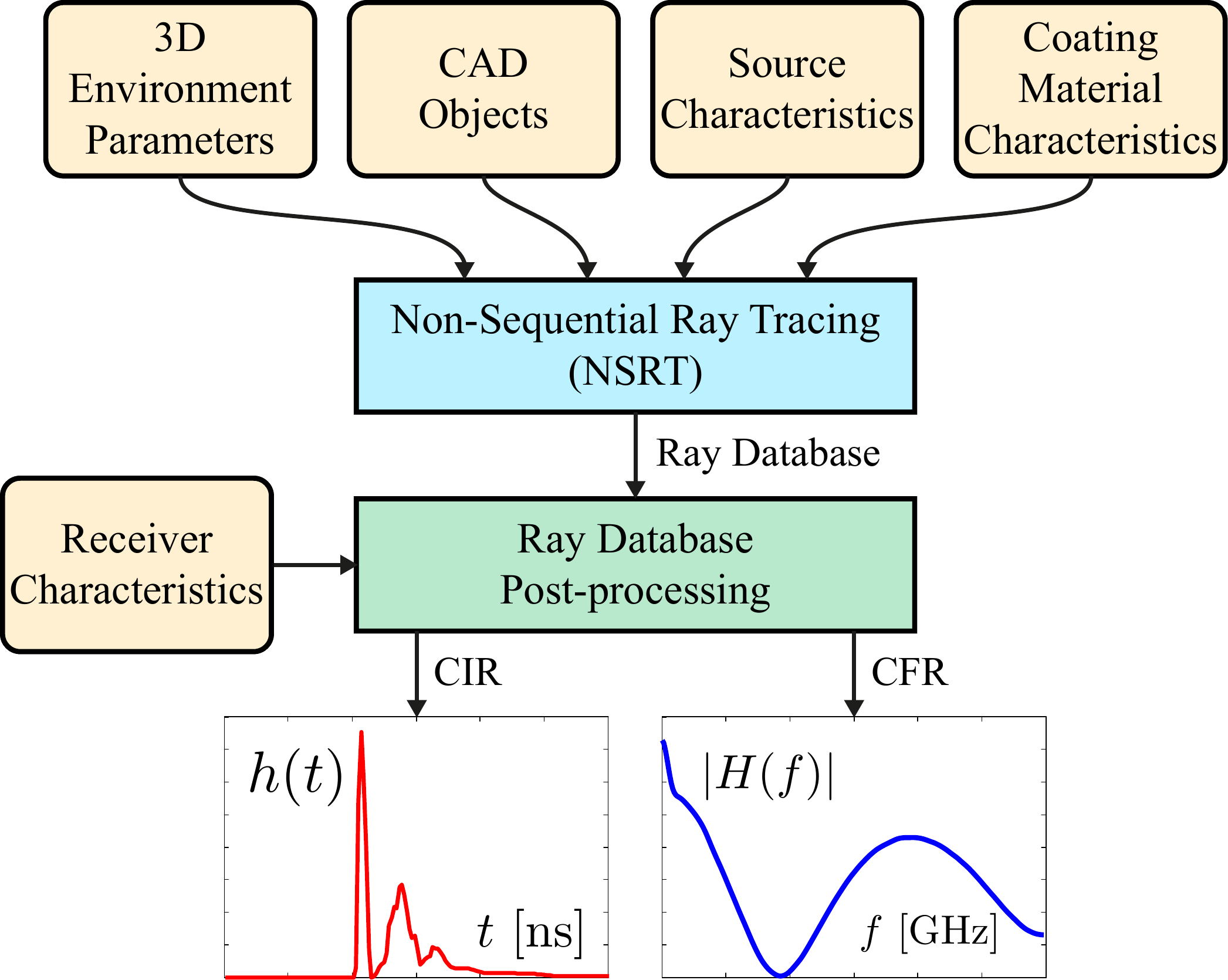}
	\caption{Block diagram for proposed MCRT based optical channel modelling method.}
	\label{fig:block_diagram}
\end{figure}

The proposed \gls{MCRT} simulations are performed with the aid of optical design software tool, Zemax OpticStudio version 20.2 \cite{zemaxopticstudio}. The benefit of using a commercial solution is the generation of reproducible, replicable and repeatable results, as the packages and libraries are standard and independent of application. The adopted software is able to perform both \gls{SRT} or \gls{NSRT} depending on the application. In \gls{NSRT}, the generated rays hit the surfaces merely based on the physical positions and the optical properties of the objects by considering the direction of the rays. Whereas, in \gls{SRT}, the rays are obliged to propagate in a predefined sequence of surfaces or objects, which is very important in imaging optics. Since the primary purpose of \gls{LiFi} systems is not only communication, but also illumination (mostly diffuse) via reading and wash lighting, the \gls{NSRT} method is adopted in this research. For our simulations, the realistic aircraft cabin environment is created by the generation of \gls{CAD} models. Then, the measurement based source and material characteristics are imported into the simulation environment. To obtain the \glspl{CIR}, a \gls{RDB} file needs to be generated via \gls{NSRT}, which contains the extensive optical and geometrical information for every single ray traced. Lastly, powerful data processing tools are employed to extract and analyse properties of the optical channel. The block diagram of the proposed \gls{MCRT} based channel modelling methodology is depicted in Fig. \ref{fig:block_diagram}.
\subsection{Generation of the Aircraft Cabin Environment}
\begin{figure*}[!t]
	\centering
	\begin{subfigure}[t]{0.71\columnwidth}
		\includegraphics[width=\columnwidth]{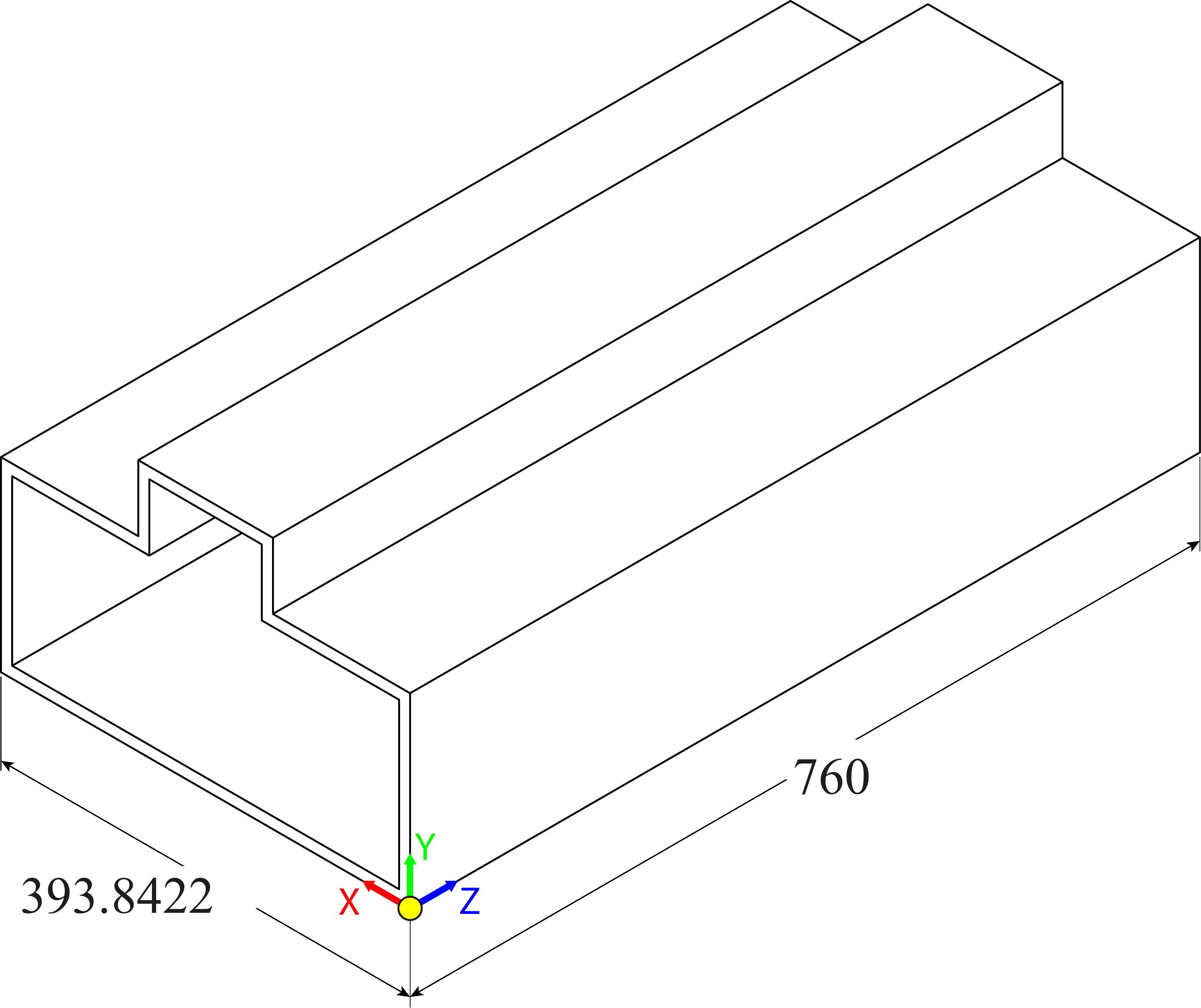}
		\caption{}
		\label{}
	\end{subfigure}~
	\begin{subfigure}[t]{0.71\columnwidth}
		\includegraphics[width=\columnwidth]{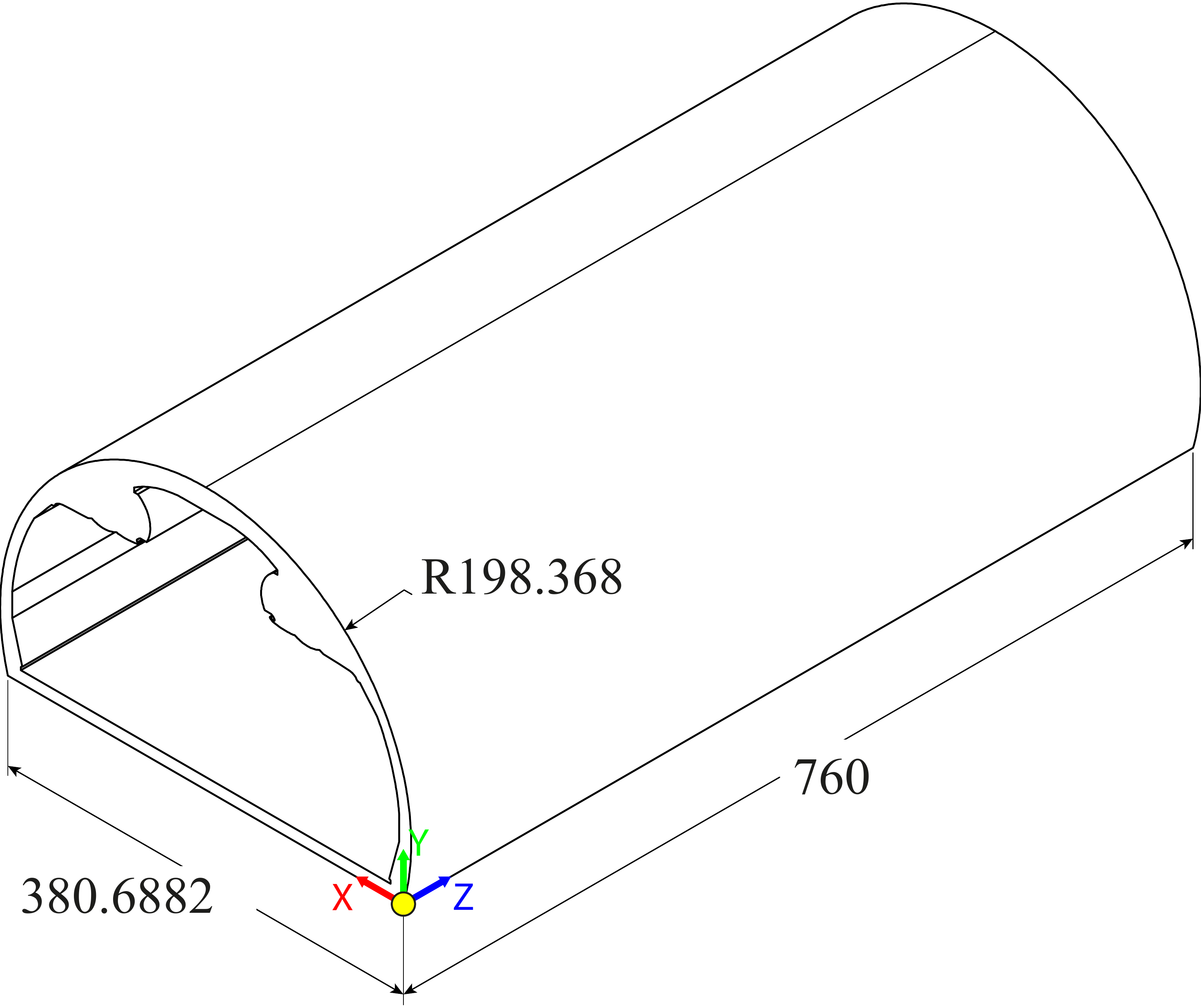}
		\caption{}
		\label{}
	\end{subfigure}
	\caption{Isometric view of the realistic (left) and simplified (right) \glsentrytext{CAD} cabin models (units are in cm). The yellow point represents the local origin of the adopted cabin models.}
	\label{fig:A320_realsimp_iso}
\end{figure*}
\begin{figure}[!t]
	\centering
	\includegraphics[width=0.7\columnwidth]{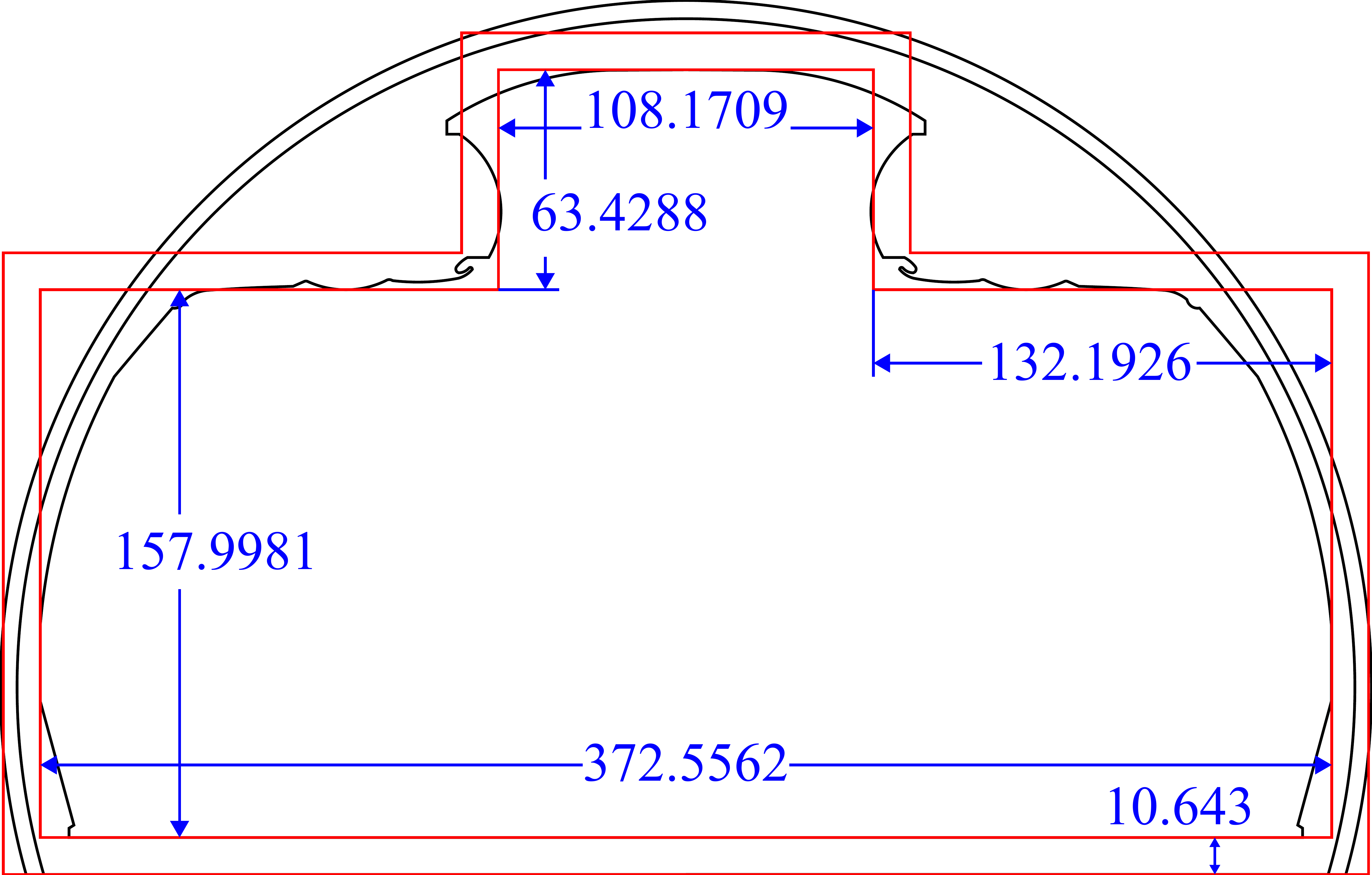}
	\caption{The geometrical details of the realistic (black) and simplified (red) narrow-body cabin models (units are in cm).}
	\label{fig:A320_realsimp_front}
\end{figure}
\begin{figure}[!t]
	\centering
	\includegraphics[width=\columnwidth]{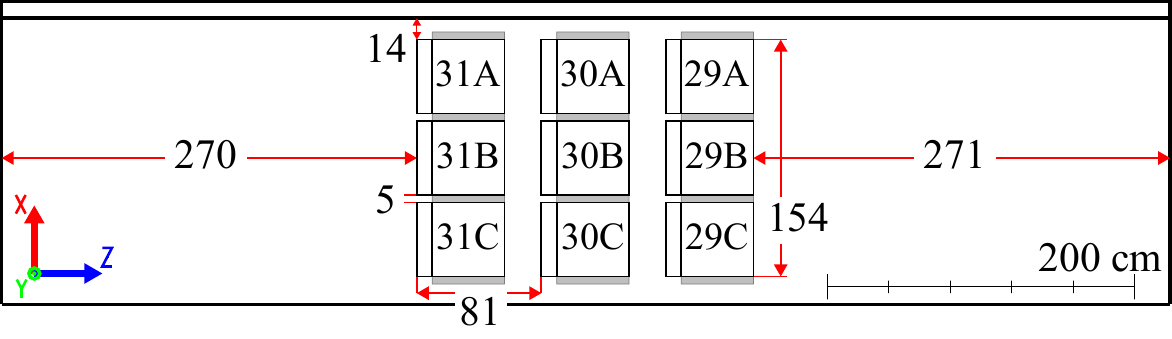}
	\caption{Top view of the seating layout (units are in cm).}
	\label{fig:seating_layout}
\end{figure}
To model the aeronautical cabin environment accurately, the global positions and orientations for the cabin and interior objects must be defined first. For the sake of clarity, the geometric parameters of the objects will be given in vector notation, which is defined \gls{w.r.t.} the global coordinate system. The origin of the global coordinate system is defined as the vector $O(0,~0,~0)$. The position orientation of each object within the simulation environment is defined by $3\times1$ location and orientation vectors $\mathbf{v}=(v_x,~v_y,~v_z)$ and $\mathbf{o}=(o_{x},~o_{y},~o_{z})$, where the $x$, $y$ and $z$ axes coordinates and rotation with respect to the $x$, $y$ and $z$ axes are given by $v_x$, $v_y$, $v_z$ and $o_x$, $o_y$, $o_z$, respectively. As can be inferred from previous section, the majority of works in the literature assumed a simplified aircraft cabin, where the curvatures and fine details of the geometry are approximated as flat surfaces. In our work, two types of cabin structures, namely realistic and simplified models, will be considered to investigate the effect of geometrical simplification on the accuracy of the channel parameters. Accordingly, a narrow-body type Airbus A320 is adopted as the main cabin structure for our realistic ray-tracing simulations. Technical details of the narrow-body A320 fuselage were obtained from \cite{A320} and the three dimensional model is regenerated using \gls{CAD} software. Then, the complex faces of the \gls{CAD} object are replaced with flat surfaces to obtain a simplified cabin model similar to \cite{dmhcob0901} and \cite{dmhcob0902}. The technical drawings and dimensions of both cabin models are depicted for isometric and cross-sectional views in Figs. \ref{fig:A320_realsimp_iso} and \ref{fig:A320_realsimp_front}, respectively. As shown in Fig. \ref{fig:A320_realsimp_iso}, the cabin structure is symmetric along the $+z$-axis, where only a section of the whole cabin is sufficient to generalize the obtained results. Thus, a portion of the cabin fuselage with the length of $7.6$ m will be employed in this research. Furthermore, only the reading lights based power contributions are considered in our optical channels, where the window shades are designed to be closed in our three dimensional cabin models. However, the variable ambient noise effect will be considered as an electrical domain shot noise at the \gls{RX} in our practical error performance simulations.
\begin{figure}[!t]
	\centering
	\begin{subfigure}[t]{.644\columnwidth}
		\includegraphics[width=\columnwidth]{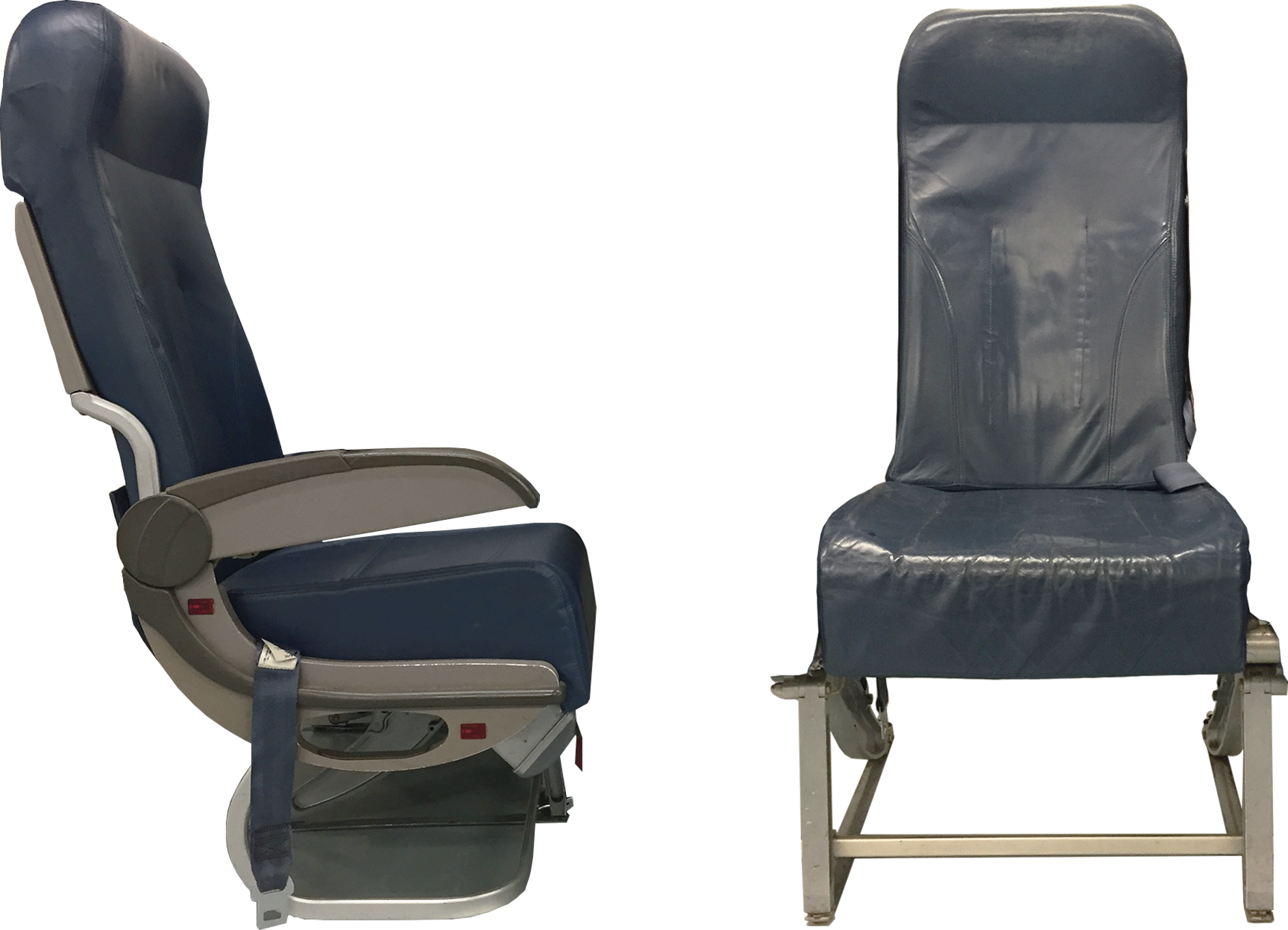}
		\caption{}
		\label{fig:real_seat}
	\end{subfigure}\\
	\begin{subfigure}[t]{.630\columnwidth}
		\hspace{2.10mm}
		\includegraphics[width=\columnwidth]{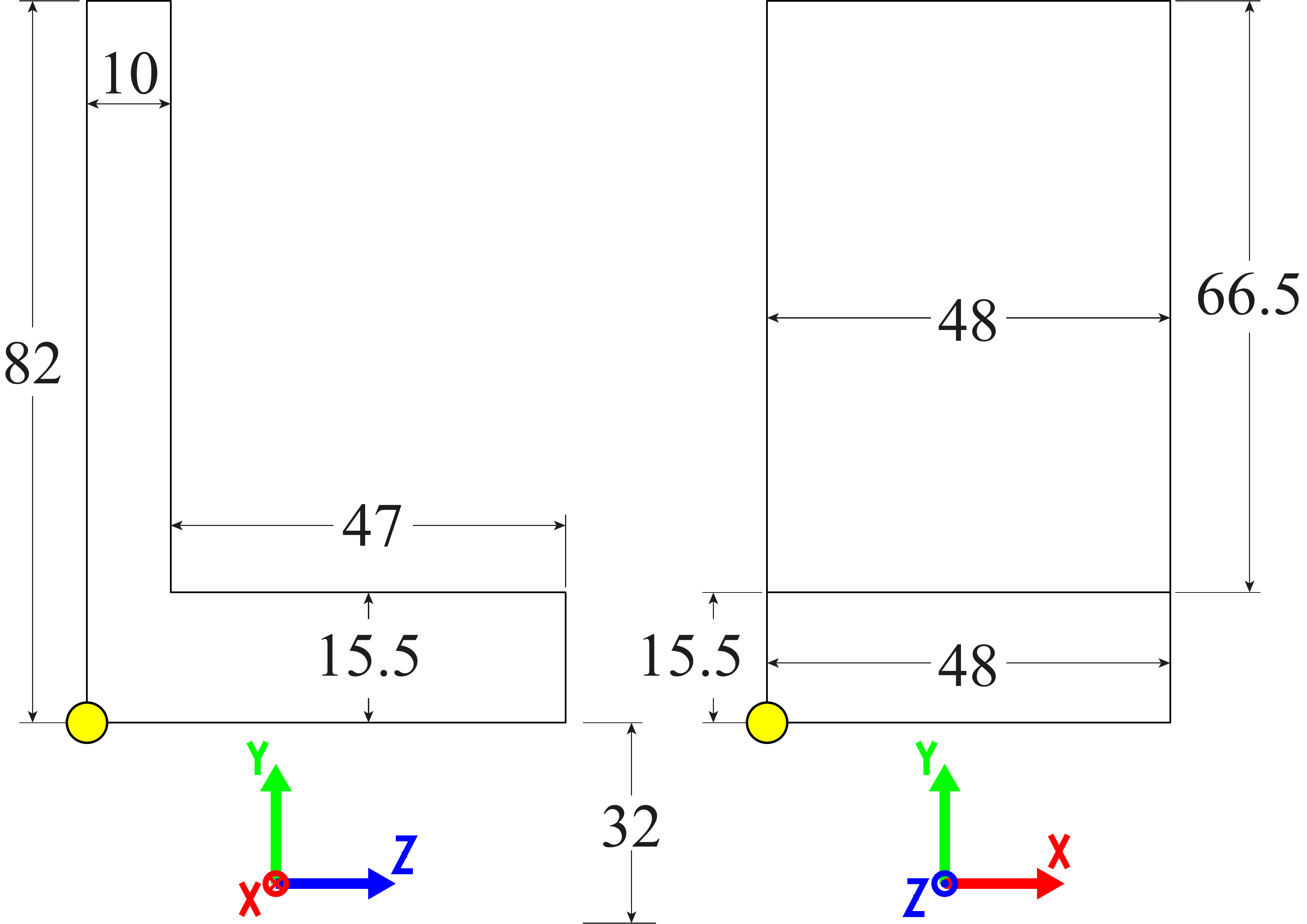}
		\caption{}
		\label{fig:model_seat}
	\end{subfigure}
	\caption{Side (left) and front (right) views of the (a) real/measured (b) recreated seats (units are in cm). The local origin of the generated seat model is given by the yellow point.}
	\label{fig:seat_model}
\end{figure}
\subsubsection{The Cabin Interior}
The majority of commercial airliners utilize narrow-beam reading lights on-board \cite{utc_reading}. Thus, the coverage of a reading light would not be wider than a single row of seats, where both the seats and passengers will serve as a natural barrier between the seat rows. Due to the symmetric structure of the cabin environment, the left hand-side, $+x$ direction, of a 6-abreast, 3-3 formation, seating will be considered to create the generalized model. Moreover, the reflection contributions from the neighbouring rows are also evaluated by taking both front and back triple seats into account. For the sake of simplicity, the rows and seats will be labelled as 29, 30, 31 and A, B, C, respectively. Details of the seating structure and dimensions are given in Fig. \ref{fig:seating_layout}.

The seat pitch is another important parameter for cabin design, which is defined as the distance between any arbitrary point on one seat and the exact same point in the seats directly in front or behind it. To maximize the number of passengers on-board, aircraft companies tend to minimize the seat pitch, which can be as low as $28$ inches. However, it has been reported in \cite{kgssb1201} that passenger well-being and comfort is strongly correlated to how far back the seat can recline. Furthermore, any seat pitch enlargements after $32$ inches are also reported not to have a significant affect on the passenger comfort, where the practical values lie between $30$ and $34$ inches for the economy class. Hence, the seat pitch is chosen to be $81$ cm ($\approx32$ inches) in our simulation environment, which falls within the standard economy class range.
\subsubsection{Passenger Seating}
The design of the seats is another important issue for passenger comfort, especially in medium and long-haul flights. The passenger seat dimensions, and further details on the seating layout, are obtained via measurements on Weber Aircraft LLC, more recently known as Safran Passenger Innovations, aircraft seating. Then, the geometrically simplified \gls{CAD} models are generated and implemented in the simulation environment. The photos of the actual seating and the technical drawings of the \gls{CAD} software generated seat models are depicted in Figs. \ref{fig:real_seat} and \ref{fig:model_seat}, respectively. Note that the local origin of the obtained seat model is depicted by the yellow point in Fig. \ref{fig:model_seat}.
\subsection{Sources, Receivers and Coating Materials}
\begin{figure}[!t]
	\centering
	\includegraphics[width=0.70\columnwidth]{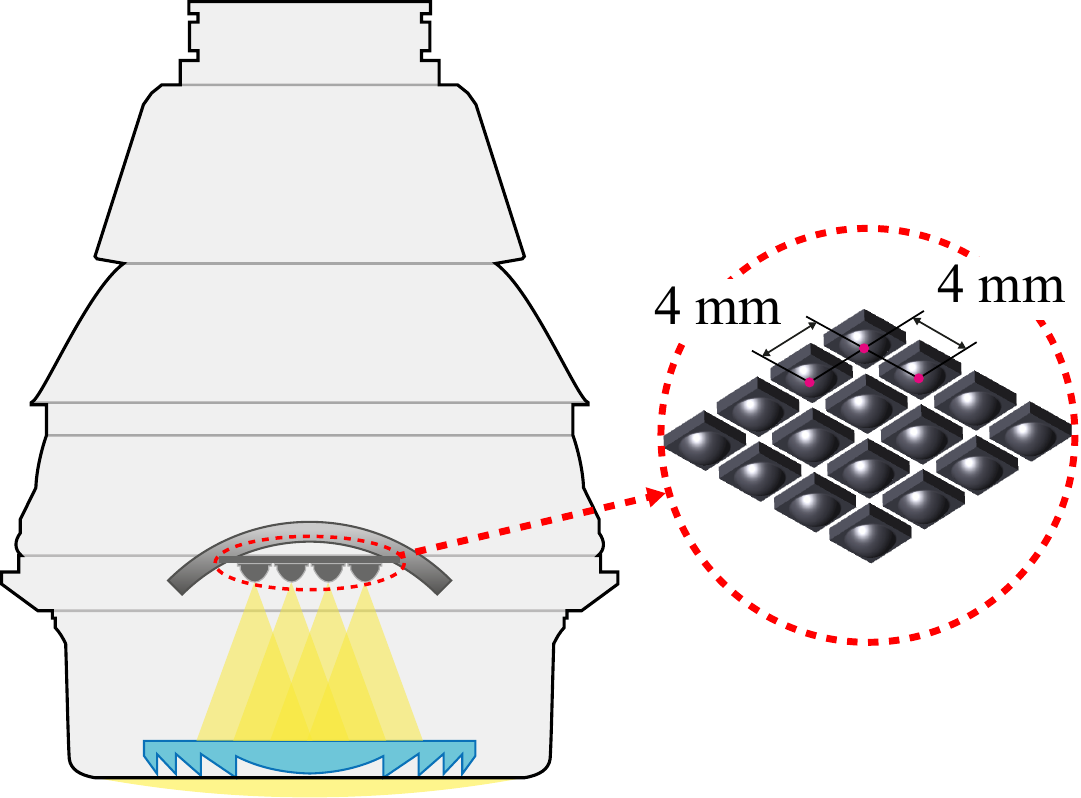}
	\caption{Typical directional reading light structure where the base is an array of OSRAM GW QSSPA1.EM high power \glsentrytext{LED} chips.}
	\label{fig:reading_light}
\end{figure}
\begin{figure}[!t]
	\centering
	\includegraphics[width=0.65\columnwidth]{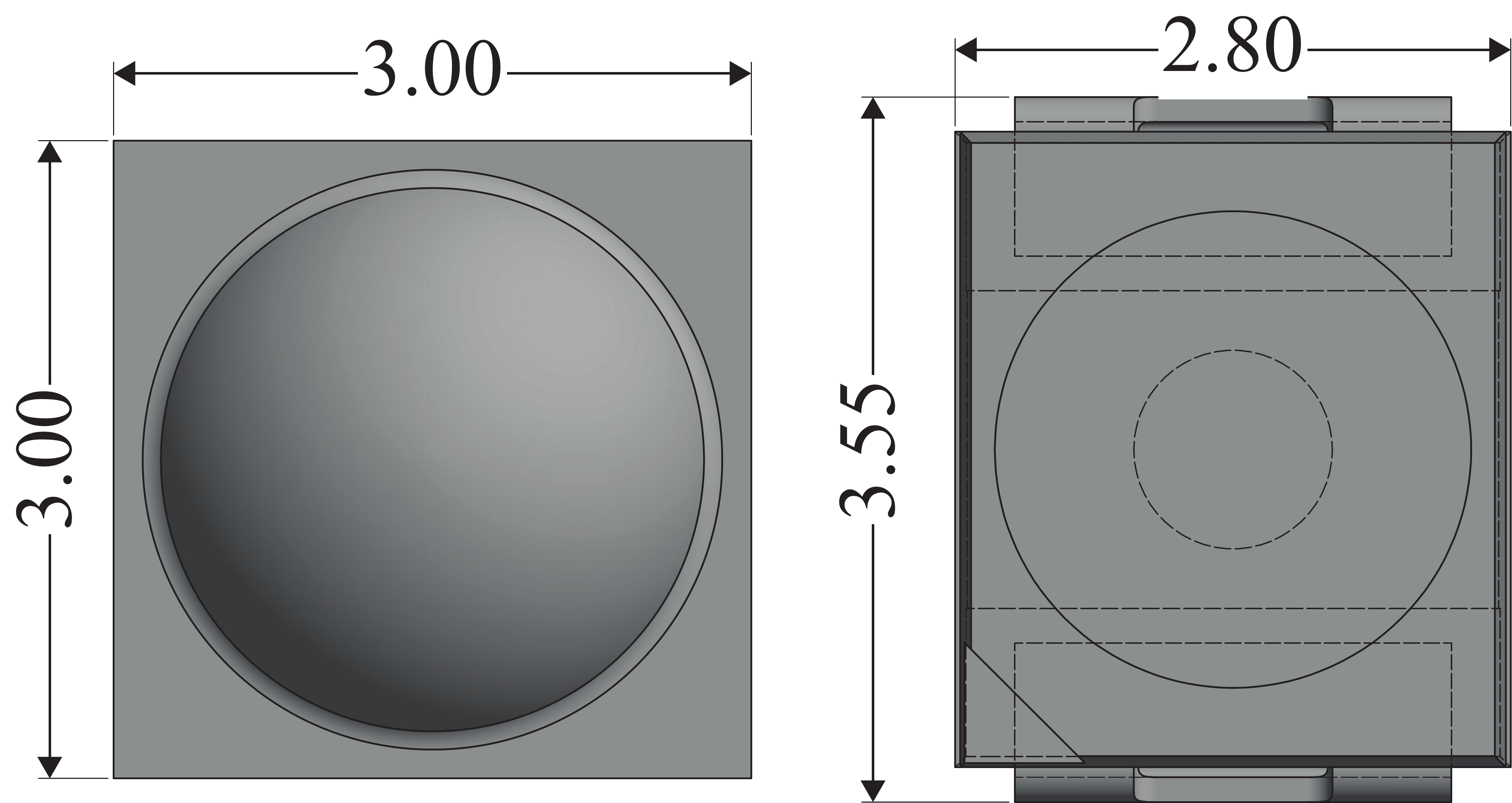}
	\caption{The technical drawings of the \glsentrytext{VL} band OSRAM GW QSSPA1.EM (left) and \glsentrytext{IR} band OSRAM SFH 4253 (right) \glsentrytext{LED} chips used in the reading light design (units are in mm).}
	\label{fig:led_technicaldrawing}
\end{figure}
The implementation of \gls{IFC} and/or \gls{IFE} systems, which require the airframe or cabin interior to be redesigned, could introduce significant costs to the current global airline industry. On the contrary, in-flight \gls{LiFi} offers the great advantage of using the existing cabin lighting structure, while only requiring very minor modifications. Accordingly, the practical implementation of on-board \gls{LiFi} mainly includes the utilization of two types of sources; reading and wash lights. The reading lights could be utilized to provide broadband data to the seats via either the \glspl{PSU} or directly to the passenger's \glspl{PED}. Similarly, the wash lights could serve for low/mid data rate applications such as sensory, \gls{IoT}, \gls{D2D} communications infrastructure etc. as well as on-board payment systems. Note that cabin illumination must be dimmed in the critical phases of flight, namely take-off and landing, for safety purposes. For overnight flights, dimming, or even turning the cabin lights off completely, should be considered for passenger comfort. More importantly, the $800-1000$ nm band is reported to be the most suitable PHY layer common mode signal region for the \gls{LiFi} applications
by \quot{IEEE 802.11bb Standardization Task Group on Light Communications}. In addition, the \gls{VL} band spectra is remained as an optional PHY mode in IEEE 802.11bb standardization process \cite{802.11bb_common}. Therefore, in this research, both the \gls{VL} and \gls{IR} bands will be used interchangeably in the \gls{DL} direction to comply with the \gls{FAA} dimming regulations as well as IEEE 802.11bb Light Communications standards. It is also important to note that the \gls{IR} band is almost always preferential for the uplink transmission for the purpose of maintaining passenger eye safety and comfort. In the following, a realistic reading light design and associated surface material characterization will be detailed to address the raised safety and communication regulations.
\begin{figure}[!t]
	\centering
	\begin{subfigure}[t]{.75\columnwidth}
		\includegraphics[width=\columnwidth]{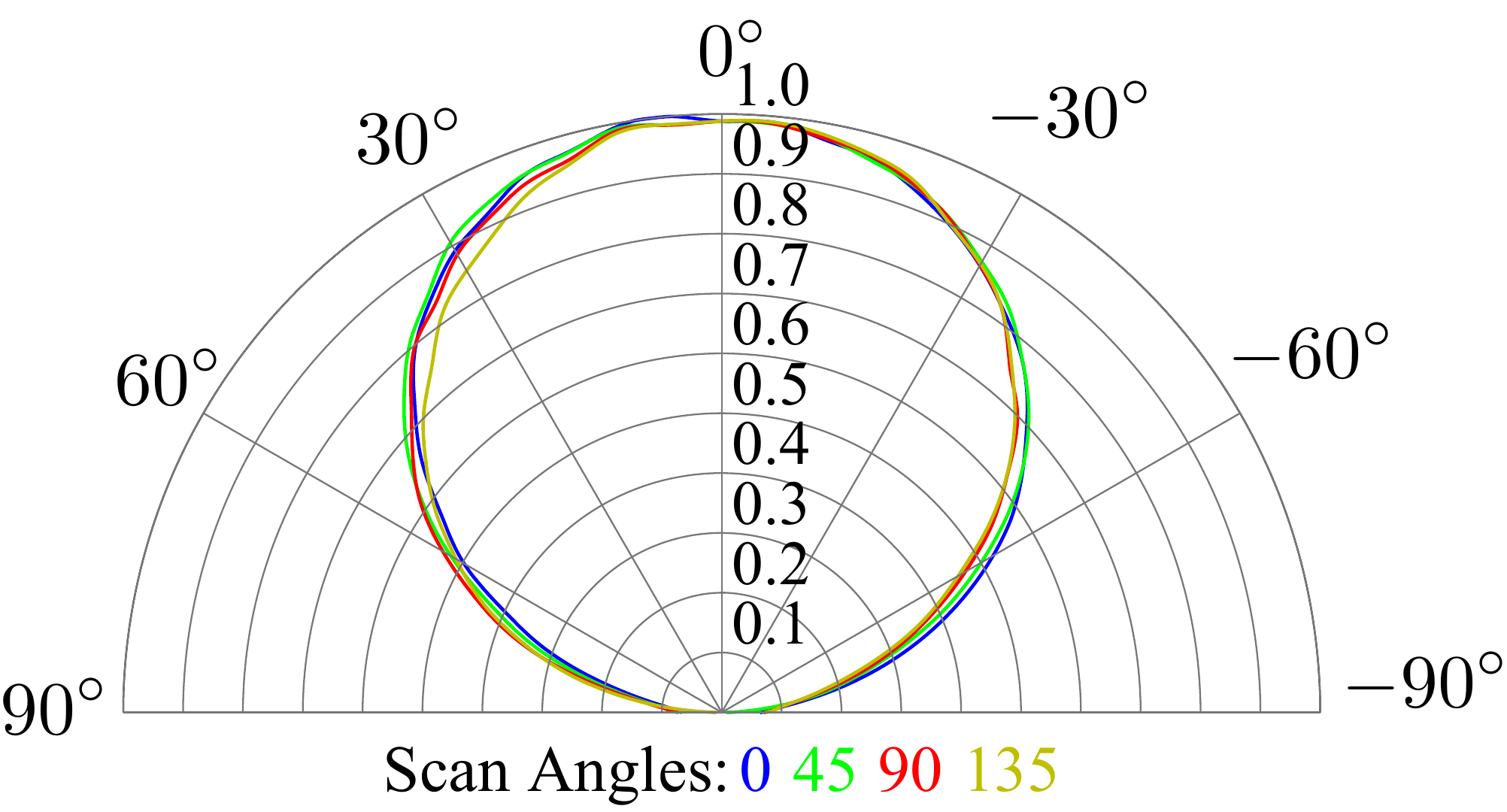}
		\caption{OSRAM GW QSSPA1.EM}
		\label{fig:direct_VL}
	\end{subfigure}\\
	\begin{subfigure}[t]{.75\columnwidth}
		\includegraphics[width=\columnwidth]{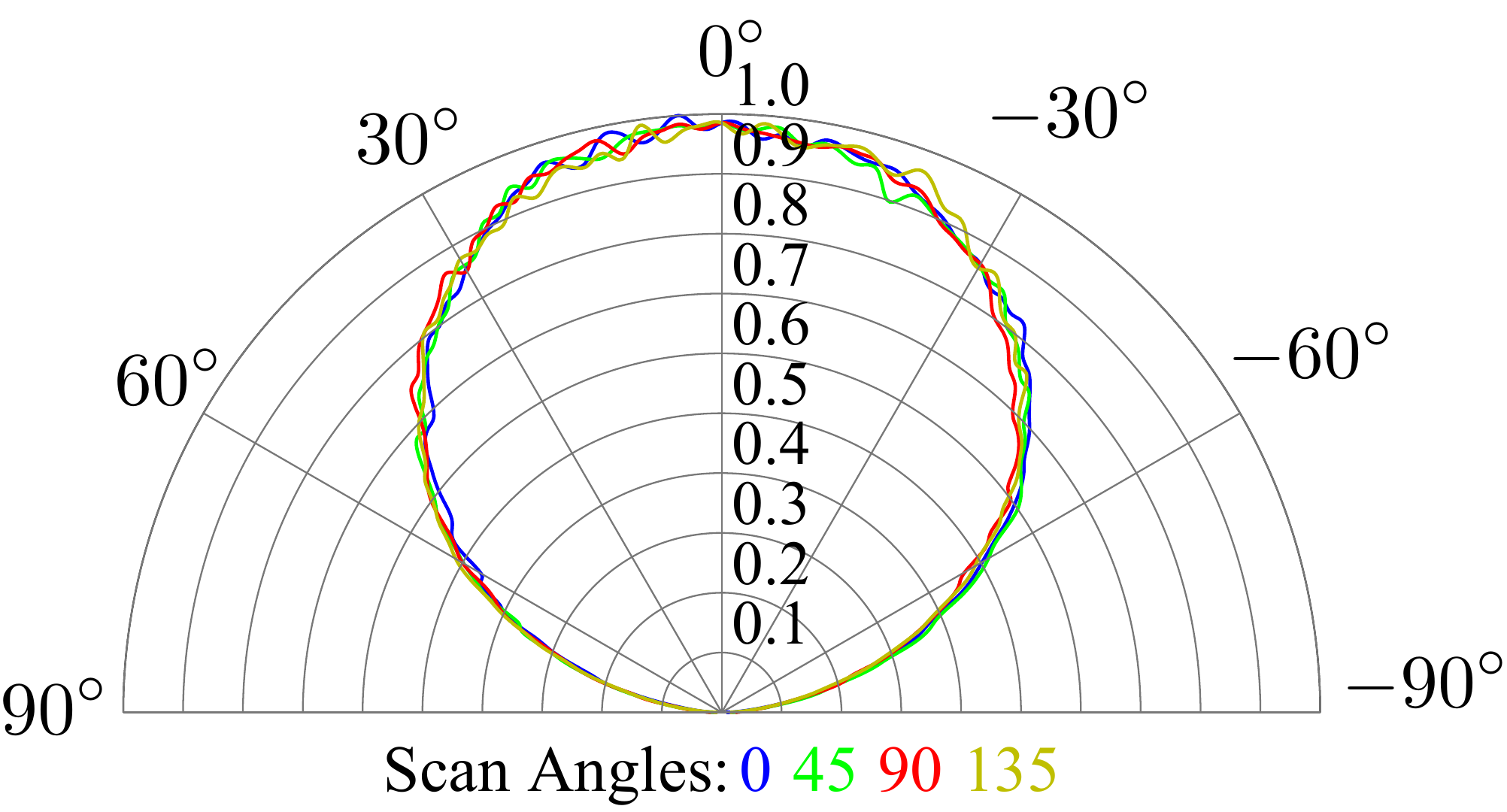}
		\caption{OSRAM SFH 4253}
		\label{fig:direct_IR}
	\end{subfigure}
	\caption{Source directivity plots of the \glsentrytext{VL} (left) and \glsentrytext{IR} (right) band \glsentrytext{LED} chips for the azimuthal angles $[0~ 45~ 90~ 135]$.}
	\label{fig:source_directivity}
\end{figure}
\subsubsection{Reading Lights}
In the aircraft interior lighting market, there are two main reading light structures based on their beam profile; highly directional and dispersed \cite{utc_reading}. Generally, to provide the best conditions for reading, highly directional lights, consisting of an \gls{LED} array as a base, a concave mirror and a lens, which could be Fresnel or bi/double-convex, are preferred. Hence, effectively a collimated beam is created as depicted in Fig. \ref{fig:reading_light}, which is directed towards the vicinity of the tray table. On the other hand, dispersed light sources contain only the \gls{LED} array base to achieve a wider spread for illumination purposes \cite{collins_reading}. As mentioned in the previous chapter, the contribution of the \gls{NLoS} paths are strictly required to design a robust communication system. Therefore, in this research, the optoelectronic characteristics of the dispersed reading lights will be investigated as potential sources for in-flight \gls{LiFi} applications.

The non-imaging optical transmit front-end structure is adopted in our study to model the reading lights, each of which will serve as a \gls{LiFi} \gls{AP} in the given system design. Accordingly, a $4\times 4$ \gls{LED} chip array is designed to form the base of the reading lights in the simulation environment which is given in Fig. \ref{fig:reading_light}. The origin point of the reading light element is chosen as the centre point of the \gls{LED}. For simplicity in the optical calculations, each \gls{LED} chip unit is assumed to be outputting $1$ W optical power that yields $16$ W power per reading light. In the designed array, off-the-shelf OSRAM GW QSSPA1.EM High Power White \gls{LED} \cite{gwqsspa1.em} and OSRAM SFH 4253 High Power \gls{IR} \gls{LED} \cite{sfh4253} chip specifications are used to model the realistic \gls{VL} and \gls{IR} band emitters, respectively. The separation between each \gls{LED} chip within the array is chosen as $4$ mm, refer to Fig. \ref{fig:reading_light}, for both sources based on their \gls{SMD} packagings. The technical drawings and the detailed dimensions of the \gls{SMD} epoxy packaging for both sources are depicted in Fig. \ref{fig:led_technicaldrawing}.
\begin{figure}[!t]
	\centering
	\begin{subfigure}[t]{.65\columnwidth}
		\includegraphics[width=1\columnwidth]{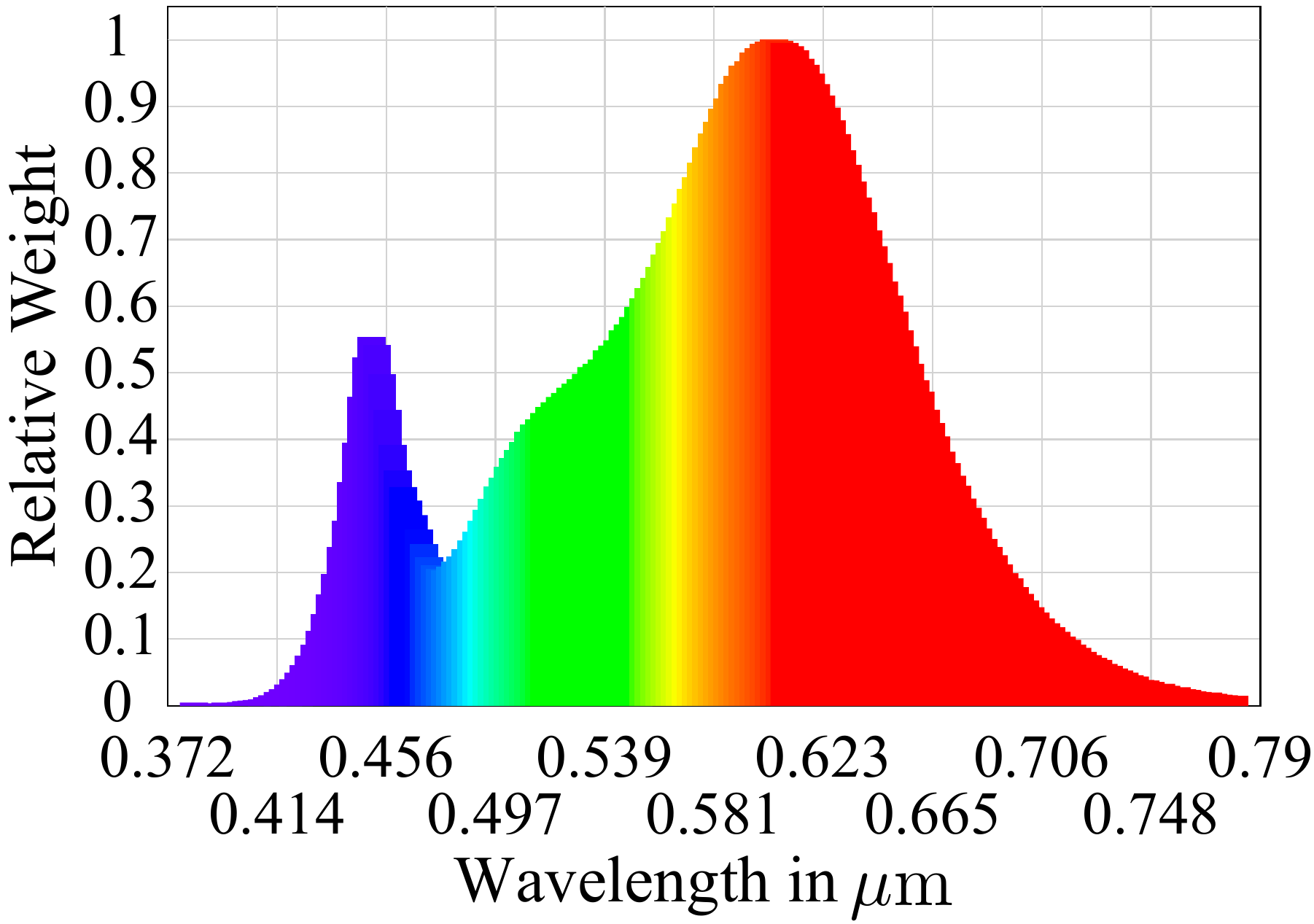}
		\caption{OSRAM GW QSSPA1.EM}
		\label{fig:spec_VL}
	\end{subfigure}\\
	\begin{subfigure}[t]{.65\columnwidth}
		\includegraphics[width=1\columnwidth]{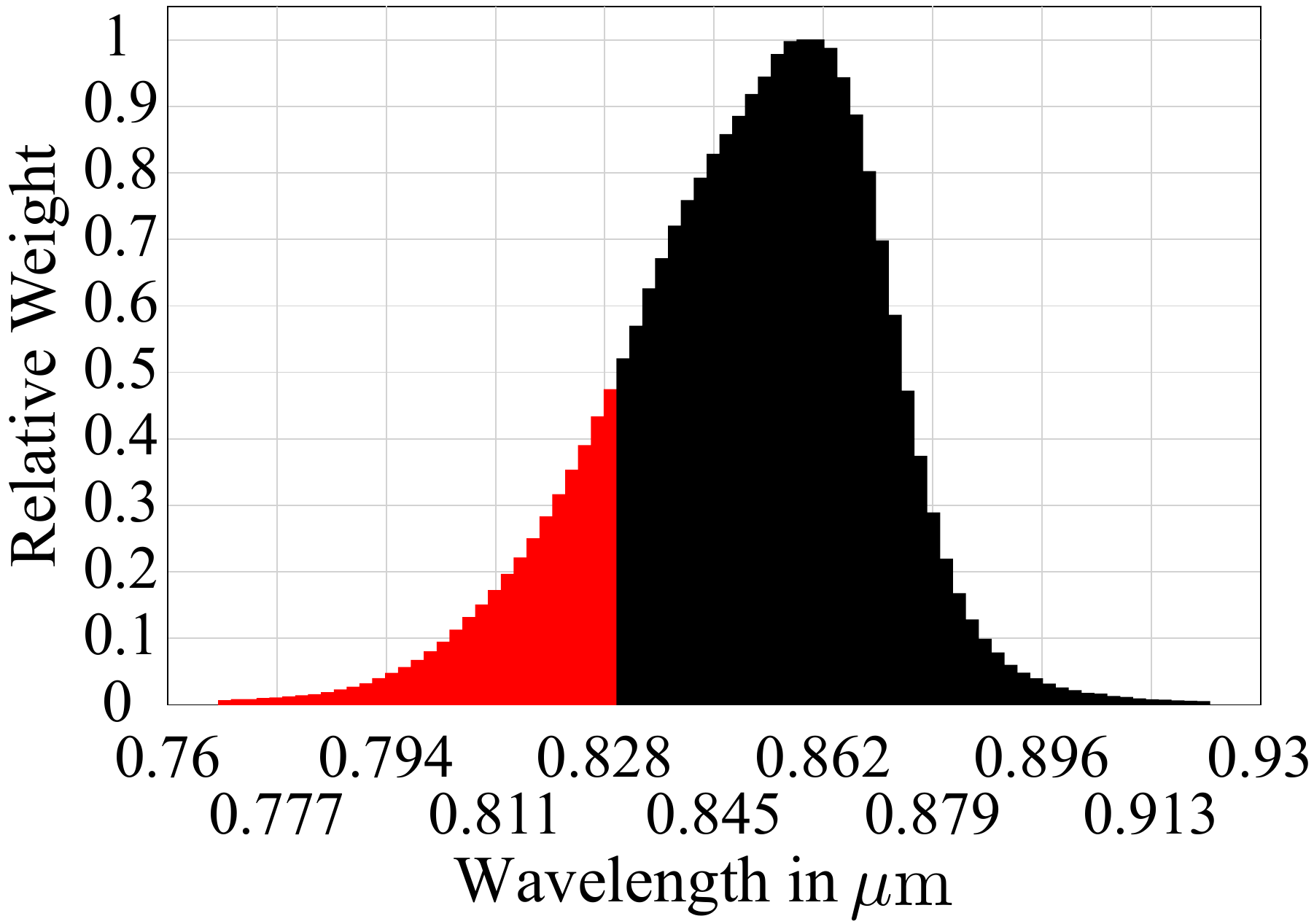}
		\caption{OSRAM SFH 4253}
		\label{fig:spec_IR}
	\end{subfigure}
	\caption{Relative radiometric colour spectrum of; (a) \glsentrytext{VL} $[0.382~0.780]~\mu\text{m}$ and (b) \glsentrytext{IR} $[0.770~0.920]~\mu\text{m}$ band \glsentrytext{LED} chips used in the MCRT simulations.}
	\label{fig:source_spectrum}
\end{figure}

To create an accurate model of the non-imaging light structure; (i) spatial, (ii) angular and (iii) spectral characterization and respective parameters are needed. The realistic spatio-angular characteristics of the sources are ensured by the manufacturer provided ray files which have been inputted into the simulation environment. The ray files contain a large number of emitted ray recordings, up to 5 million, to project source characteristics without requiring complex internal parameters. The emission patterns of the chosen sources are given by the source directivity plots provided in Fig. \ref{fig:source_directivity}. Accordingly, the angle axis shows the polar angle for a source located in the $+z$ direction and the colours represent different azimuthal angle scans, $0^\circ$, $45^\circ$, $90^\circ$ and $135^\circ$, for the spherical coordinate system. As can be seen from the figure, both sources have a Lambertian-like emission pattern with strong axial symmetry, which corresponds to \gls{FWHM} of $120^\circ$, in other words, half power semi angle of $\Phi_{\text{h}}=60^\circ$. However, note that the ideal diffuse (Lambertian) emitter model assumes a point source, whereas our realistic modelling contains spatial information about the source emitting profiles.

As our channel modelling technique operates in the granularity level of a \quot{light ray}, non-monochromatic sources must be broken into discrete wavelengths to encounter refraction, reflection, diffraction and absorption effects. The radiometric spectral characteristics of the sources could be defined in the simulation environment by \gls{SPCD} files. Accordingly, the \gls{SPCD} file format in OpticStudio contains the relative spectral distribution coefficients with their corresponding wavelengths. The values in the spectral file will be used to determine the relative frequency of the rays with the associated spectra. Hence, a higher relative weight will increase the likelihood that the emergence of a ray with a given wavelength in \gls{MCRT}. Since each ray carries a fraction of the total power of the source in \gls{MCRT}, the higher weight would result in a larger optical power in the given wavelength, which is also the case in real life measurements. The relative radiometric colour spectrum plots for the adopted sources are provided in Fig. \ref{fig:source_spectrum}. As can be seen from figure, the relative spectral distribution function of the adopted white (visible band) \gls{LED}, $f_\text{v}(\lambda)$, against wavelength $\lambda$, consists of two local maximums; at the $\lambda_S^\text{blue}=450$ nm and $\lambda_S^\text{yellow}=604$ nm, due to the blue base and yellow phosphor coating. The spectral characterization of the white source is defined by $66$ and $136$ measurement based data points for blue and yellow spectra, respectively. Furthermore, in our realistic model, the \gls{CCT} of the white \glspl{LED} is chosen to be 3000 K. Similarly, the relative radiometric spectral distribution for the adopted \gls{IR} source is given by Fig. \ref{fig:spec_IR}. The relative spectral distribution function of the \gls{IR} \gls{LED}, $f_\text{i}(\lambda)$, has a global maxima at the $\lambda_S^\text{IR}=860$ nm. The spectral characterization of the \gls{IR} source is also ensured by the $76$ spectral data points, where the further resolution is obtained by spline interpolation if needed.
\begin{figure}[!t]
	\centering
	\begin{subfigure}[t]{.5\columnwidth}
		\includegraphics[width=\columnwidth]{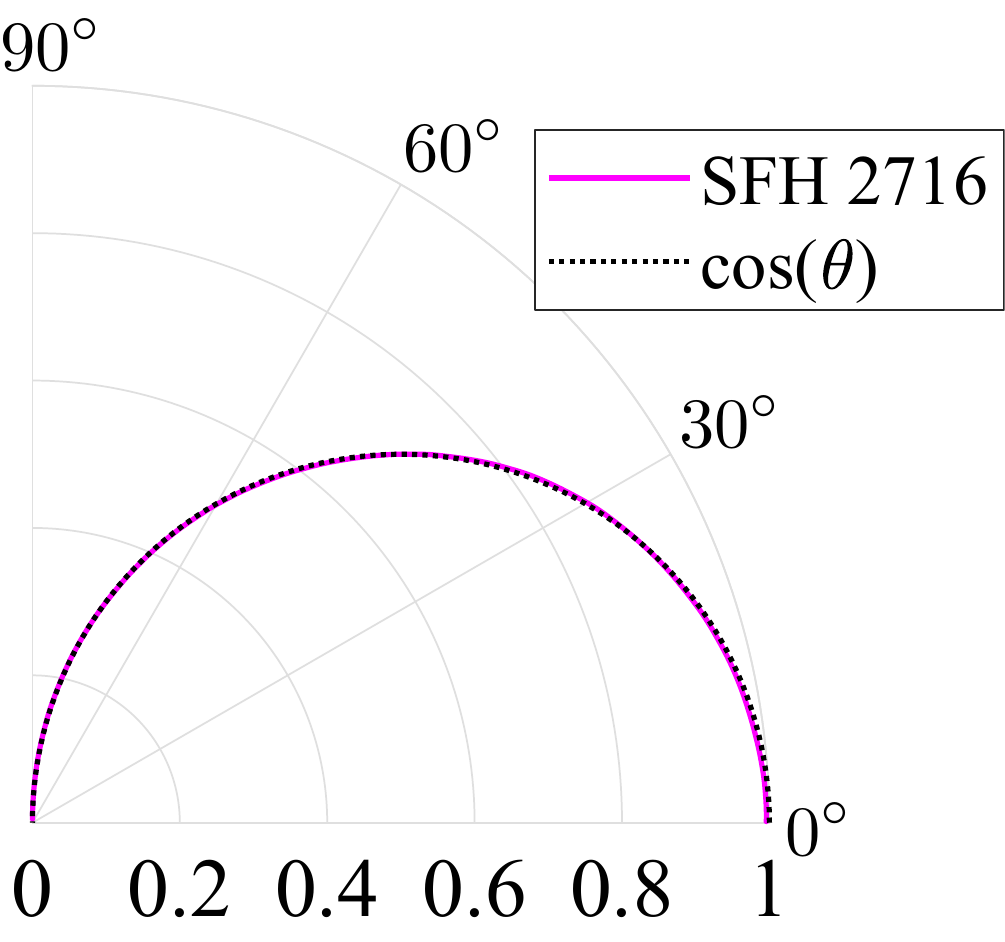}
		\caption{OSRAM SFH 2716}
		\label{fig:det_direct_VL}
	\end{subfigure}~
	\begin{subfigure}[t]{.5\columnwidth}
		\includegraphics[width=\columnwidth]{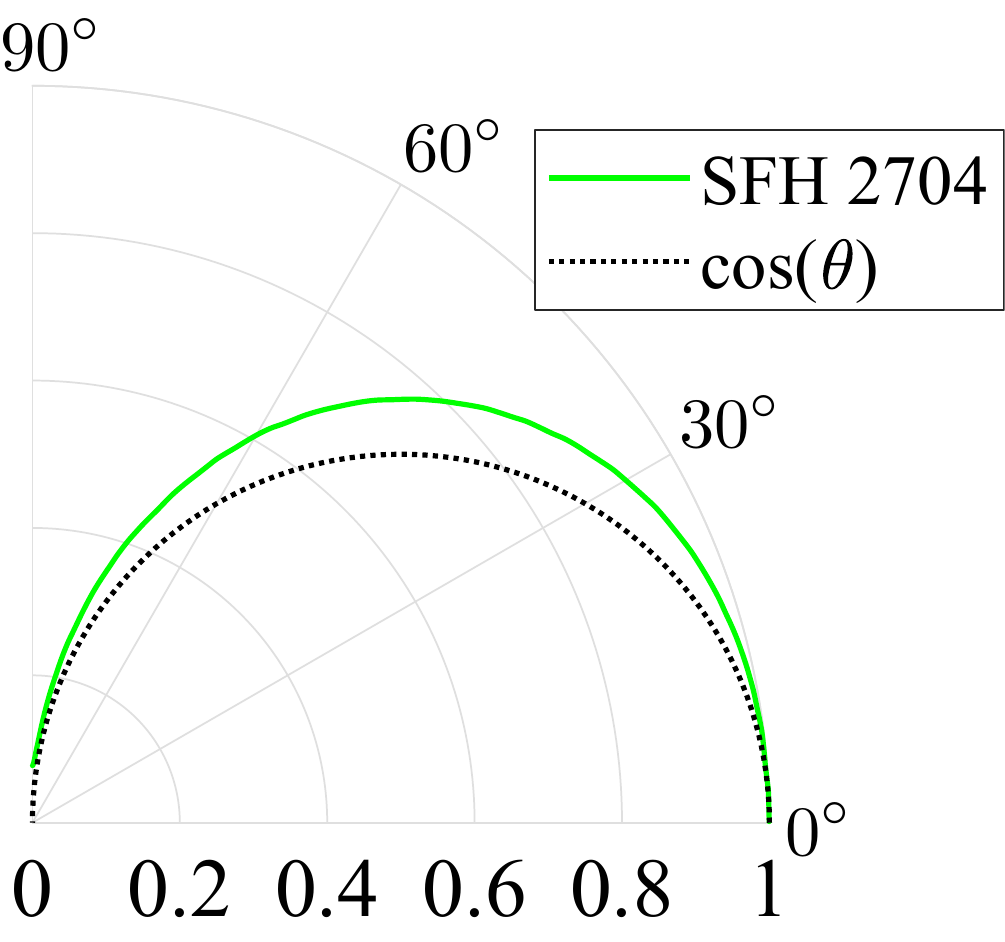}
		\caption{OSRAM SFH 2704}
		\label{fig:det_direct_IR}
	\end{subfigure}
	\caption{Relative angular responsivity characteristic plots for the adopted \glsentrytext{VL} (left) and \glsentrytext{IR} (right) band detectors. The ideal cosine responsivity curve is given by black dotted line as a benchmark.}
	\label{fig:detector_directivity}
\end{figure}
\begin{figure}[!t]
	\centering
	\begin{subfigure}[t]{.5\columnwidth}
		\includegraphics[width=\columnwidth]{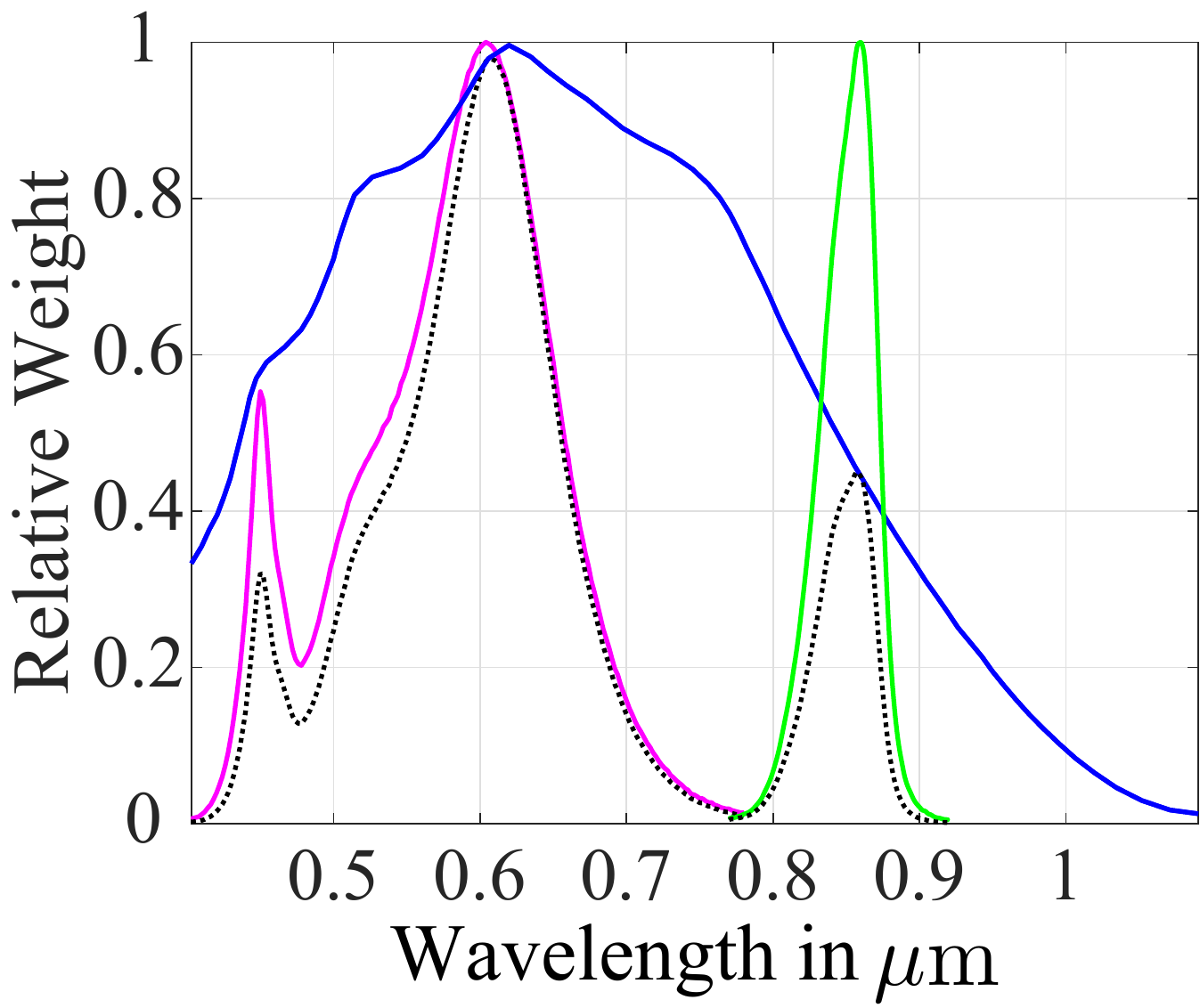}
		\caption{OSRAM SFH 2716 with $\lambda_R^\text{VL}=620$ nm (solid blue)}
		\label{fig:det_spec_VL}
	\end{subfigure}~
	\begin{subfigure}[t]{.5\columnwidth}
		\includegraphics[width=\columnwidth]{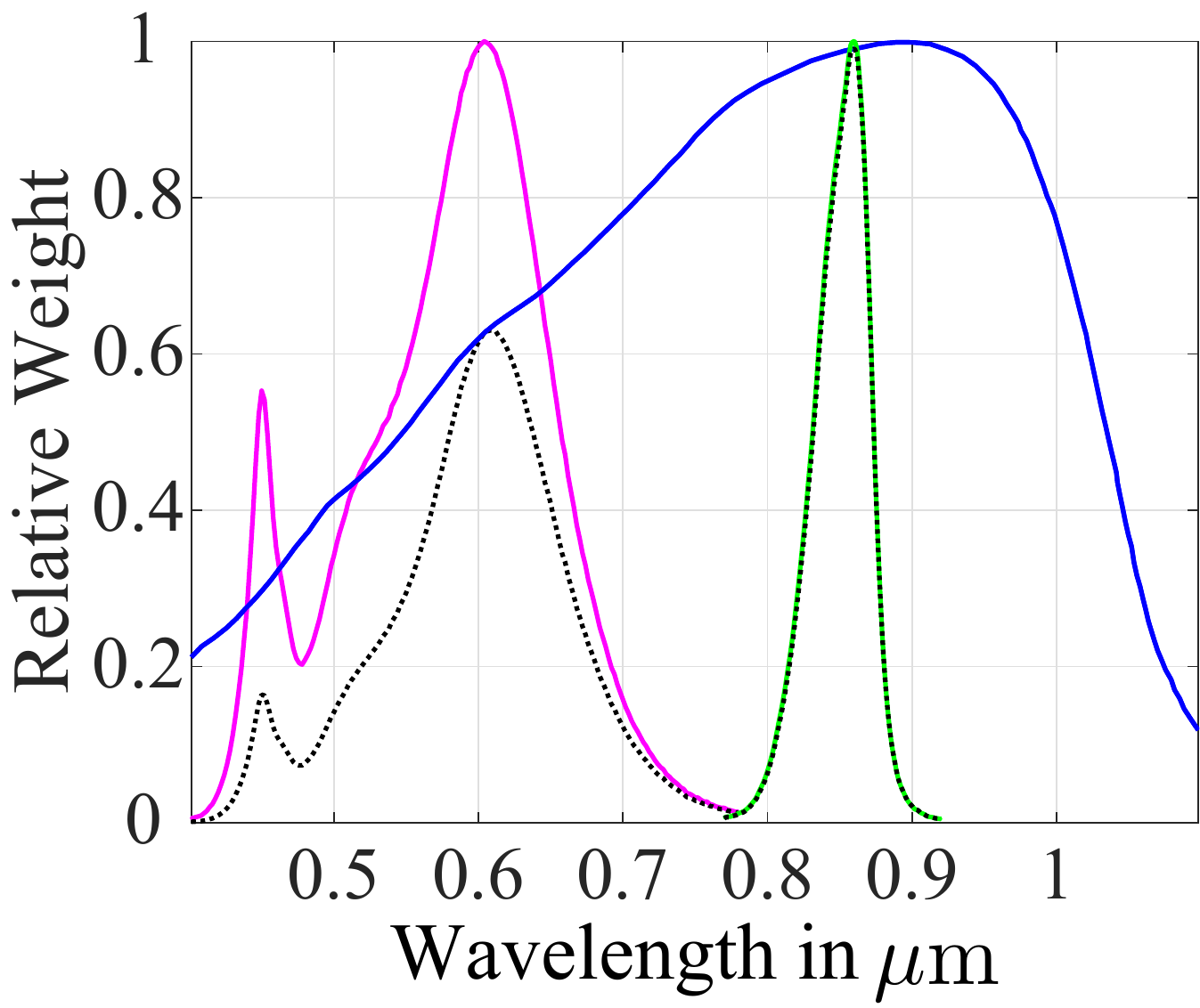}
		\caption{OSRAM SFH 2704 with $\lambda_R^{\text{IR}}=900$ nm (solid blue)}
		\label{fig:det_spec_IR}
	\end{subfigure}
	\caption{The relative spectral response curves for the adopted detectors, $g_\text{v}(\lambda)$ and $g_\text{i}(\lambda)$ (solid blue) with the relative spectral distributions for the OSRAM GW QSSPA1.EM, $f_\text{v}(\lambda)$ (solid magenta) and OSRAM SFH 4253, $f_\text{i}(\lambda)$ (solid green).}
	\label{fig:det_spec_VL_IR}
\end{figure}
\begin{figure}[!t]
	\centering
	\includegraphics[width=1.0\columnwidth]{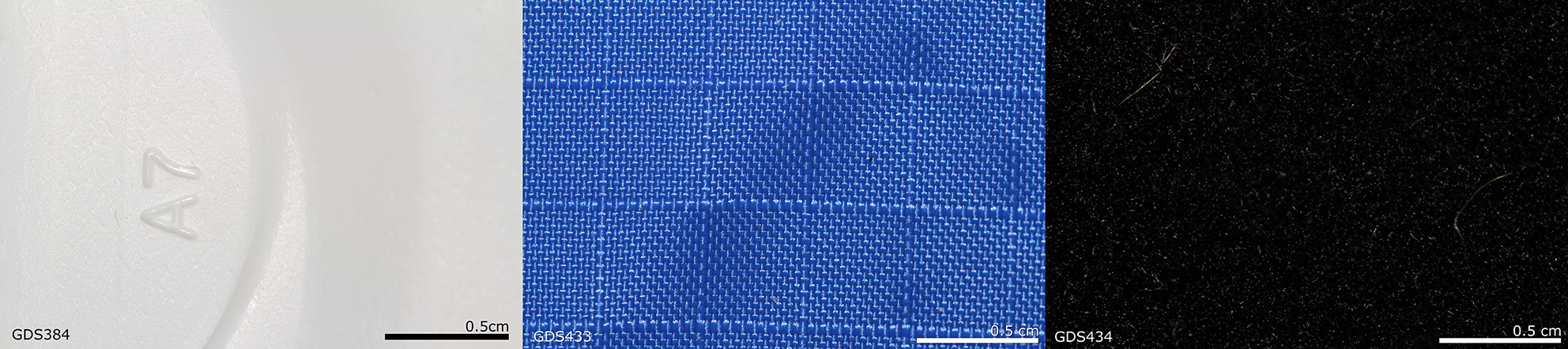}
	\caption{The coating materials used in the cabin interior design; white opaque plastic (left), blue ripstop nylon fabric (middle) and black polyester pile carpet (right) \cite{USGS}.}
	\label{fig:coating_materials}
\end{figure}
\begin{figure*}[!t]
	\centering
	\begin{subfigure}[t]{.56\columnwidth}
		\includegraphics[width=\columnwidth]{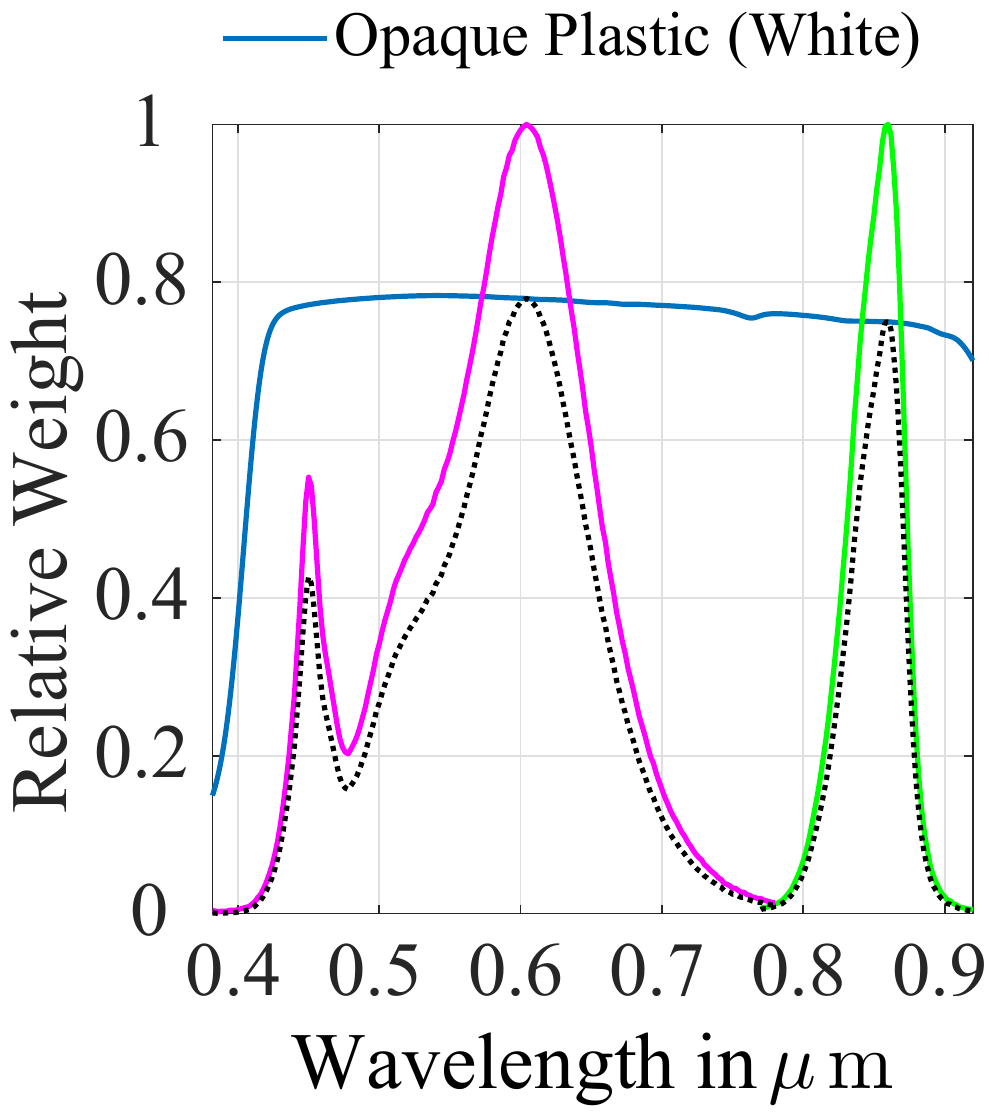}
		\caption{}
		\label{}
	\end{subfigure}~
	\begin{subfigure}[t]{.56\columnwidth}
		\includegraphics[width=\columnwidth]{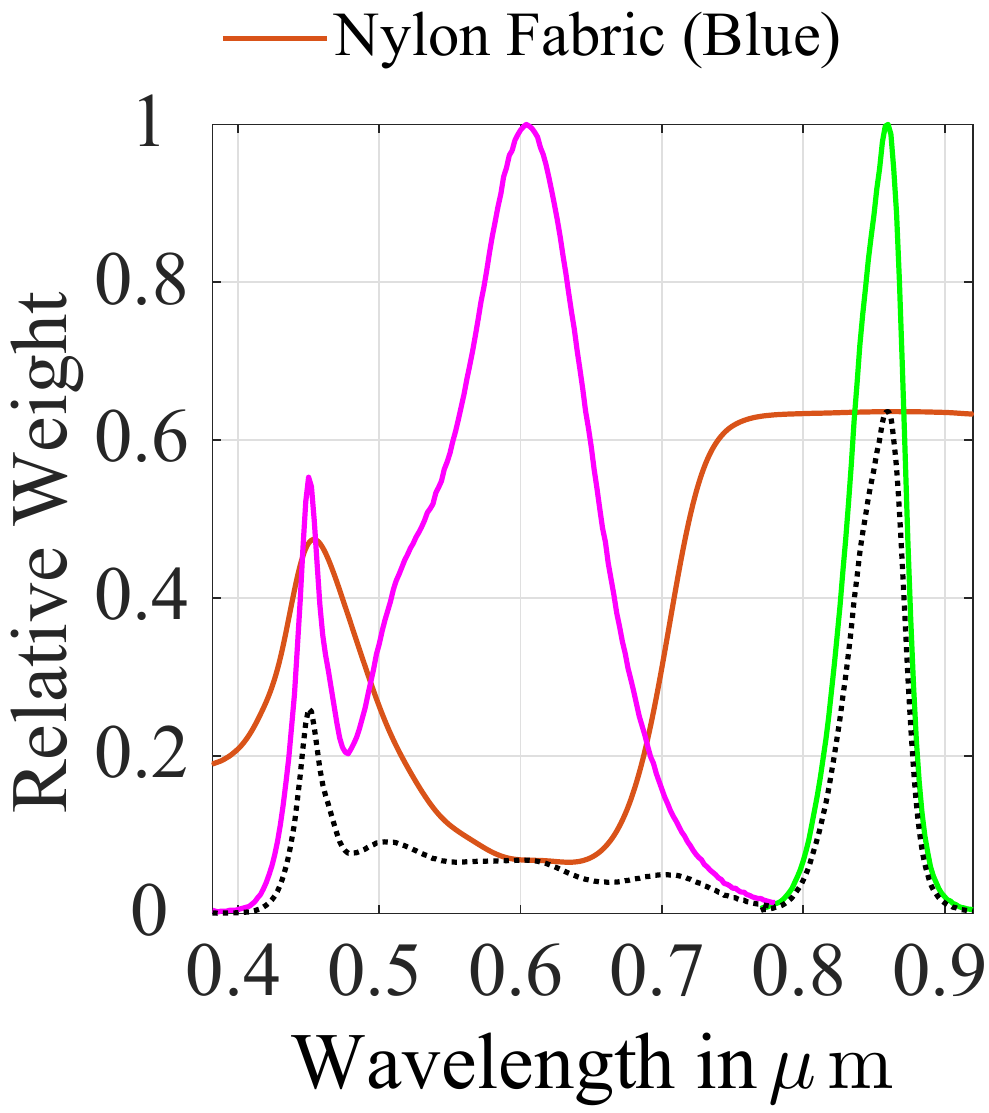}
		\caption{}
		\label{}
	\end{subfigure}~
	\begin{subfigure}[t]{.56\columnwidth}
		\includegraphics[width=\columnwidth]{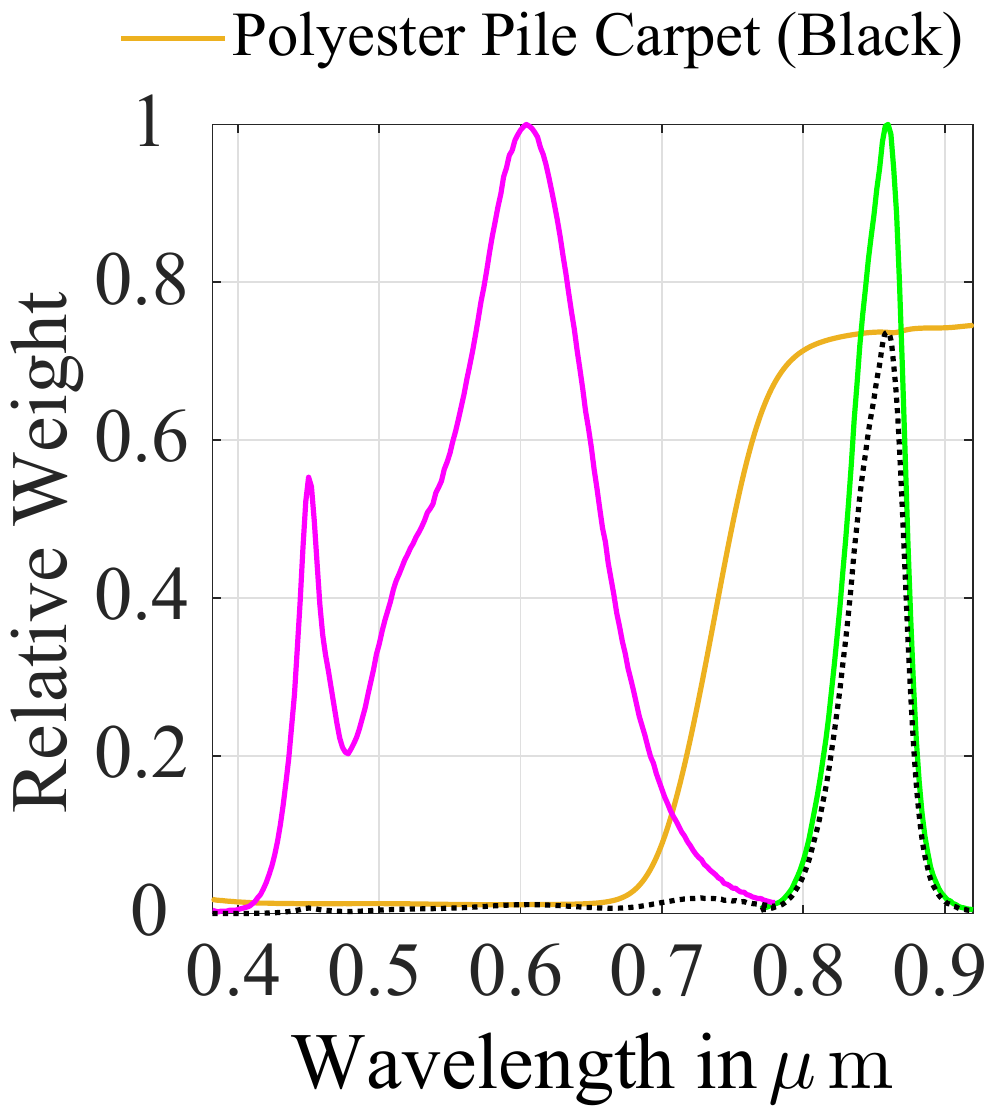}
		\caption{}
		\label{}
	\end{subfigure}
	\caption{Relative spectral reflectivity values of the coating materials with respect to the adopted VL and IR band source characteristics. The $f_{\text{v}}(\lambda)$ and $f_{\text{i}}(\lambda)$ are depicted as magenta and green solid lines, respectively. The resultant characteristics after multiplication of relative spectral reflectivity of the coating materials and relative spectral distributions of the sources are given by black dotted lines under the respective curves.}
	\label{fig:spectral_coating}
\end{figure*}
\subsubsection{Receivers}
Our \gls{MCRT} approach is capable of generating a complete optical channel analysis by taking the realistic receiver characteristics into consideration. Similar to the source modelling, spatial, angular and spectral specifications are also required for realistic detector design. As discussed in the previous subsection, the \gls{VL} and \gls{IR} band emission characteristics are adopted at the \gls{TX}. Thus, the utilization of two detectors is also required in order to match the spectral responsivity characteristics between the \glspl{PD} and their intended sources. In our simulations, two silicone PIN \glspl{PD}; OSRAM SFH 2716 with the peak sensitivity at $\lambda_R^\text{VL}=620$ nm and OSRAM SFH 2704 with the peak sensitivity at $\lambda_R^{\text{IR}}=900$ nm are adopted as the \gls{VL} and \gls{IR} band detectors, respectively. Accordingly, rectangular shaped bare (without an optical filter) \glspl{PD} with a $1~\text{cm}^2$ active area are created in the simulation environment. The relative angular responsivity curves of the OSRAM SFH 2716 and OSRAM SFH 2704 are depicted in Figs. \ref{fig:det_direct_VL} and \ref{fig:det_direct_IR}, respectively. In figures, the ideal cosine responsivity curve for a bare detector, in \cite{kb9701}, is also given as the benchmark. It can be inferred from Figs. \ref{fig:det_direct_VL} and \ref{fig:det_direct_IR} that the angular responsivity of our adopted \gls{IR} band detectors follow a slightly modified cosine profile. Note that the angular characteristics of both receive \glspl{PD} are defined by $100$ measurement based samples. Without loss of generality, the spectral and angular detection profile of the \glspl{PD} are assumed to be homogeneous \gls{w.r.t.} the spatial domain. Finally, the obtained profiles are inputted to our simulation environment to model the spatio-angular characteristics of the detector elements.

Similar to source spectra, the detector spectral characteristics are also fed into the simulation environment for the chosen detection devices. The relative spectral responsivity functions for OSRAM SFH 2716, $g_\text{v}(\lambda)$, and OSRAM SFH 2704, $g_\text{i}(\lambda)$, detectors, which are represented by $84$ and $136$ measured data points, are plotted in Figs. \ref{fig:det_spec_VL} and \ref{fig:det_spec_IR}, respectively. The solid blue lines in Figs. \ref{fig:det_spec_VL} and \ref{fig:det_spec_IR} depict the respective detector characteristics. Moreover, magenta and green lines represent the OSRAM GW QSSPA1.EM and OSRAM SFH 4253 relative spectral distribution functions which are denoted by $f_\text{v}(\lambda)$ and $f_\text{i}(\lambda)$, respectively. The overall spectral response of the \gls{VL} and \gls{IR} band detectors \gls{w.r.t.} the \gls{VL} and \gls{IR} band sources are calculated by $f_\text{v}(\lambda) g_\text{v}(\lambda)$, $f_\text{i}(\lambda) g_\text{i}(\lambda)$ and $f_\text{i}(\lambda) g_\text{v}(\lambda)$, $f_\text{i}(\lambda) g_\text{i}(\lambda)$, respectively and also plotted by the dotted black lines under the magenta and green curves. As can be seen from Figs. \ref{fig:det_spec_VL} and \ref{fig:det_spec_IR}, each source requires its matching/tuned detector at the \gls{RX} to harvest the highest collected power. Consequently, in our simulations \gls{VL}, $\lambda_S^\text{blue}=450$ nm and $\lambda_S^\text{yellow}=604$ nm, and \gls{IR}, $\lambda_S^\text{IR}=860$ nm, band sources are matched with their corresponding detectors with the peak wavelengths of $\lambda_R^\text{VL}=620$ nm and $\lambda_R^{\text{IR}}=900$ nm, respectively.
\subsubsection{Coating Materials}
The \gls{NLoS} path contributions within dense and complex environments play an important role in the overall channel conditions. In the \gls{IFOWC} literature, the majority of the research has been dedicated to the \gls{IR} spectrum sources and their relative reflection characteristics. However, the spectral profiles between the \gls{VL} and \gls{IR} band source-detector pair differ significantly in practice, which is depicted in Fig. \ref{fig:det_spec_VL_IR}. Specifically, the reflectivity of the coating materials within the \gls{IR} band become almost flat due to the inherent narrow-band characteristics of the \gls{IR} sources, where the optical power is concentrated in the vicinity of a single wavelength. However, the optical power of the \gls{VL} sources are spread across multiple wavelengths, generally blue and yellow or red, green and blue, which makes them wide-band sources. Hence, the effective optical properties of the interior surfaces \gls{w.r.t.} the \gls{VL} and \gls{IR} bands must be modelled accurately.
\begin{figure*}[!t]
	\centering
	\begin{subfigure}[t]{0.541926\columnwidth}
		\includegraphics[width=\columnwidth]{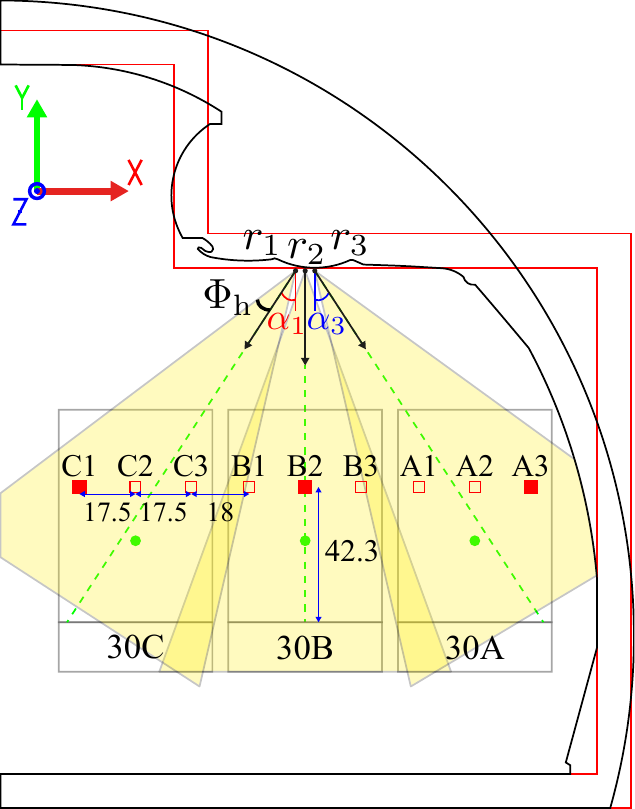}
		\caption{}
		\label{fig:sim_env_a}
	\end{subfigure}~
	\begin{subfigure}[t]{0.693\columnwidth}
		\includegraphics[width=\columnwidth]{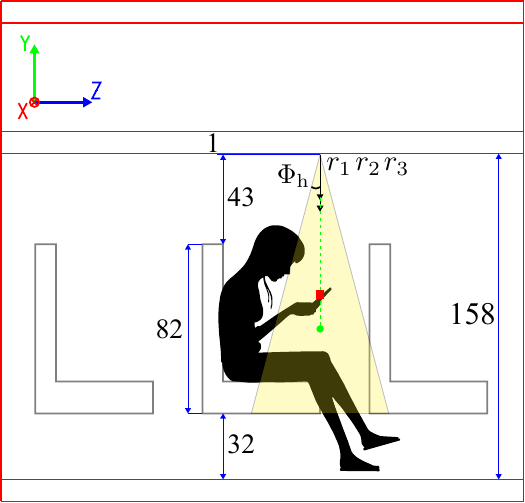}
		\caption{}
		\label{fig:sim_env_b}
	\end{subfigure}
	\caption{Details of the reading light based \glsentrytext{CIR} simulations; front (left) and side (right) views of the cabin (units are in cm). Green circles and red squares represent the centre point for the upper section of the seats and \glsentrytext{CIR} measurement points in each seat, respectively.}
	\label{fig:sim_env}
\end{figure*}

To capture the characteristics of realistic cabin interior coatings, the spectral reflectance measurement data is used in our simulations. Accordingly, the reflection and absorption characteristics of the aeronautical cabin interior coating profiles are obtained from the \gls{USGS} High Resolution Spectral Library Version 7 \cite{USGS}. The main surface coating for the cabin interior, including the side walls, ceiling and head luggage compartments, is typically chosen to be \gls{FRP} in real world applications to meet the flammability requirements. In our simulations, the light coloured plastic, depicted in Fig. \ref{fig:coating_materials}, is adopted to accurately imitate the characteristics of the main interior surfaces of the cabin. Thus, the white opaque plastic material and its optical properties are imported into the simulation environment. Note that the interior faces of both the realistic and simplified cabin models are assumed to be coated with the same material. Secondly, a blue coloured nylon ripstop fabric coating has been imported to model the real passenger seat dress covering, which is shown in Fig. \ref{fig:real_seat}. Lastly, the cabin flooring is chosen to be made out of black polyester pile carpet, to accurately model a real world commercial airline interior. The visual representations of the adopted coating materials and their spectral reflectivity characteristics are given in Figs. \ref{fig:coating_materials} and \ref{fig:spectral_coating}, respectively. In Fig. \ref{fig:spectral_coating}, the portion of the reflectivity spectrum that corresponds to the adopted source spectral distributions, $f_\text{v}(\lambda)$ and $f_\text{i}(\lambda)$, are given for the chosen materials separately. Similar to the source spectral plots, the result of a multiplication between the source characterization and the coating material reflectivity functions is plotted by dotted black lines under the respective source curve. It can easily be inferred from Fig. \ref{fig:spectral_coating} that the reflectivity characteristics of the white opaque plastic is almost flat for both \gls{VL} and \gls{IR} wavelength regions. Conversely, the reflectivity profiles of the blue ripstop nylon fabric and black polyester pile carpet are highly absorbent in the \gls{VL} band as opposed to the \gls{IR} band. Moreover, the effect of the coating materials on the overall response is also different from each other, as anticipated, since the source profiles are significantly contrasting. Note that the representation of coating characteristics by an average reflectance value, as vastly practised in the literature, would introduce a recursive error in the channel modelling. Explicitly, the error will be amplified in each reflection of a light ray during \gls{CIR} calculations. In our simulations, the coating materials are represented by $519$ data points, for the $0.382-0.780$ $\mu$m region to match the spectra of adopted \gls{VL} and \gls{IR} band sources.
\section{MCRT Simulations for In-flight LiFi Channels}
In this section, details of the \gls{MCRT} based channel simulation methodology will be provided. Then, a comprehensive time and frequency domain analyses on the obtained optical channels and the inferred results will be presented for different wavelengths, cabin models and \gls{UE} locations. The \gls{MCRT} simulations will be conducted for a practical in-flight \gls{LiFi} communications scenario, where the data bearing optical power is transferred from the reading lights to the \gls{PED}. Accordingly, a three dimensional simulation environment with aircraft cabin, auxiliary interior elements, sources, receivers and coating material models, as detailed in the previous section, are utilized in \gls{MCRT} simulations. The detailed illustration of the simulation environment and arrangement of the components is given in Fig. \ref{fig:sim_env}. As can be seen from Figs. \ref{fig:A320_realsimp_iso} and \ref{fig:sim_env}, both cabin models are centred along the $x$ and $y$-axes in simulations such that the simplified cabin represents the rough approximation of the complex geometric curvatures in the realistic model. Three reading lights, $r_1$, $r_2$ and $r_3$, are assumed to be mounted on the \gls{PSU} with a $3$ cm between them in the $x$-axis. To maximize the delivered optical power to their intended seats, the reading lamps are directed towards the centre of mass for the upper section of the seat, where the centre point is depicted as a green circle in Figs. \ref{fig:sim_env_a} and \ref{fig:sim_env_b}. Depending on the manufacturer and cabin model type, the reading light adjustment could be an option. However, due to space limitations, only the fixed adjustment is simulated in this work without loss of generality. To deliver the maximum optical power within each individual seating area for the fixed case, the selection of initial $z$-axis rotation values $\alpha_1$, $\alpha_2$ and $\alpha_3$, which are introduced to the $r_1$, $r_2$ and $r_3$, respectively, can be given by
\begin{align}
	\alpha_1 &\geq -\arctan(53/84)\approx -32.25^\circ, \nonumber \\ \alpha_2 &= \arctan(0/84)=0^\circ, \nonumber \\ \alpha_3 &\leq \arctan(53/84) \approx 32.25^\circ.
\end{align}
\noindent As can be seen from the expression above, the main objective of the reading lights is to provide sufficient illumination in the vicinity of the passenger tray table. To achieve that goal, we aim to deliver the maximum optical power to the centre point of an imaginary line which intersects the seats from top to bottom (refer to green circles in Fig. \ref{fig:sim_env}). For the sake of implementation simplicity, initial rotation angles of $\alpha_1=-33^\circ$, $\alpha_2=0^\circ$ and $\alpha_3=33^\circ$ are adopted in our \gls{MCRT} simulations by choosing the least integer angle greater than or equal to the given limits. The elevation of the receiver elements, depicted as red squares in Figs. \ref{fig:sim_env_a} and \ref{fig:sim_env_b}, is chosen with the aid of human models to represent the average height of the hand-held \glspl{PED}. A total of three mobile terminal locations, C1, B2 and A3, are chosen to investigate the \gls{LoS} and/or \gls{NLoS} channel effects as well as cell centre and edge performances. Unlike the imaging ray tracing applications, where the generated rays hit each surface only once in a premeditated sequence, non-imaging \gls{NSRT} is the main technique used in our simulations. The reason behind this is the illumination purposes of the reading lamps and unordered reflections from complex surface geometries with various reflectivity profiles. Also, the interior surfaces are designed to introduce spectrally dependent reflectivity and scattering if a ray strikes a surface. In every bounce of the light ray, the specular components of the reflections receive zero energy, which means the total energy is set to be equally divided among $\nu_\text{s}$ scattered rays. Thus, the number of total rays must be traced after $\kappa^\text{th}$ order reflection becomes, $n_\text{R}=(\nu_\text{s})^{\kappa}$. In the last iteration, the maximum number of rays that are traced by the simulator is calculated by, $n_\text{R,max}=(\nu_\text{s})^{\kappa_\text{max}}$. In our simulations, $\nu_\text{s}=5$ is chosen to accurately represent purely diffuse reflections in a computationally efficient manner. Another important \gls{MCRT} parameter is the \quot{minimum relative ray intensity}, which decides when to terminate the iteration of tracing a particular ray. Thus, it terminates a trace when the ray intensity is below or equal to a fraction of the initial intensity. In our simulations, the \quot{minimum relative ray intensity} is chosen as $10^{-4}$ and $10^{-5}$ for \gls{IR} and \gls{VL} spectra simulations, respectively. Note that the overall reflectivity of the coating materials in the \gls{VL} band is significantly lower compared to the \gls{IR}, thus, a lower minimum relative ray intensity value must be chosen to capture the details of the \gls{VL} channel. Therefore, the maximum number of reflections captured by our \gls{MCRT} simulations, $\kappa_\text{max}$, becomes  $4$ and $6$ for \gls{IR} and \gls{VL} bands, respectively. Lastly, the number of rays generated per \gls{LED} chip is chosen to be $5$ million in our simulations, which yields to a total of $80$ million rays per reading light in \gls{IR} band simulations. Since the generation of white light requires yellow and blue components, where each of which are chosen to be represented by $5$ million rays. Hence, the total of $160$ million rays are generated per white \gls{LED} based reading light.
\subsection{Optical Channel Characterization}
The multiple-bounce (multipath) \gls{CIR} between the source $S$ and the receiver $R$ could be expressed by our \gls{MCRT} simulation results as follows:
\begin{align}
	h(t;S,R)=\sum_{i=1}^{i_\text{hit}}P_i \delta(t-t_i),
	\label{eq:CIR}
\end{align}\noindent where $P_i$, $i_\text{hit}$ and $t_i$ denote the received incoherent irradiance, total number of rays that hit the receiver and the elapsed time for $i^\text{th}$ ray to reach the receiver, respectively. It is important to note that all the incoming rays that strike the detector surface with different irradiance values and time delays. In order to reduce the individual rays related observation errors and to obtain meaningful information from the ray scatter, data binning/clustering on $h(t;S,R)$ is applied, which yields the discrete-time optical (physical) \gls{CIR} as follows:
\begin{align}
	h[n;S,R]=\sum_{n=0}^{N_\text{b}-1}\widetilde{P}_n \delta(n-t_n), \quad \text{for } n \in \{ 0,1,\cdots,N_\text{b}-1 \}.
	\label{eq:CIR_discrete}
\end{align}\noindent Accordingly, the number of bins is given by 
\begin{align}
	N_\text{b}=\left\lceil \frac{t_\text{L}-t_1}{\Delta w} \right\rceil,
\end{align}
\noindent where the time of arrival for the first and last rays are denoted by $t_1$ and $t_\text{L}$, respectively. In addition, $\Delta w$ is width of the bins. Hence, the bin edge for the $n^\text{th}$ bin could also be given by $t_n=t_1 + n\Delta w$. Moreover, the cumulated irradiance value within the given bin interval is calculated by $\widetilde{P}_n=\sum\limits_{\forall i}P_i$, where $\forall i \in \left[ t_n~ t_{n+1} \right]$, if $n = N_\text{b}-1$ and $\forall i \in \left[ t_n~ t_{n+1} \right)$, otherwise. After obtaining the discrete-time \gls{CIR}, the important parameters for channel characterization could be devised. The optical \gls{CFR} is described in terms of the \gls{CIR} obtained in \eqref{eq:CIR_discrete} as follows:
\begin{align}
	H(f;S,R) = \int\limits_{-\infty}^{\infty}h(t;S,R)e^{-j 2\pi f t}\text{d}t \approx \sum\limits_{n=0}^{N_\text{b}-1}h[n;S,R]e^{-j\frac{2\pi kn}{N}},
	\label{eq:h_freqresp}
\end{align}\noindent where the sampling frequency could be calculated as $\Delta f = 1/\Delta w$. Hence, the frequency axis of the \gls{DFT}, becomes $k \in \frac{\Delta f}{N}\circ\left\{ -\frac{N}{2}, -\frac{N}{2}+1,\cdots,\frac{N}{2}-1 \right\}$. The number of subcarriers in the \gls{DFT} operation could also be found by $N=2^{\lceil \log_2  \left(N_\text{b}\right) \rceil}$. It is also important to note from the above expressions that the temporal domain accuracy is directly related to the bin width, $\Delta w$, where the resulting discrete-time \gls{CIR} closely approximates the actual channel when the bin width approaches zero, $\lim\limits_{\Delta w \rightarrow 0} h[n;S,R] \approx h(t;S,R)$. Another important parameter, the \gls{DC} channel gain or total optical power of the impulse response can also be calculated by using \eqref{eq:h_freqresp},
\begin{align}
	H(0;S,R)&=\int_{-\infty}^{\infty}h(t;S,R)\text{d}t \nonumber \\
&\approx\sum_{n=-\infty}^{\infty}h[n;S,R]=\sum\limits_{n=0}^{N_\text{b}-1}\sum\limits_{\kappa=0}^{\kappa_\text{max}}h^{(\kappa)}[n;S,R].
\end{align}\noindent By using the above expression, the average transmitted and received optical powers could be linked by using the \gls{DC} channel gain as follows: $P_R=H(0;S,R)P_S$, which yields the path loss of
\begin{align}
	\text{PL} = -10\log_{10}H(0;S,R)\quad \text{in dB}.
\end{align}
\noindent The \gls{RMS} delay spread and mean delay are two important measures to define the multipath richness of the channel, which also shows the impact of \gls{ISI} on the system performance. Hence, the \gls{RMS} delay spread could be calculated by using the second and zeroth moments of $h[n;S,R]$ as follows:
\begin{align}
	\tau_\text{RMS}=\sqrt{\frac{\int\limits_{-\infty}^{\infty}(t-\bar{\tau})^2 h^2(t;S,R)\text{d}t}{\int\limits_{-\infty}^{\infty}h^2(t;S,R)\text{d}t}} = \sqrt{\frac{\sum\limits_{n=0}^{N_\text{b}-1}(n-\bar{\tau})^2 h^2[n;S,R]}{\sum\limits_{n=0}^{N_\text{b}-1}h^2[n;S,R]}},
	\label{eq:RMS_delayspread}
\end{align}\noindent where the mean delay is given in terms of the zeroth and first raw moments of $h[n;S,R]$ by
\begin{align}
	\bar{\tau}=\frac{\int\limits_{-\infty}^{\infty}th^2(t;S,R)\text{d}t}{\int\limits_{-\infty}^{\infty}h^2(t;S,R)\text{d}t} =\frac{\sum\limits_{n=0}^{N_\text{b}-1}nh^2[n;S,R]}{\sum\limits_{n=0}^{N_\text{b}-1}h^2[n;S,R]}
	\label{eq:mean_delay}
\end{align}\noindent Lastly, the dominance of the \gls{LoS} link is also an important factor to evaluate the contribution of \gls{LoS} and \gls{NLoS} paths, which indicates how flat the channel is. Accordingly, the \quot{flatness factor} is calculated as the ratio between \gls{LoS} and \gls{NLoS} channel powers as follows:
\begin{align}
\rho &= \frac{P_{\text{LoS}}}{P_{\text{LoS}}+P_{\text{NLoS}}}=\frac{\int\limits_{-\infty}^{\infty}h^{(0)}(t;S,R)\text{d}t}{\sum\limits_{\kappa=0}^{\kappa_\text{max}}\int_{-\infty}^{\infty}h^{(\kappa)}(t;S,R)\text{d}t} =\frac{\sum\limits_{n=0}^{N_\text{b}-1}h^{(0)}[n;S,R]}{H(0;S,R)}.
\end{align}
\begin{table}[!t]
	\centering 
	\resizebox{1.0\linewidth}{!}{
		\renewcommand{\arraystretch}{1.3} 
		\begin{tabular}{c|c}
			\hline
			Cabin Models & \begin{tabular}[c]{@{}c@{}}A320 simplified \\ A320 realistic\end{tabular} \\ \hline
			Cabin Model Positions (cm) & \begin{tabular}[c]{@{}c@{}}$\mathbf{p}_{\text{simp}}=(0,~0,~0)$ \\ $\mathbf{p}_{\text{real}}=(6.577,~0,~0)$\end{tabular} \\ \hline
			Seat Model Positions (cm) & \begin{tabular}[c]{@{}c@{}}
				$\mathbf{p}_{\text{31C}}=(215.199,~42.643,~270)$ \\ $\mathbf{p}_{\text{31B}}=(268.199,~42.643,~270)$ \\ $\mathbf{p}_{\text{31A}}=(321.199,~42.643,~270)$ \\
				$\mathbf{p}_{\text{30C}}=(215.199,~42.643,~351)$ \\ $\mathbf{p}_{\text{30B}}=(268.199,~42.643,~351)$ \\ $\mathbf{p}_{\text{30A}}=(321.199,~42.643,~351)$ \\ 
				$\mathbf{p}_{\text{29C}}=(215.199,~42.643,~432)$ \\ $\mathbf{p}_{\text{29B}}=(268.199,~42.643,~432)$ \\ $\mathbf{p}_{\text{29A}}=(321.199,~42.643,~432)$ \end{tabular} \\ \hline
			Reading Light Positions (cm) & \begin{tabular}[c]{@{}c@{}}$\mathbf{p}_{r_1}=(289.199,~167.643,~408)$ \\ $\mathbf{p}_{r_2}=(292.199,~167.643,~408)$ \\ $\mathbf{p}_{r_3}=(295.199,~167.643,~408)$\end{tabular} \\ \hline
			Reading Light $z$-axis Orientations & \begin{tabular}[c]{@{}c@{}}$\alpha_1=-33^\circ$ \\ $\alpha_3=33^\circ$\end{tabular} \\ \hline
			Number of Chips per Reading Light & $16$ ($4\times 4$) \\ \hline
			\begin{tabular}[c]{@{}c@{}}Number of Generated Rays per \\ LED Chip / Reading Light\end{tabular} & \begin{tabular}[c]{@{}c@{}}VL Band: $10\times 10^6$ / $160\times 10^6$ \\ IR Band: $5\times 10^6$ / $80\times 10^6$ \end{tabular} \\ \hline
			Power per Reading Light ($P_\text{t}$) & $16$ W \\ \hline
			Model of the LED Chips & \begin{tabular}[c]{@{}c@{}}VL Band: OSRAM GW QSSPA1.EM \\ IR Band: OSRAM SFH 4253 \end{tabular} \\ \hline
			FWHM of the LED Chips & $120^\circ$ \\ \hline
			PD Positions (cm) & \begin{tabular}[c]{@{}c@{}}
				$\mathbf{p}_{\text{C1}}=(221.699,~100.443,~408)$ \\ $\mathbf{p}_{\text{C2}}=(239.199,~100.443,~408)$ \\ $\mathbf{p}_{\text{C3}}=(256.699,~100.443,~408)$ \\
				$\mathbf{p}_{\text{B1}}=(274.699,~100.443,~408)$ \\ $\mathbf{p}_{\text{B2}}=(292.199,~100.443,~408)$ \\ $\mathbf{p}_{\text{B3}}=(309.699,~100.443,~408)$ \\ 
				$\mathbf{p}_{\text{A1}}=(327.699,~100.443,~408)$ \\ $\mathbf{p}_{\text{A2}}=(345.199,~100.443,~408)$ \\ $\mathbf{p}_{\text{A3}}=(362.699,~100.443,~408)$ \end{tabular} \\ \hline
			Model of the PDs & \begin{tabular}[c]{@{}c@{}}VL Band: OSRAM SFH 2716 \\ IR Band: OSRAM SFH 2704\end{tabular} \\ \hline
			Effective Area of the PDs & $1~\text{cm}^2$ \\ \hline
			FWHM of the PDs & \begin{tabular}[c]{@{}c@{}}OSRAM SFH 2716: $120^\circ$ \\ OSRAM SFH 2704: $132^\circ$\end{tabular} \\ \hline
			Coating Materials & \begin{tabular}[c]{@{}c@{}}White Opaque Plastic ($\nu_\text{s}=5$)\\ Blue Ripstop Nylon ($\nu_\text{s}=5$)\\ Black Polyester Pile Carpet ($\nu_\text{s}=5$)\end{tabular} \\ \hline
			Time Resolution/ Bin Width ($\Delta w$) & $0.2$ ns \\ \hline
		\end{tabular}
	}
	\caption{Details of the parameters used in the MCRT simulations.}
	\label{table:MCRT_param}
\end{table}
\subsection{MCRT Simulation Results}
In this subsection, the proposed \gls{MCRT} simulation results will be presented for both \gls{IR} and \gls{VL} spectra. The complete set of parameters used in these simulations are provided in Table \ref{table:MCRT_param}.
\subsubsection{IR Band Results}
\begin{figure*}[!t]
	\centering
	\begin{subfigure}[t]{.54\columnwidth}
		\includegraphics[width=\columnwidth]{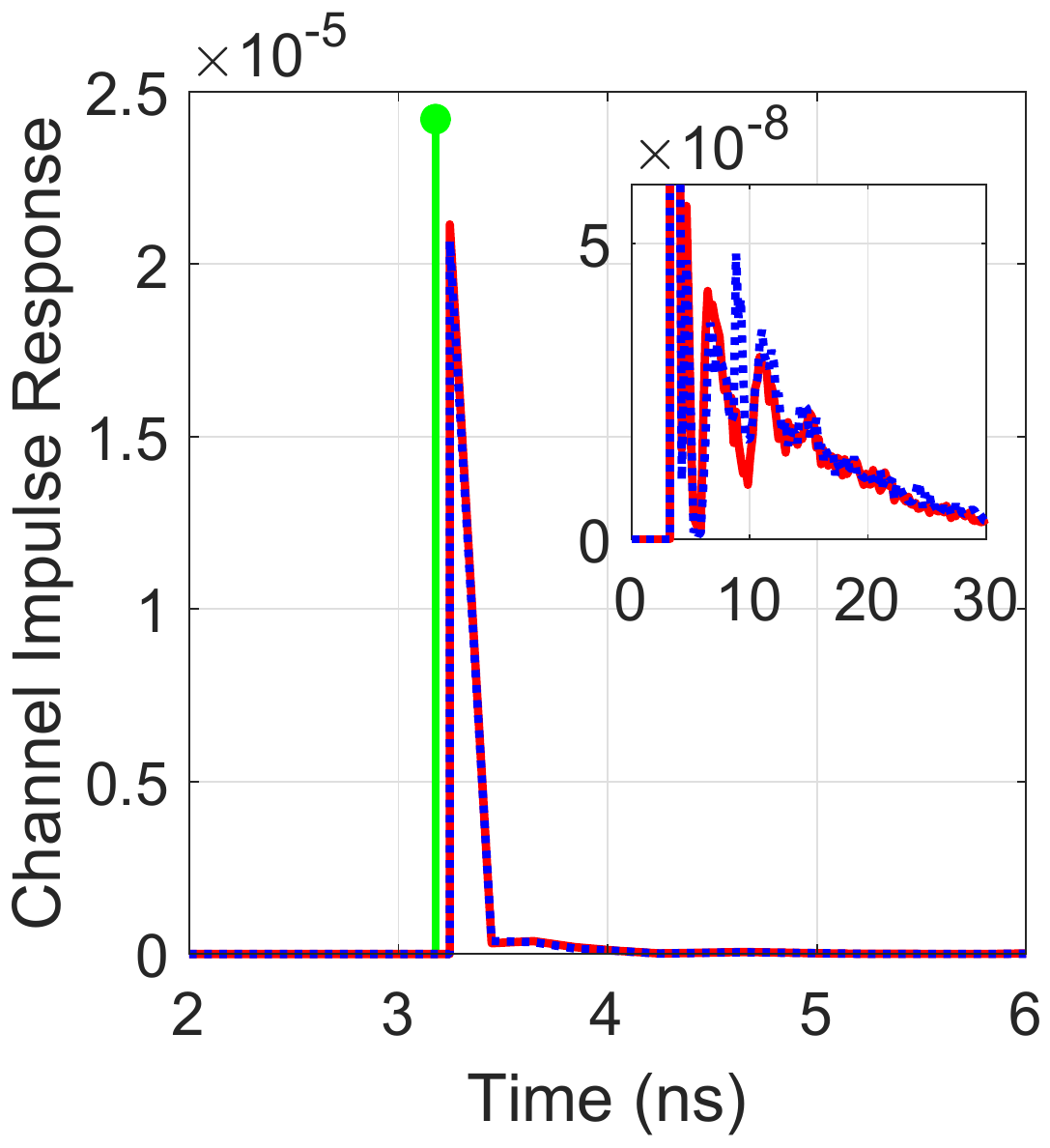}
		\caption{$h(t;r_1,\text{C1})$}
		\label{}
	\end{subfigure}~
	\begin{subfigure}[t]{.54\columnwidth}
		\includegraphics[width=\columnwidth]{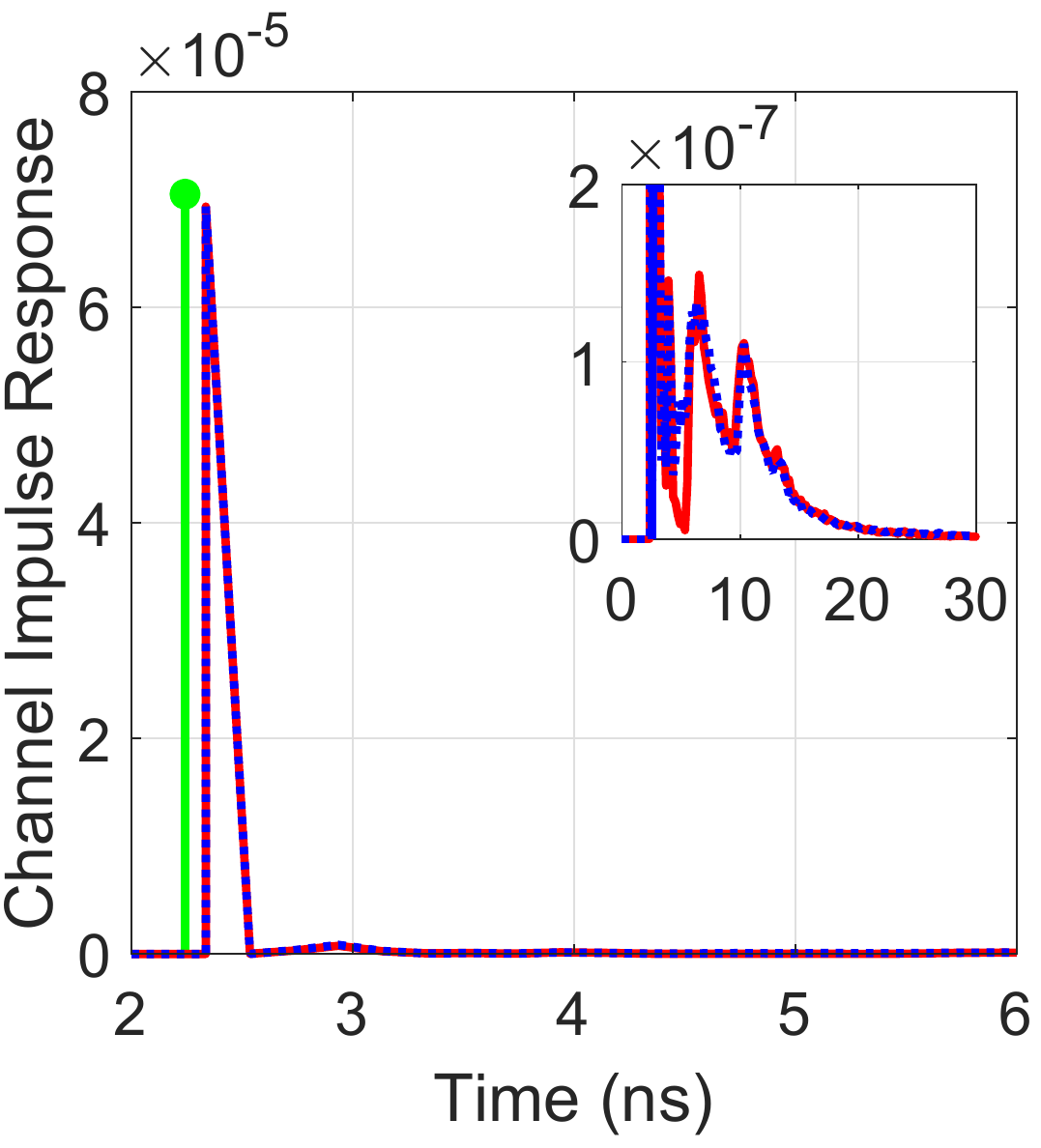}
		\caption{$h(t;r_2,\text{B2})$}
		\label{}
	\end{subfigure}~
	\begin{subfigure}[t]{.54\columnwidth}
		\includegraphics[width=\columnwidth]{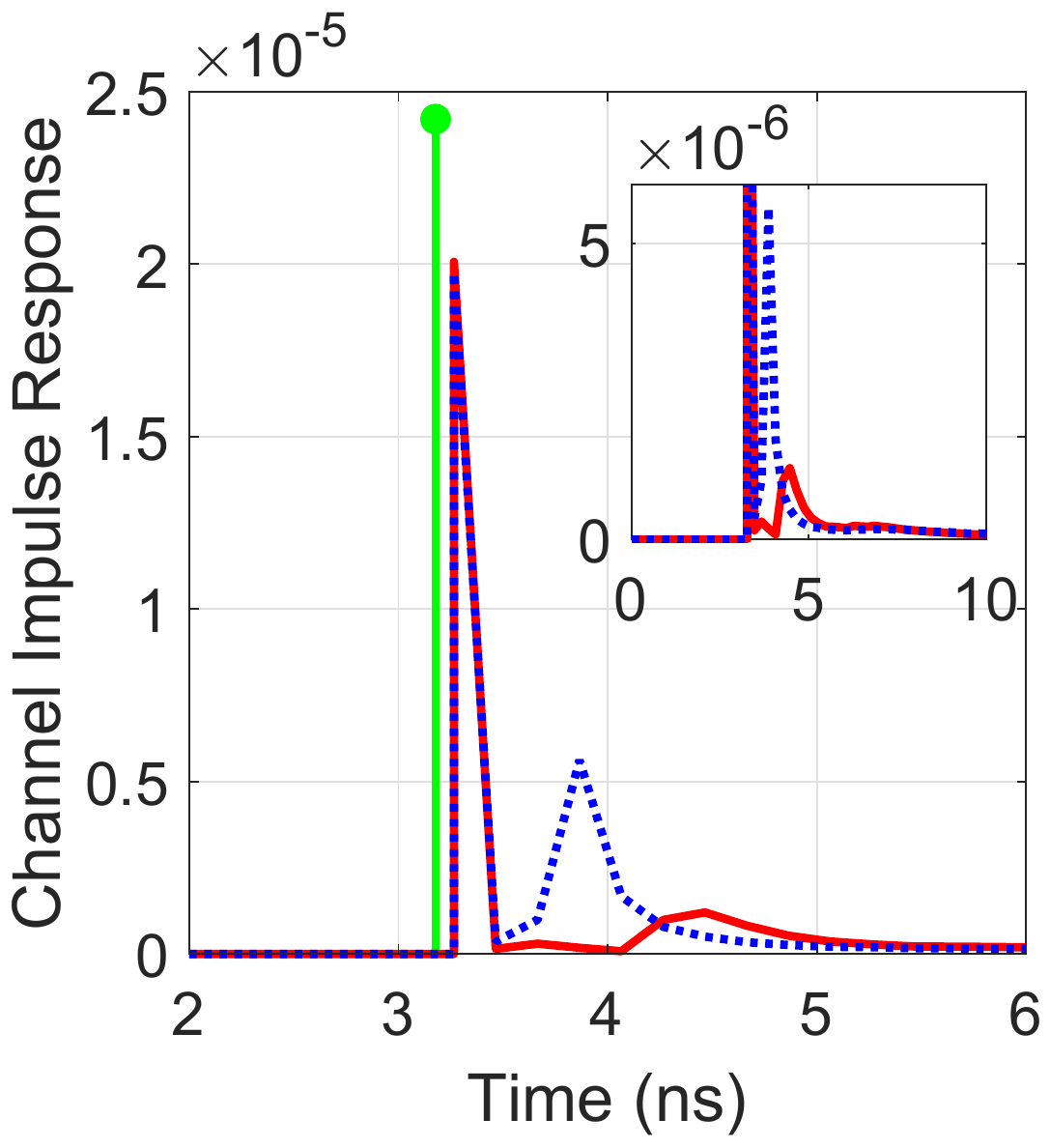}
		\caption{$h(t;r_3,\text{A3})$}
		\label{}
	\end{subfigure}\\
	\begin{subfigure}[t]{.54\columnwidth}
		\includegraphics[width=\columnwidth]{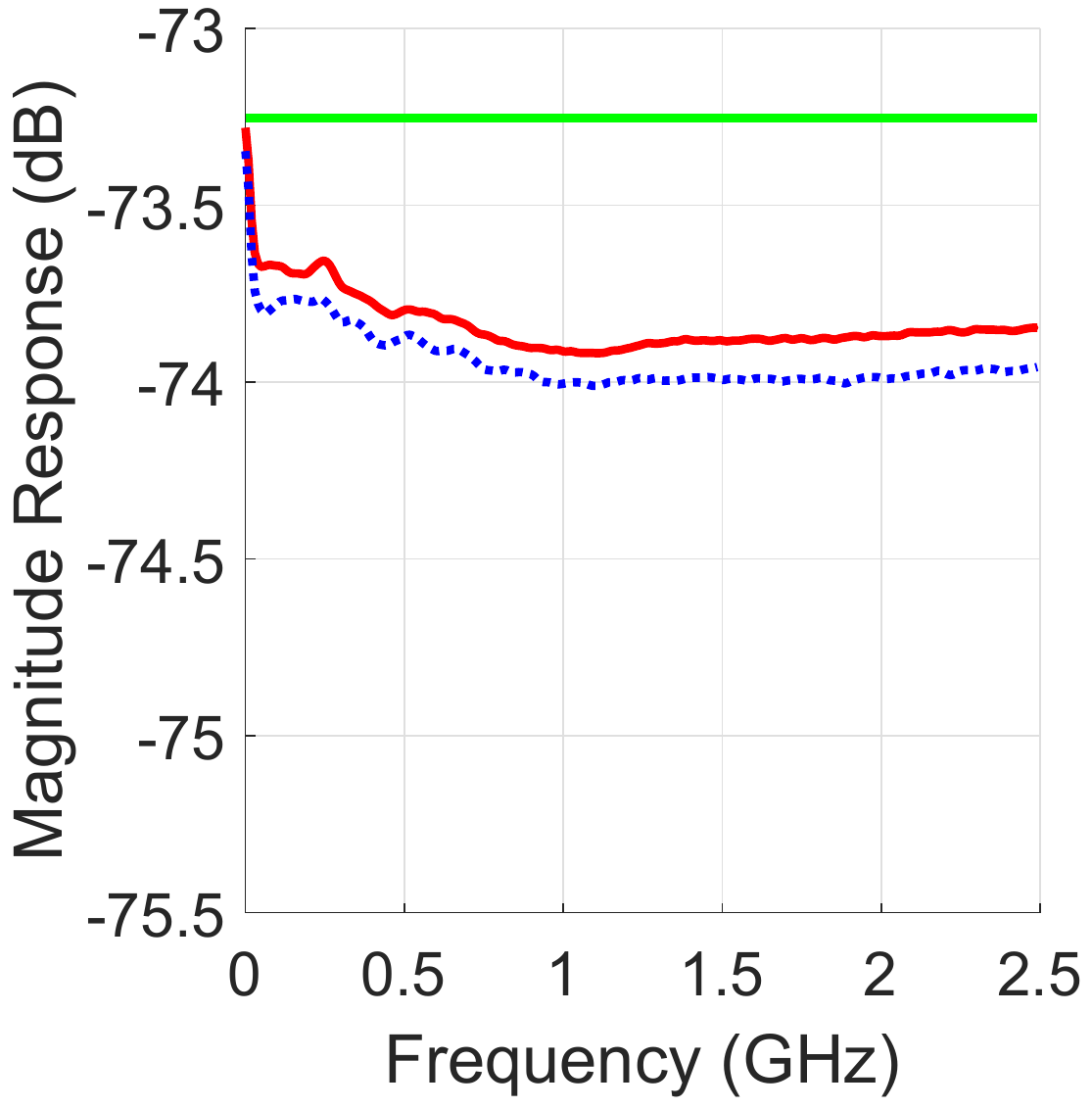}
		\caption{$\lvert H(f;r_1,\text{C1}) \rvert$}
		\label{}
	\end{subfigure}~
	\begin{subfigure}[t]{.54\columnwidth}
		\includegraphics[width=\columnwidth]{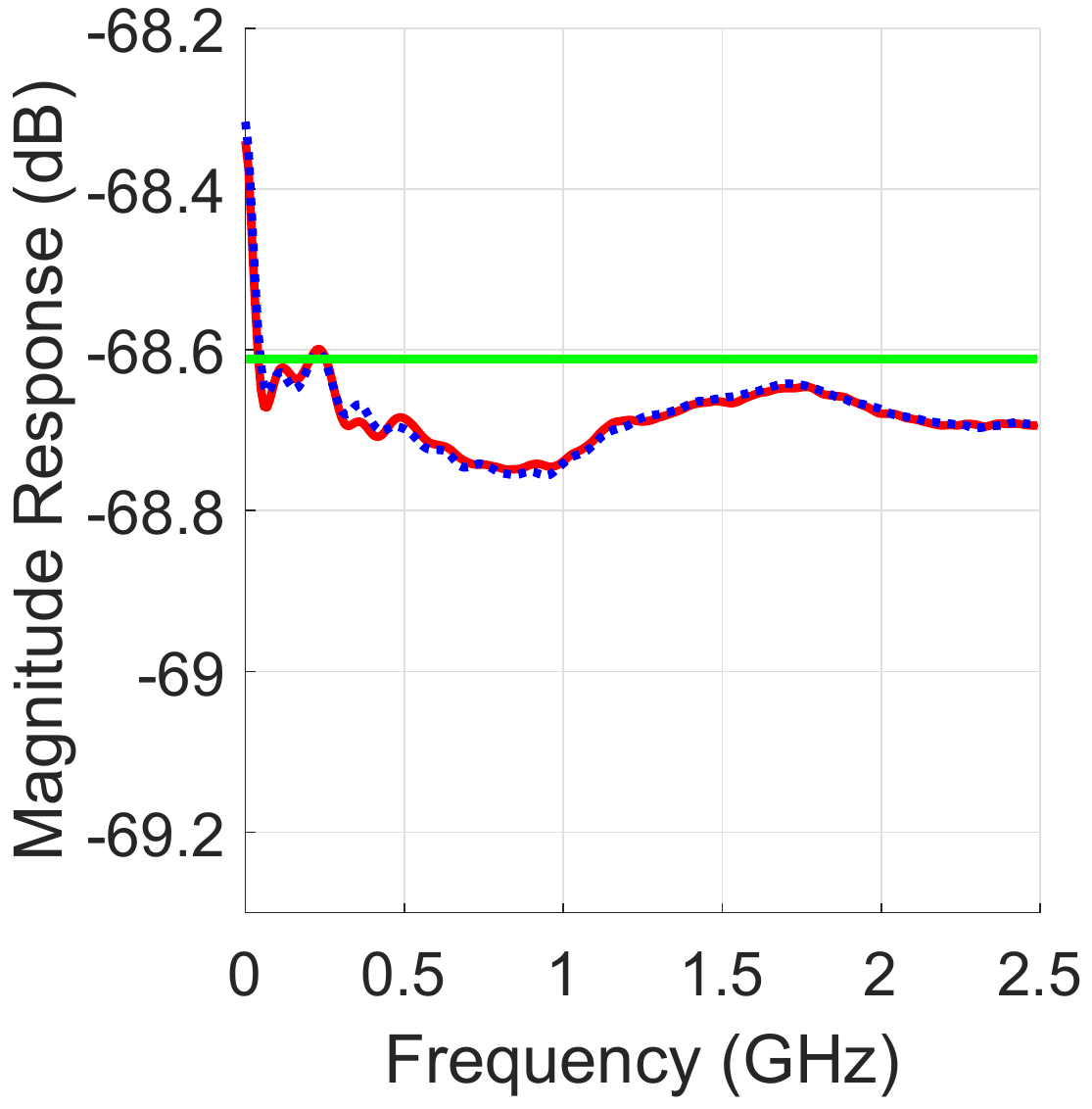}
		\caption{$\lvert H(f;r_2,\text{B2}) \rvert$}
		\label{}
	\end{subfigure}~
	\begin{subfigure}[t]{.54\columnwidth}
		\includegraphics[width=\columnwidth]{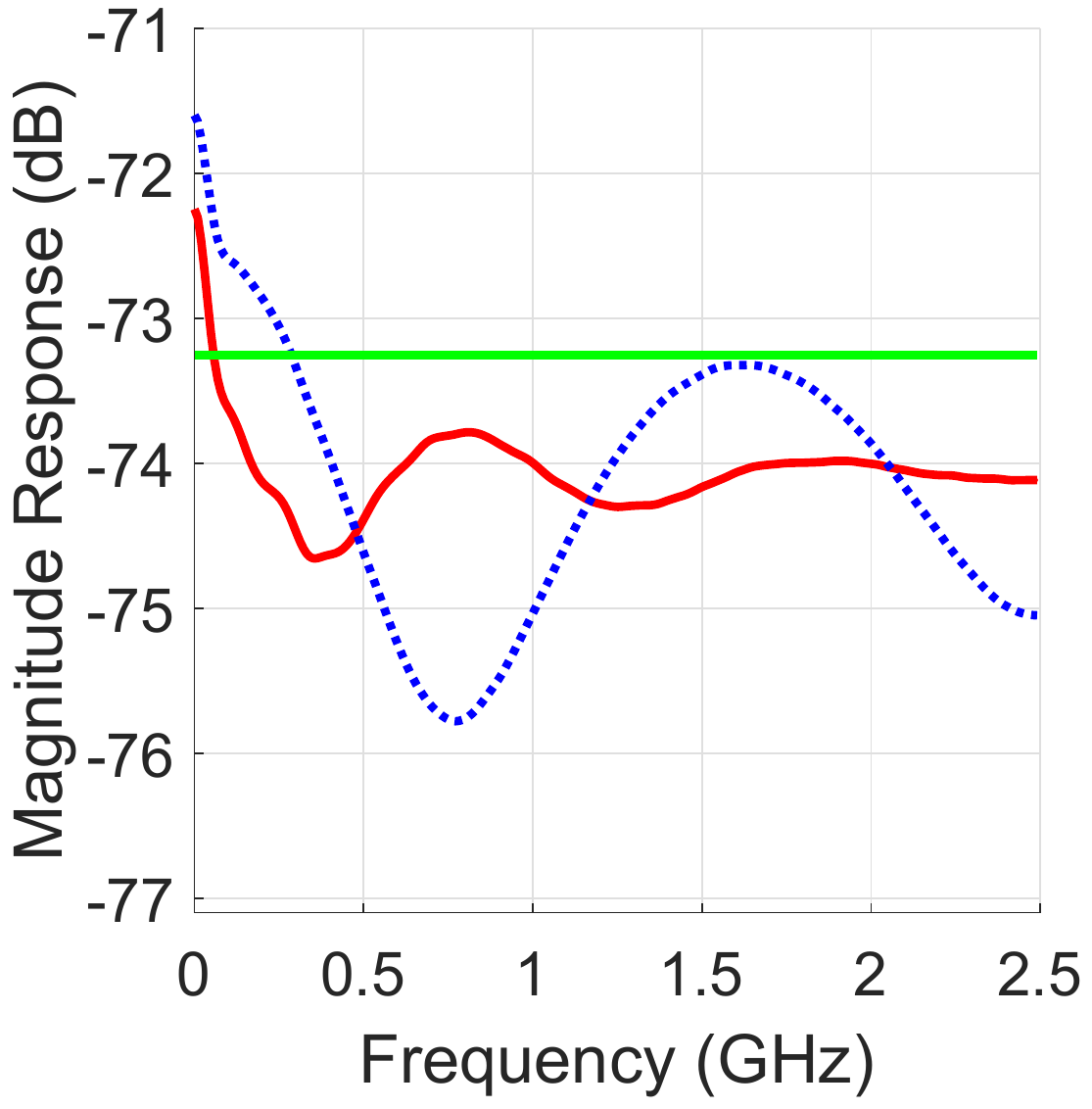}
		\caption{$\lvert H(f;r_3,\text{A3}) \rvert$}
		\label{}
	\end{subfigure}
	\caption{IR band in-flight LiFi CIR and CFR simulation results obtained by proposed MCRT based method for seated simplified cabin model (\protect\includegraphics[height=0.2cm]{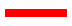}) and seated realistic cabin model (\protect\includegraphics[height=0.2cm]{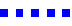}). The analytical LoS channel (\protect\includegraphics[height=0.2cm]{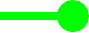}) is presented as a benchmark.}
	\label{fig:CIR_IR_seated}
\end{figure*}

The \gls{IR} band \gls{CIR} and \gls{CFR} plots for both realistic and simplified cabin models including the passenger seating are given in Fig. \ref{fig:CIR_IR_seated}. The rows in $2\times3$ subplot matrix represent the domain of the plot e.g., time or frequency, where the columns are the three measurement points for seats A, B and C. Moreover, the channel characterization parameters are also obtained and presented in Table \ref{table:CIR_IR_seated} in conjunction with the figures, where the number of rays that hit each \glspl{PD} is given by $i_\text{hit}$. The time and frequency domain simulation results for the adopted cabin models are presented and compared with \gls{LoS} analytical model in Figs. \ref{fig:CIR_IR_seated}(a)-(f). Accordingly, up to $4$-bounce, $0\leq\kappa\leq 4$, \gls{CIR} simulation results, $h(t;r_1,\text{C1})$, $h(t;r_2,\text{B2})$ and $h(t;r_3,\text{A3})$, are presented for adopted \gls{UE} locations C1, B2 and A3, respectively. Moreover, the frequency response plots are obtained for \gls{FFT} size of $N=512$ for the channel bandwidth of $B_\text{CH}=5$ GHz. Note that the Figs. \ref{fig:CIR_IR_seated}(d)-(f) are only presented for the frequency range of $f~\in[0~B_\text{CH}/2]$, since the frequency response of the \gls{LiFi} data signals are conjugate symmetric due to the \gls{IM/DD} technique.
\begin{table*}[!t]
	\centering 
	\resizebox{0.75\linewidth}{!}{
		\renewcommand{\arraystretch}{1.3} 
		\begin{tabular}{|c|c|c|c|c|c|c|c|c|}
			\hline
			\textbf{} & \multicolumn{4}{c|}{\textbf{Simplified Cabin (IR Band)}} & \multicolumn{4}{c|}{\textbf{Realistic Cabin (IR Band)}} \\ \hline
			$S,R$ & $i_\text{hit}$ & $H[0;S,R]$ & $\tau_\text{RMS}$ (ns) & $\rho$ & $i_\text{hit}$ & $H[0;S,R]$ (W) & $\tau_\text{RMS}$ (ns) & $\rho$ \\ \hline
			$r_1,\text{C1}$ & $101430$ & $2.405\text{E}^{-5}$ & $0.084$ & $0.859$ & $102793$ & $2.369\text{E}^{-5}$ & $0.094$ & $0.872$ \\ \hline
			$r_2,\text{B2}$ & $192648$ & $7.503\text{E}^{-5}$ & $0.055$ & $0.924$ & $197009$ & $7.545\text{E}^{-5}$ & $0.053$ & $0.919$ \\ \hline
			$r_3,\text{A3}$ & $272609$ & $3.052\text{E}^{-5}$ & $0.241$ & $0.638$ & $318557$ & $3.543\text{E}^{-5}$ & $0.252$ & $0.549$ \\ \hline
		\end{tabular}
	}
	\caption{The proposed MCRT simulation results to characterize the $4$-bounce IR channels for both the simplified and realistic cabin models.}
	\label{table:CIR_IR_seated}
\end{table*}

In Figs. \ref{fig:CIR_IR_seated}(a) and (d), the \gls{MCRT} based \gls{CIR} and \gls{CFR} results are given for both cabin models along with the \gls{LoS} analytical model as a benchmark. For the \gls{UE} location C1, the magnitude of only the \gls{LoS} component obtained by the \gls{MCRT} simulations for simplified and realistic cabins become approximately $2.12\times10^{-5}$ and $2.07\times10^{-5}$, respectively. In comparison with the \gls{LoS} analytical channel model, simplified and realistic cabin models based simulation results yield $12.60\%$ and $14.63\%$ less \gls{LoS} magnitude, respectively. The reason for this is the angular and spectral characteristics of both the optical front-end and interior coating in simulations, which are omitted in the analytical model. Specifically, the angle of emergence and incidence values are significantly larger for points C1 and A3 compared to B2. Note that a small time difference, around $0.07$ ns, between the analytical \gls{LoS} model and the simulations occurs due to the simulated real world imperfections in the \gls{MCRT}. Furthermore, the effect of higher order reflections, $\kappa>0$, to the channel characteristics could also be assessed by using the flatness factor. Hence, the $\rho$ for simplified and realistic cabins based simulation results becomes $0.859$ and $0.872$, which is $14.1\%$ and $12.79\%$ less compared to the \gls{LoS} analytical model, respectively. Therefore, it can be inferred that both the analytical model and simplified cabin are able to capture the details of the channel when the \gls{NLoS} components are not as significant. This is also confirmed by the frequency response plot in Fig. \ref{fig:CIR_IR_seated}(d), where the mean magnitude response difference of $0.58$ and $0.69$ dB is observed between the simplified and realistic cabin simulations with the \gls{LoS} analytical model, respectively. Also, the peak-to-peak magnitude response difference for simplified and realistic cabin simulations is given by $0.64$ and $0.66$ dB, respectively. It can be inferred both from both the time and frequency domain results that the effect of higher order reflections are quite minor in seat C due to the large physical distance from the reflector surfaces e.g., side wall, which yields a \gls{LoS} dominated link characteristics.

In Figs. \ref{fig:CIR_IR_seated}(b) and (e), the \gls{MCRT} based \gls{CIR} and \gls{CFR} results are obtained for the \gls{UE} location B2, respectively. Accordingly, \gls{LoS} components of the simulation results becomes approximately $6.93\times10^{-5}$ for both cabin models. This means that the \gls{LoS} component for the simulation results are $1.62\%$ less for both cabin models compared to the \gls{LoS} analytical model. However, the \gls{DC} channel gain of simplified and realistic cabin models increased by $67.95\%$ and $68.60\%$, respectively, compared to the \gls{UE} location C1. The main reason behind this significant increase is the decreased source-receiver distance, where it becomes the minimum for the \gls{TX}-\gls{RX} pair of $r_2-\text{B2}$ among all \gls{UE} locations. Another important set of factors to note is the spectral dependencies of the optical front-end and coating as well as the angles of emergence and incidence. However, compared to C1, where the effect of an imperfect angular response is significant, both the emergence and incidence angles are almost zero in B2. Thus, we can conclude that the magnitude difference between the \gls{LoS} component of the simulation results and the analytical model is mainly caused by the spectral dependencies. Moreover, the increase in the flatness factor from C1 to B2 is also significant. Accordingly, the flatness factor increases by $7.57\%$ and $5.39\%$ for simplified and realistic cabin models, respectively. Although the optical power that comes from the side-wall reflections is increased in B2 compared to C1, the dominant \gls{LoS} component makes the \gls{NLoS} contributions negligible. Also, we can observe that the mean magnitude response of the simplified and realistic cabin simulations become approximately $0.07$ dB less than the analytical \gls{LoS} channel model. The peak-to-peak magnitude response difference for simplified and realistic cabin simulations is given by $0.41$ and $0.44$ dB, respectively. Therefore, we can conclude that the received optical power for the simulation point B2 is going to be the highest among all the other points due to the minimum source-receiver separation, which effectively means a higher communications capacity. In addition, the frequency flat/\gls{LoS} channel assumption for seat B is also accurate due to the weak \gls{NLoS} contribution.
\begin{figure*}[!t]
	\centering
	\begin{subfigure}[t]{.54\columnwidth}
		\includegraphics[width=\columnwidth]{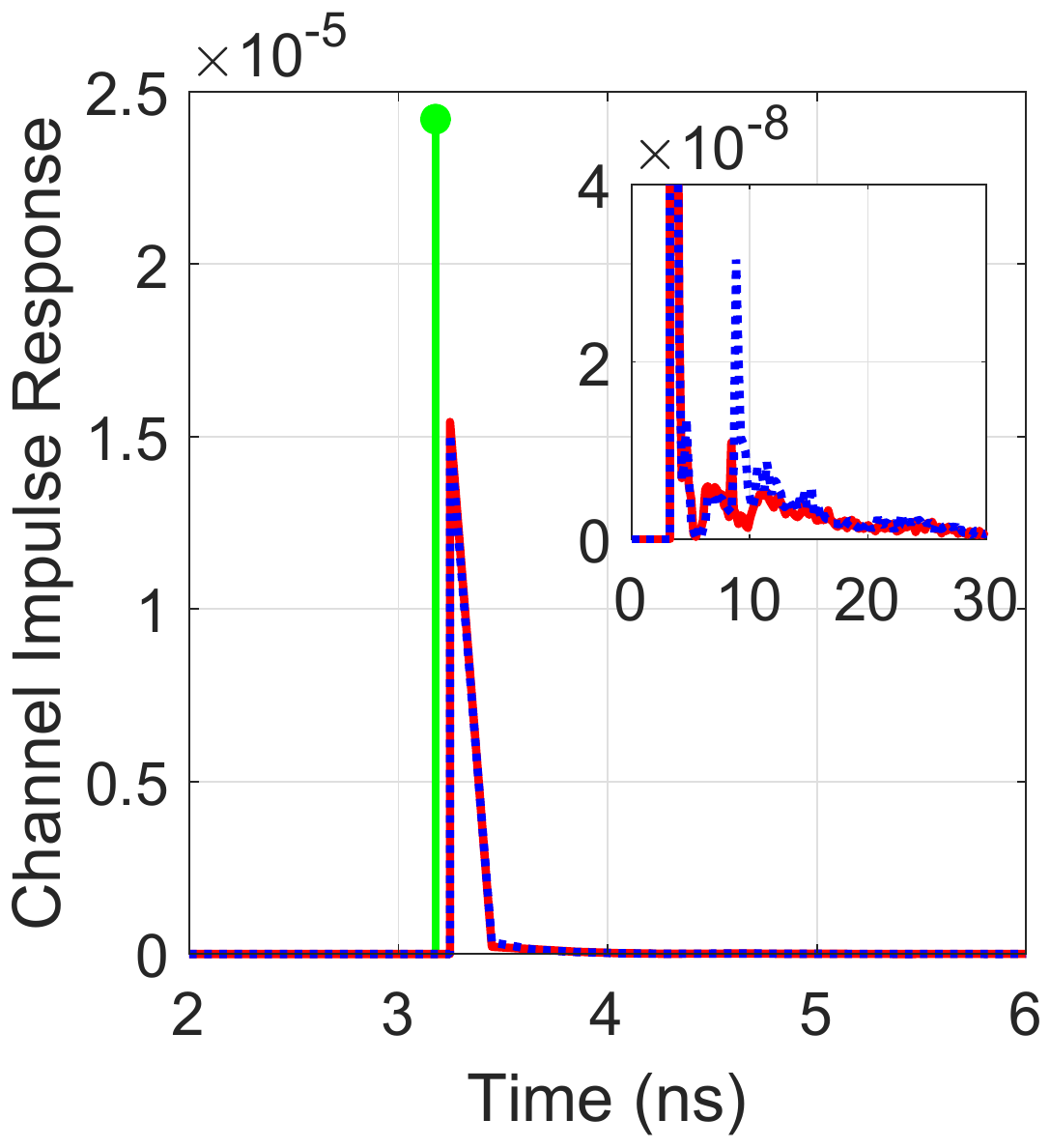}
		\caption{$h(t;r_1,\text{C1})$}
		\label{}
	\end{subfigure}~
	\begin{subfigure}[t]{.54\columnwidth}
		\includegraphics[width=\columnwidth]{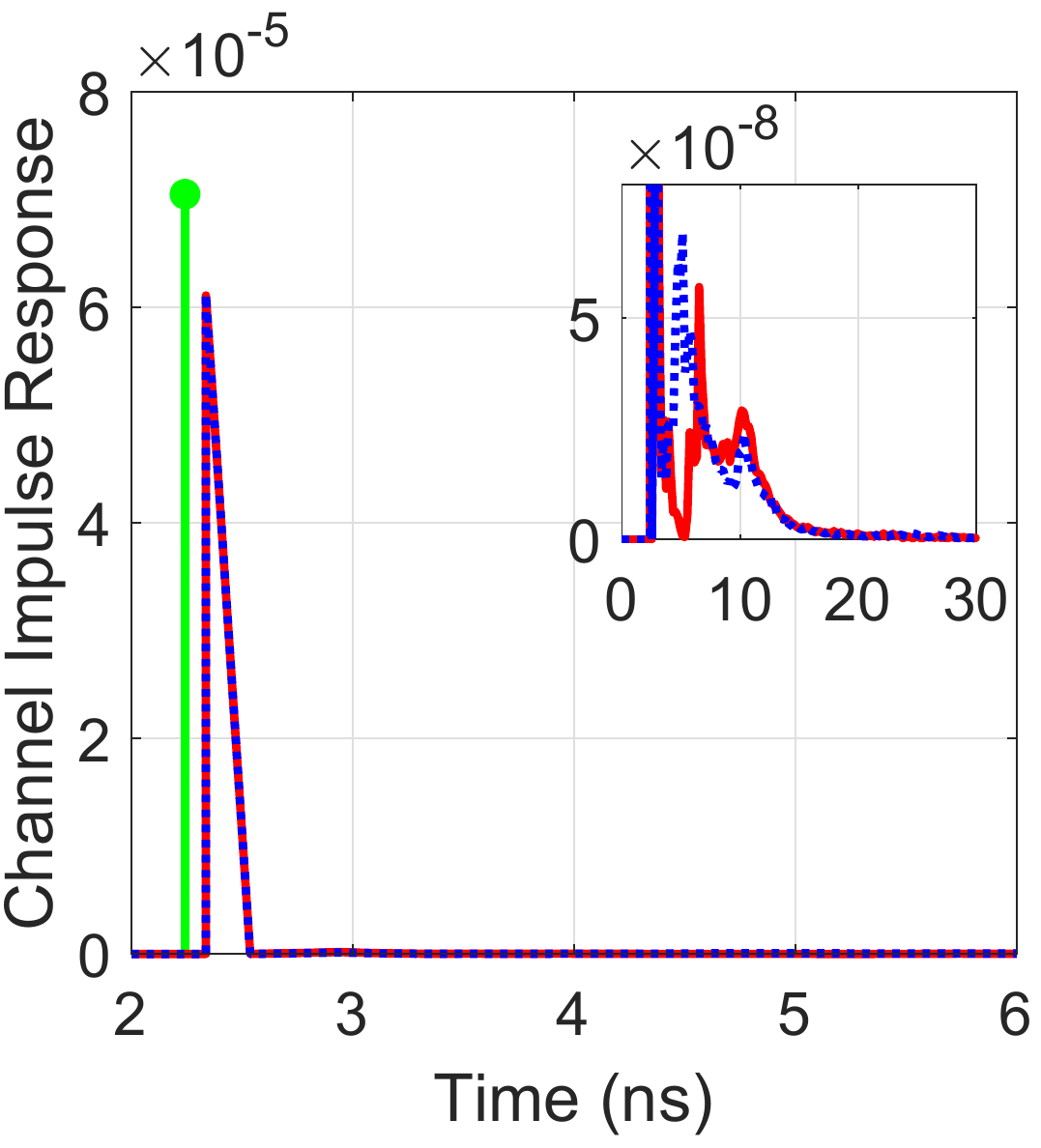}
		\caption{$h(t;r_2,\text{B2})$}
		\label{}
	\end{subfigure}~
	\begin{subfigure}[t]{.54\columnwidth}
		\includegraphics[width=\columnwidth]{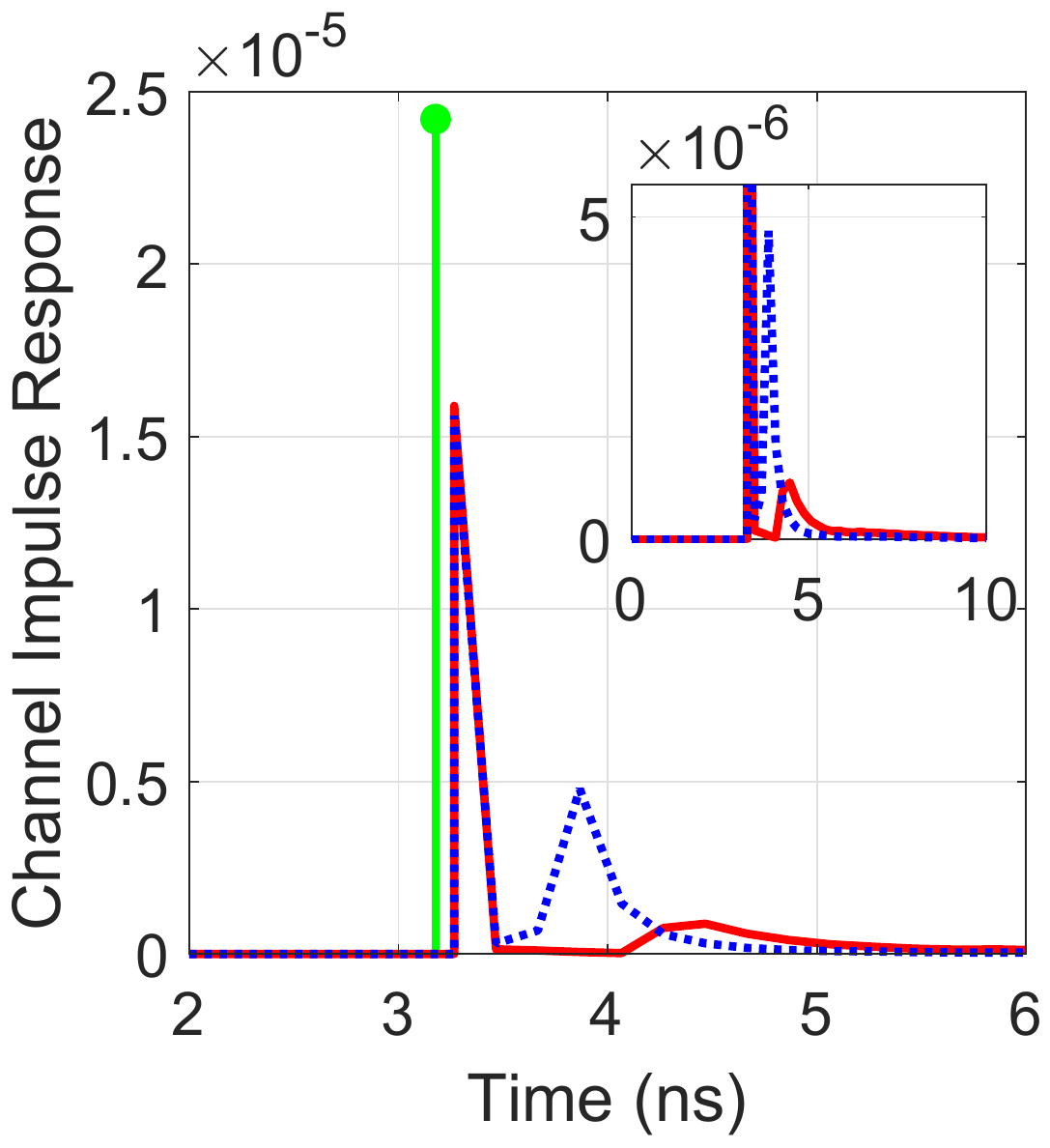}
		\caption{$h(t;r_3,\text{A3})$}
		\label{}
	\end{subfigure}\\
	\begin{subfigure}[t]{.54\columnwidth}
		\includegraphics[width=\columnwidth]{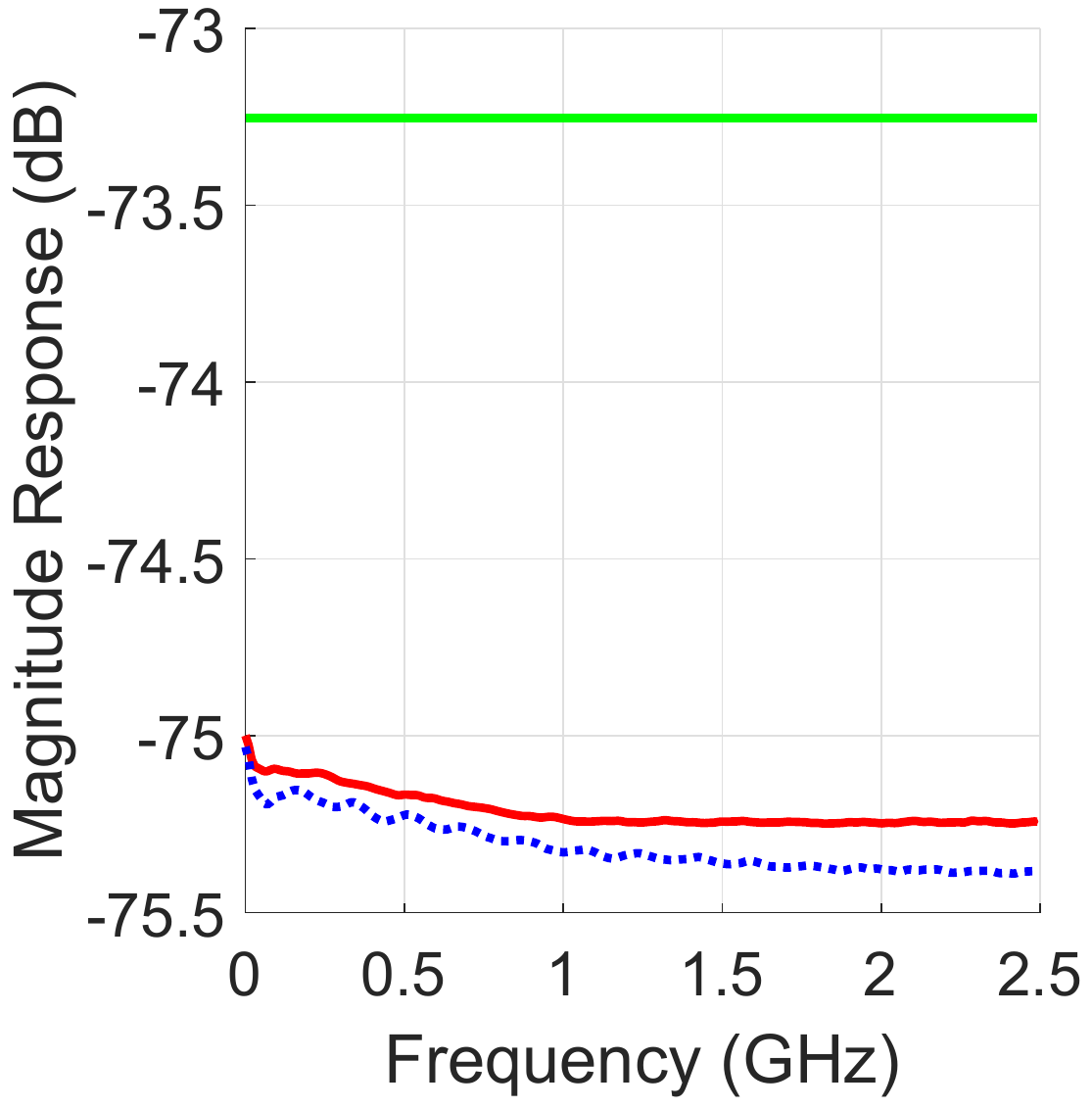}
		\caption{$\lvert H(f;r_1,\text{C1}) \rvert$}
		\label{}
	\end{subfigure}~
	\begin{subfigure}[t]{.54\columnwidth}
		\includegraphics[width=\columnwidth]{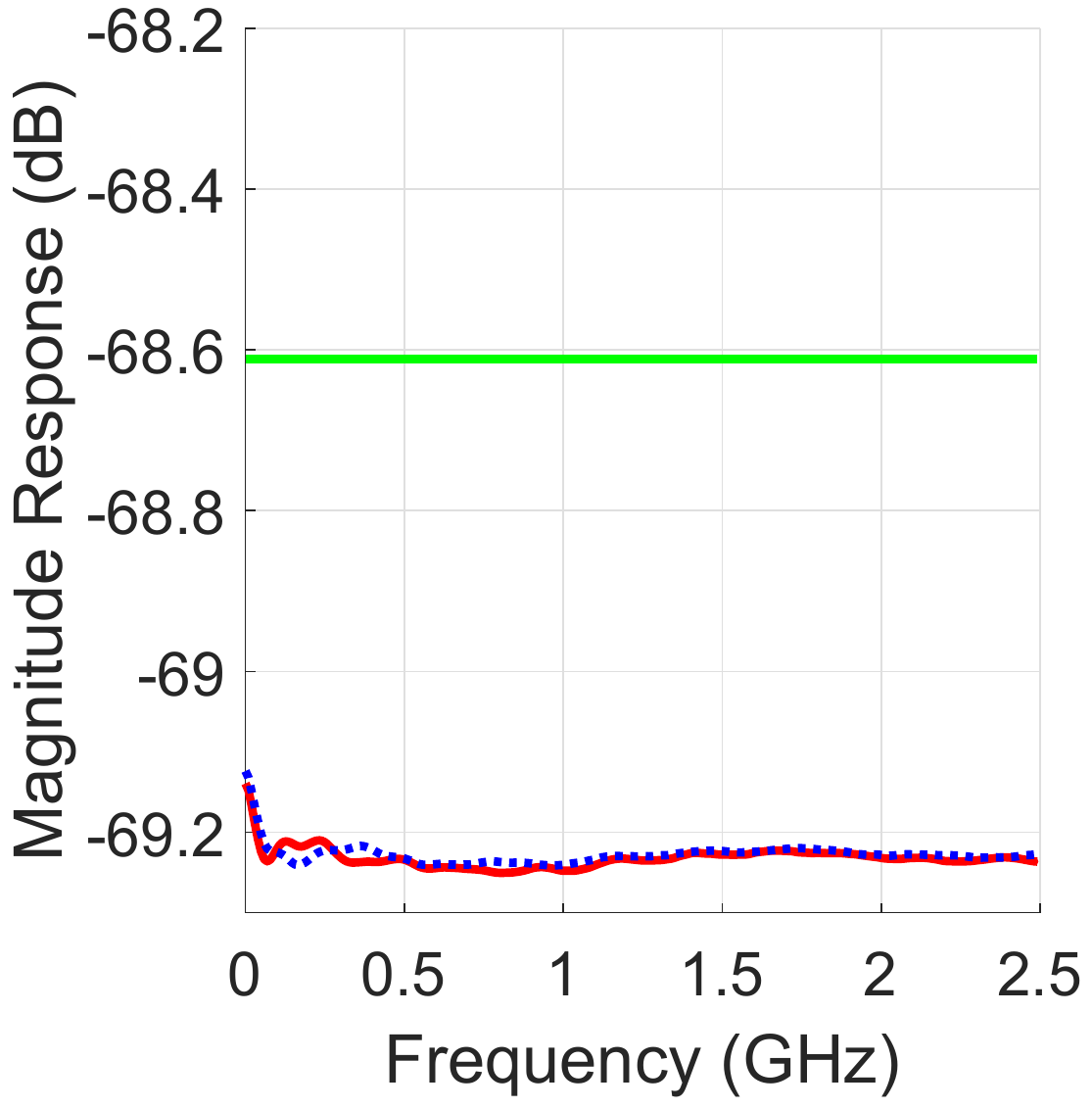}
		\caption{$\lvert H(f;r_2,\text{B2}) \rvert$}
		\label{}
	\end{subfigure}~
	\begin{subfigure}[t]{.54\columnwidth}
		\includegraphics[width=\columnwidth]{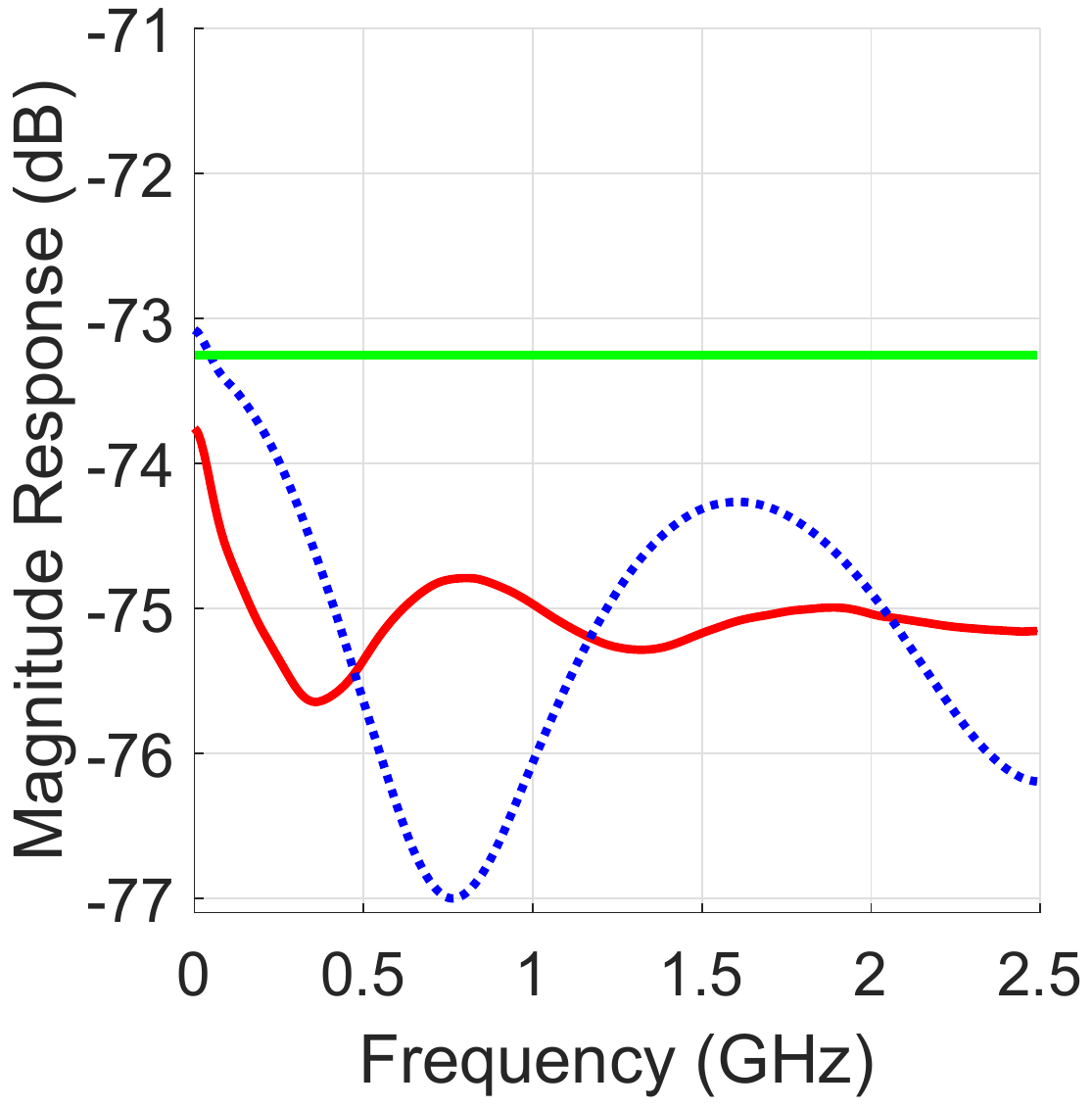}
		\caption{$\lvert H(f;r_3,\text{A3}) \rvert$}
		\label{}
	\end{subfigure}
	\caption{VL band in-flight LiFi CIR and CFR simulation results obtained by proposed MCRT based method for seated simplified cabin model (\protect\includegraphics[height=0.2cm]{figs/icon/cabinsimplified_icon.pdf}) and seated realistic cabin model (\protect\includegraphics[height=0.2cm]{figs/icon/cabinrealistic_icon.pdf}). The analytical LoS channel (\protect\includegraphics[height=0.2cm]{figs/icon/los_icon.pdf}) is presented as a benchmark.}
	\label{fig:CIR_VL_seated}
\end{figure*}
\begin{table*}[!t]
	\centering 
	\resizebox{0.75\linewidth}{!}{
		\renewcommand{\arraystretch}{1.3} 
		\begin{tabular}{|c|c|c|c|c|c|c|c|c|}
			\hline
			\textbf{} & \multicolumn{4}{c|}{\textbf{Simplified Cabin (VL Band)}} & \multicolumn{4}{c|}{\textbf{Realistic Cabin (VL Band)}} \\ \hline
			$S,R$ & $i_\text{hit}$ & $H[0;S,R]$ & $\tau_\text{RMS}$ (ns) & $\rho$ & $i_\text{hit}$ & $H[0;S,R]$ (W) & $\tau_\text{RMS}$ (ns) & $\rho$ \\ \hline
			$r_1,\text{C1}$ & $271763$ & $1.619\text{E}^{-5}$ & $0.021$ & $0.929$ & $277731$ & $1.608\text{E}^{-5}$ & $0.031$ & $0.937$ \\ \hline
			$r_2,\text{B2}$ & $542237$ & $6.242\text{E}^{-5}$ & $0.015$ & $0.979$ & $535559$ & $6.265\text{E}^{-5}$ & $0.014$ & $0.976$ \\ \hline
			$r_3,\text{A3}$ & $1075022$ & $2.155\text{E}^{-5}$ & $0.150$ & $0.709$ & $1060614$ & $2.519\text{E}^{-5}$ & $0.191$ & $0.607$ \\ \hline
		\end{tabular}
	}
	\caption{The proposed MCRT simulation results to characterize the $6$-bounce VL channels for both the simplified and realistic cabin models.}
	\label{table:CIR_VL_seated}
\end{table*}

In Figs. \ref{fig:CIR_IR_seated}(c) and (f), the simulation results are obtained by using both the simplified and realistic cabin models and compared with the benchmark for point A3. Similar to C1 results, the \gls{LoS} component magnitudes for the simplified and realistic cabin simulations become $2.01\times10^{-5}$ and $1.98\times10^{-5}$, respectively. This means that the simplified and realistic \gls{LoS} simulation components are $17.08\%$ and $18.37\%$ lower compared to the analytical model. This biggest difference among all points also stems from the combination of imperfect angular and spectral profiles of the elements. Furthermore, it can clearly be seen that the secondary reflection for both simplified and realistic \gls{CIR} emerged after $3.8$ ns. Note that compared to point B2, the \gls{RMS} delay spread has increased by $338.18\%$ and $375.47\%$ for simplified and realistic cabin simulations, respectively. Similarly, the flatness factor in point A3 becomes approximately $30.95\%$ and $40.26\%$ for simplified and realistic cabin simulation results, respectively. From Fig. \ref{fig:CIR_IR_seated}(f), the mean magnitude response for both the simplified and realistic cabin simulations becomes $0.81$ and $0.88$ dB less than the \gls{LoS} analytical model, respectively. Furthermore, the peak-to-peak magnitude response difference for simplified and realistic cabin simulation results become $2.41$ and $4.18$ dB, respectively. More importantly, the $3$ dB cut-off frequency for the \gls{IR} band \gls{LiFi} optical channel in realistic cabin becomes $498$ MHz. Consequently, it can be said that the location of the \gls{UE} impacts two of the most important parameters of the communication channel; (i) received signal power and, (ii) the effective bandwidth in such complex environments.
\subsubsection{VL Band Results}
The time and frequency domain analysis of the in-flight \gls{LiFi} channel for the \gls{VL} band source and receiver pair are given for the adopted cabin models and \gls{UE} locations in Figs. \ref{fig:CIR_VL_seated}(a)-(f). Unlike the \gls{IR} band results, the \gls{VL} band \gls{LiFi} channels are simulated up to $6$-bounces, $0\leq \kappa \leq 6$, and the detailed parameters are presented in Table \ref{table:CIR_VL_seated}. The higher order reflections compared to the \gls{IR} band simulations were required as the absorption rate of the coating materials are higher in the \gls{VL} spectra. It is also important to note that the \gls{LoS} analytical channel expression, given in \cite{kb9701}, is independent of the operation wavelength, which will be granting a convenient direct comparison between \gls{IR} and \gls{VL} band results.

The \gls{CIR} and \gls{CFR} comparisons for the \gls{UE} location C1 are depicted in Figs. \ref{fig:CIR_VL_seated}(a) and (d), respectively. As can be seen from figures, the \gls{LoS} channel components of $1.54\times10^{-5}$ and $1.51\times10^{-5}$ are obtained by simplified and realistic cabin simulations, respectively. Thus, these numbers yield $36.27\%$ and $37.74\%$ smaller \gls{LoS} component for simplified and realistic cabin simulations compared to analytical model, respectively. The underlying reason for this difference is also realistic spectral and angular characteristics which are taken into consideration in our simulations. Even though point C1 is furthest from the side wall, the \gls{NLoS} components still comprises approximately $7.1\%$ and $6.3\%$ of the total received optical power for simplified and realistic cabin simulations, respectively. In the frequency domain, the mean value of the magnitude responses for both the simplified and realistic cabin simulations become $1.96$ and $2.06$ dB smaller compared to analytical benchmark, respectively. Moreover, the peak-to-peak magnitude response difference of $0.25$ and $0.36$ dB are obtained by simplified and realistic simulation results, respectively. Therefore, we can infer from the obtained results that, due to the higher optical imperfections presented in \gls{VL} band simulations, the received signal power differs from the analytical expectation more than that of the \gls{IR} band results.

In Figs. \ref{fig:CIR_VL_seated}(b) and (e), both the impulse and frequency response results and comparisons are given for both cabin models along with the \gls{LoS} analytical expression for the \gls{UE} location of B2. Accordingly, the simplified and realistic simulation results yield a \gls{LoS} component of $6.11\times10^{-5}$ and $6.12\times10^{-5}$, respectively. Hence, the presented values yield a $13.32\%$ and $13.23\%$ smaller \gls{LoS} component in comparison with the benchmark, respectively. As can be seen from Table \ref{table:CIR_VL_seated}, the \gls{NLoS} contribution to total received optical power is $2.1\%$ and $2.4\%$ for simplified and realistic cabin simulation results, respectively. The mean magnitude response difference between simulations against the benchmark becomes $0.62$ dB for both cabin models. Furthermore, the peak-to-peak magnitude response difference for simplified and realistic cabin simulations is given by $0.11$ and $0.12$ dB, respectively.  Again, the \gls{UE} location in seat B yields the highest achievable received optical signal power compared to other two seats.

Lastly, in Figs. \ref{fig:CIR_VL_seated}(c) and (f), both the time and frequency domain results and comparisons are given for both cabin models when the \gls{UE} is located in point A3. The \gls{LoS} components obtained by simplified and realistic cabin simulations is given by $1.59\times10^{-5}$ and $1.56\times10^{-5}$, respectively. Thus, the simplified and realistic cabin simulation results yield $34.33\%$ and $35.52\%$ lower \gls{LoS} component compared to the benchmark, respectively. It is important to note that the contribution of the spectral characteristics of the source, receiver and coating is higher in the \gls{VL} band as the detector introduces an almost ideal cosine profile. It has been shown by narrow-band \gls{IR} results that the realistic source and receivers do not match closely with the flat spectral profile. Therefore, the \gls{VL} band sources, whose power is spread over a wider spectral range, present a greater mismatch. In terms of \gls{RMS} delay spread, the simplified and realistic simulations yield $900\%$ and $1264\%$ increase for the simulation point A3 compared to B2, respectively. Also, the portion of the \gls{NLoS} channel contribution became almost $30\%$ and $40\%$ in the simplified and realistic cabin applications, respectively. In the frequency domain, the difference of the mean magnitude response between the simplified and realistic simulations with the benchmark becomes $1.83$ and $1.91$ dB, respectively. Also, the peak-to-peak magnitude response values for simplified and realistic simulation results also given by $1.89$ and $3.92$ dB, respectively. Note that the $3$ dB cut-off frequency for the \gls{VL} band \gls{LiFi} optical channel in realistic cabin becomes $566$ MHz.
\section{Performance of DCO-OFDM Under The Channel Impairments}
In this section, the performance of the practical in-flight \gls{LiFi} systems are evaluated by employing a widely adopted multi-carrier optical transmission technique, \gls{DCO-OFDM} \cite{da1301}. Accordingly, the \gls{BER} performance of \gls{DCO-OFDM} is investigated under both optical domain multipath propagation and \gls{LED} non-linearity based clipping.
\begin{table}[!t]
	\centering
	\caption{The set of parameters used in the BER simulations.}
	\label{table_berpar}
	\resizebox{0.95\columnwidth}{!}{
		\renewcommand{\arraystretch}{1.3} 
		\begin{tabular}{|l|l|l|}
			\hline
			\textbf{Parameter} 			 & \textbf{Description}                                    		  & \textbf{Value} \\ \hline\hline
			$B_{\text{CH}}$               			 & The channel bandwidth                       & $5$ GHz              \\ \hline
			$N$               			 & Number of subcarriers                       & $512$              \\ \hline
			$M$               			 & Order of the QAM constellation                       & $4$ and $64$              \\ \hline
			$\beta_\text{dB}$               			 & DC bias value                       & $19.19$ dB              \\ \hline
			$L$               			 & Number of channel taps                       & $7$              \\ \hline
			$N_{\text{CP}}$               			 & Length of the CP                       & $7$              \\ \hline
			$I_{\textrm{min}}$               			 & The lower limit for the $I_\textrm{f}$                       & $100$ mA \cite{gwqsspa1.em,sfh4253}              \\ \hline
			$I_{\textrm{max}}$               			 & The upper limit for the $I_\textrm{f}$                       & $700$ mA \cite{gwqsspa1.em,sfh4253}             \\ \hline
		\end{tabular}
	}
\end{table}
	
As a first step in \gls{DCO-OFDM}, the binary user data vector, which contains the information to be transmitted, is parsed into $(N-2)\log_2(M)$ elements. The parameter $M$ denotes the order of the \gls{QAM} modulation. Then, each element is one-to-one mapped into a complex valued $M$-\gls{QAM} symbol, $X[k]$, where $k \in \{1,2,\cdots, \frac{N}{2}-1\}$. Please note that in order to obtain a real valued signal after the \gls{IFFT} operation, Hermitian symmetry must be imposed on the frequency domain symbols, $X[N-k]=X[k]^\ast,~\forall k \in \{0,1,\cdots,N-1\}$, where $X[0]=X[N/2]=0$. Hence, only $\frac{N}{2}-1$ subcarriers carry useful information in \gls{DCO-OFDM}. For $N>64$, the time domain signal after the \gls{IFFT}, will follow \gls{iid} Gaussian distribution, $x[n]\sim \mathcal{N}\left( 0, \sigma^2 \right),~\forall n\in \{0,1,\cdots,N-1\}$ according to the central limit theorem. Since, we have adopted the standard definitions of the \gls{IFFT}/\gls{FFT} pair at the \gls{TX} and \gls{RX}, respectively, $\sigma^2 = \frac{1}{N}$.
	
The main transmission technique in \gls{LiFi} systems is referred to as \gls{IM/DD}, where the information is carried via the instantaneous light intensity. Thus, a \gls{DC} bias, $\beta$, must be introduced prior to transmission as the light intensity cannot take a negative value. Furthermore, double-sided clipping is also needed to make the transmit signal compliant with the dynamic range of the \glspl{LED}. Hence, the resultant truncated Gaussian distributed time-domain signal could be expressed by
\begin{align}
\hat{x}[n]=\big[x[n]+\beta\big]^{I_\text{max}}_{I_\text{min}} = \left\{\begin{array}{ll}
		I_\text{min}, & \text{if } x[n] < I_\text{min}-\beta \\
		I_\text{max}, & \text{if } x[n] > I_\text{max}-\beta \\
		x[n]+\beta, & \text{otherwise}
	\end{array}\right.,
\end{align}
\noindent where $I_\text{min}\leq \beta \leq I_\text{max}$ and $\beta = r\sigma$. The parameter $r$ denotes the bias proportionality factor, which adjusts the amount of bias in terms of the standard deviation of the unbiased and unclipped $x[n]$. After the \gls{DC} bias addition, the time domain shifted signal follows the distribution $x[n]+\beta \sim \mathcal{N}\left( r\sigma,(r^2+1)\sigma \right)$. In literature, the \gls{DC} bias in decibels is also defined by $\beta_\text{dB}=10\log_{10}\left(r^2+1\right)$. To simulate the  realistic transmit \gls{LED} front-end electrical properties of the adopted \gls{VL} (OSRAM GW QSSPA1.EM) and \gls{IR} (OSRAM SFH 4253) band sources, the forward current value is constrained to be within the $I_f \in \left[ I_\text{min}~I_\text{max} \right] = [100~700]$ mA range \cite{gwqsspa1.em,sfh4253}. Moreover, in order to locate the time domain signal at the middle of the dynamic range and minimize the non-linear clipping effect of the \glspl{LED}, $\beta = 400$ mA ($\beta_\text{dB}=19.19$ dB) is adopted in our simulations.

It is important to note that the non-linear and memoryless \gls{LED} clipping effect could be represented as a linear process with a deterministic attenuation factor, $A$, and a random time domain additive clipping noise, $c[n]$, by Bussgang's theorem as follows \cite{dsh1202,cke1201}:
\begin{align}
	z[n] = \hat{x}[n]-\beta = Ax[n] + c[n] ,
	\label{eq:clipping_noise}
\end{align}
\noindent where $\text{E}\{z[n]\}=\text{E}\{c[n]\}=\mu_z=P_\text{opt}-\beta$. The optical power of the biased and double sided clipped signal could be calculated by
\begin{align}
P_\text{opt} &= \text{E}\{ \hat{x}[n] \} = \frac{\sigma}{\sqrt{2\pi}}\left( e^{-(I_\text{min}-\beta)^2/2\sigma^2} - e^{-(I_\text{max}-\beta)^2/2\sigma^2} \right) \nonumber \\
&+\left( I_\text{max}-\beta \right) Q \left( \frac{I_\text{max}-\beta}{\sigma} \right) + \left( \beta-I_\text{min} \right) Q \left( \frac{I_\text{min}-\beta}{\sigma} \right) + I_\text{min}.
\label{eq:DCO_optpower}
\end{align}
\noindent If we subtract $\mu_z$ from both sides of \eqref{eq:clipping_noise},
\begin{align}
	z^\prime[n] = z[n]-\mu_z = Ax[n] + c^\prime[n] ,
\end{align}
\noindent where $z^\prime[n]$, $c^\prime[n]$ and $x[n]$ are all zero-mean random variables. Thus, the value of $A$ could be deduced from the above expressions by
\begin{align}
A = \frac{\text{E}\{ x[n]y^\prime[n] \}}{\sigma} = Q \left( \frac{I_\text{min}-\beta}{\sigma} \right)-Q \left( \frac{I_\text{max}-\beta}{\sigma} \right).
\end{align}
\noindent As can be seen from the above expression, the attenuation factor could simply be represented as the area under the unclipped portion of the biased time domain signal $x[n]+\beta$. It is important to note that any \gls{DC} bias introduced at the \gls{TX} and/or \gls{RX} won't effect the frequency domain symbols since it will fall on to the unused $0^\text{th}$ subcarrier. Therefore, the electrical power of the time-domain clipping noise could be represented by
\begin{align}
P_c = \sigma_c^2 = P_\text{elec}-P_\text{opt}^2-A^2\sigma^2 ,
\end{align}
\noindent where the electrical power of the biased and double-sided clipped signal could be expressed as follows:
\begin{align}
&P_\text{elec} = \text{E}\{ \hat{x}[n]^2\} = \nonumber \\
&\frac{\sigma}{\sqrt{2\pi}}\left( \left(I_\text{min}+\beta\right)e^{-(I_\text{min}-\beta)^2/2\sigma^2} - \left(I_\text{max}+\beta\right)e^{-(I_\text{max}-\beta)^2/2\sigma^2} \right) \nonumber \\
& +\left( \beta^2+\sigma^2-I_\text{min}^2 \right)Q\left( \frac{I_\text{min}-\beta}{\sigma} \right) \nonumber \\
&+ \left( I_\text{max}^2-\beta^2-\sigma^2 \right)Q\left( \frac{I_\text{max}-\beta}{\sigma} \right)+I_\text{min}^2 .
\end{align}

At the receiver, the multipath optical wireless channel distorted signal is received by a \gls{PD}. Furthermore, the \gls{AWGN} also plays a role in the electrical domain signal, $y[n]=h[n]\circledast \hat{x}[n] + w[n]$, obtained after the \gls{CP} removal. Note that the effective noise term at the \gls{RX}, $w[n]$, consists of the addition of shot and thermal noises, where the shot noise emerges as a result of ambient light sources and information bearing signal itself. In the case where high ambient light power at the \gls{PD} is significantly larger than the transmit signal power, the shot noise becomes signal independent. Therefore, the high intensity shot noise at the \gls{RX} could be modelled as a summation of independent low power Poisson processes, which could be approximated as a zero mean Gaussian distribution. Consequently, the effective noise could be modelled as \gls{AWGN}, $w[n]\sim\mathcal{N}\left(0,\sigma_w^2 \right)$, where $\sigma_w^2 = \sigma_\text{shot}^2 + \sigma_\text{thermal}^2$. Independent from the shot noise, the thermal noise emerges due to the random motions of the electrons in the front-end circuitry.
\begin{figure}[!t]
	\centering
	\begin{subfigure}[t]{.5\columnwidth}
		\includegraphics[width=\columnwidth]{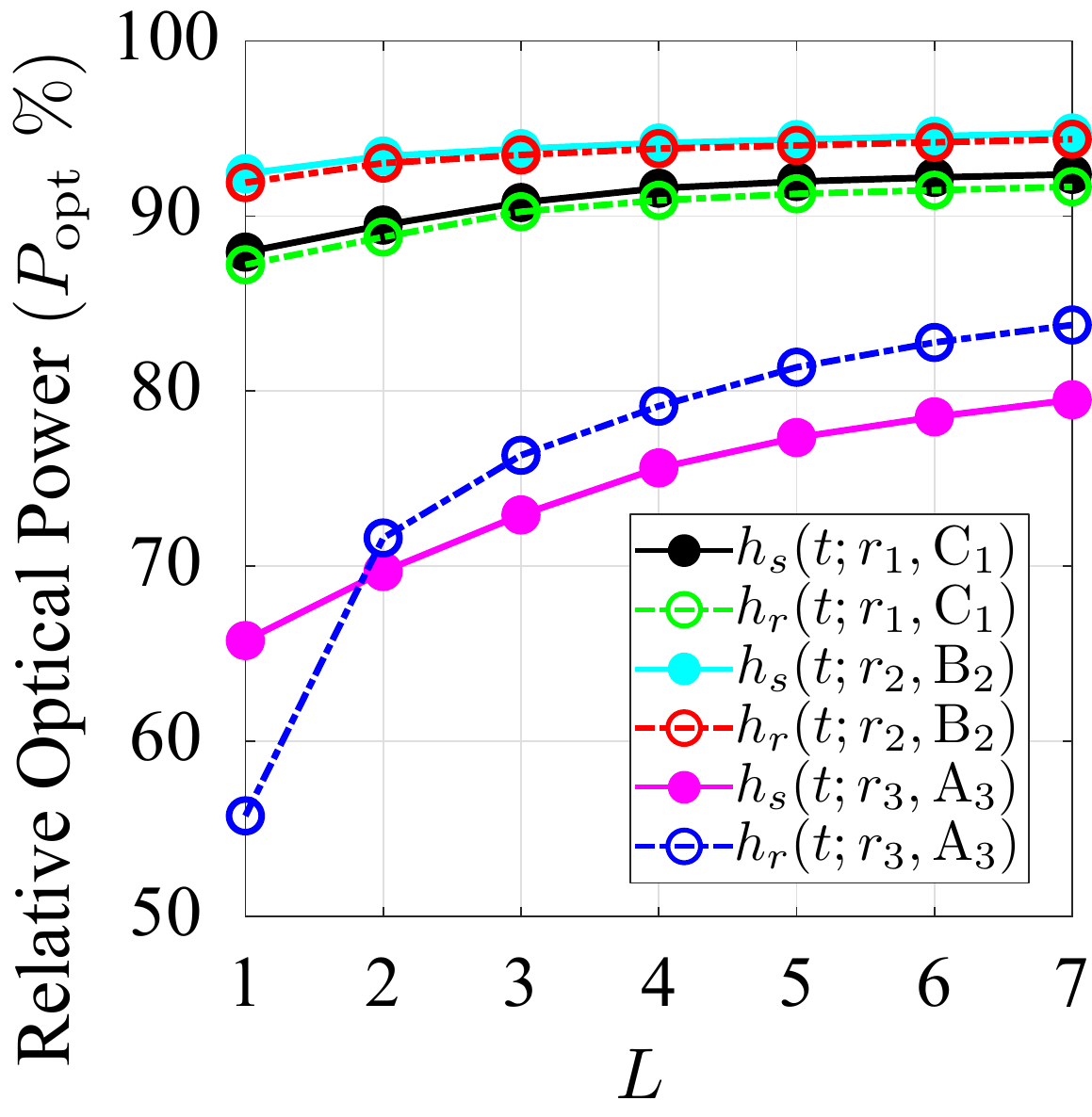}
		\caption{IR band}
		\label{fig:}
	\end{subfigure}~
	\begin{subfigure}[t]{.5\columnwidth}
		\includegraphics[width=\columnwidth]{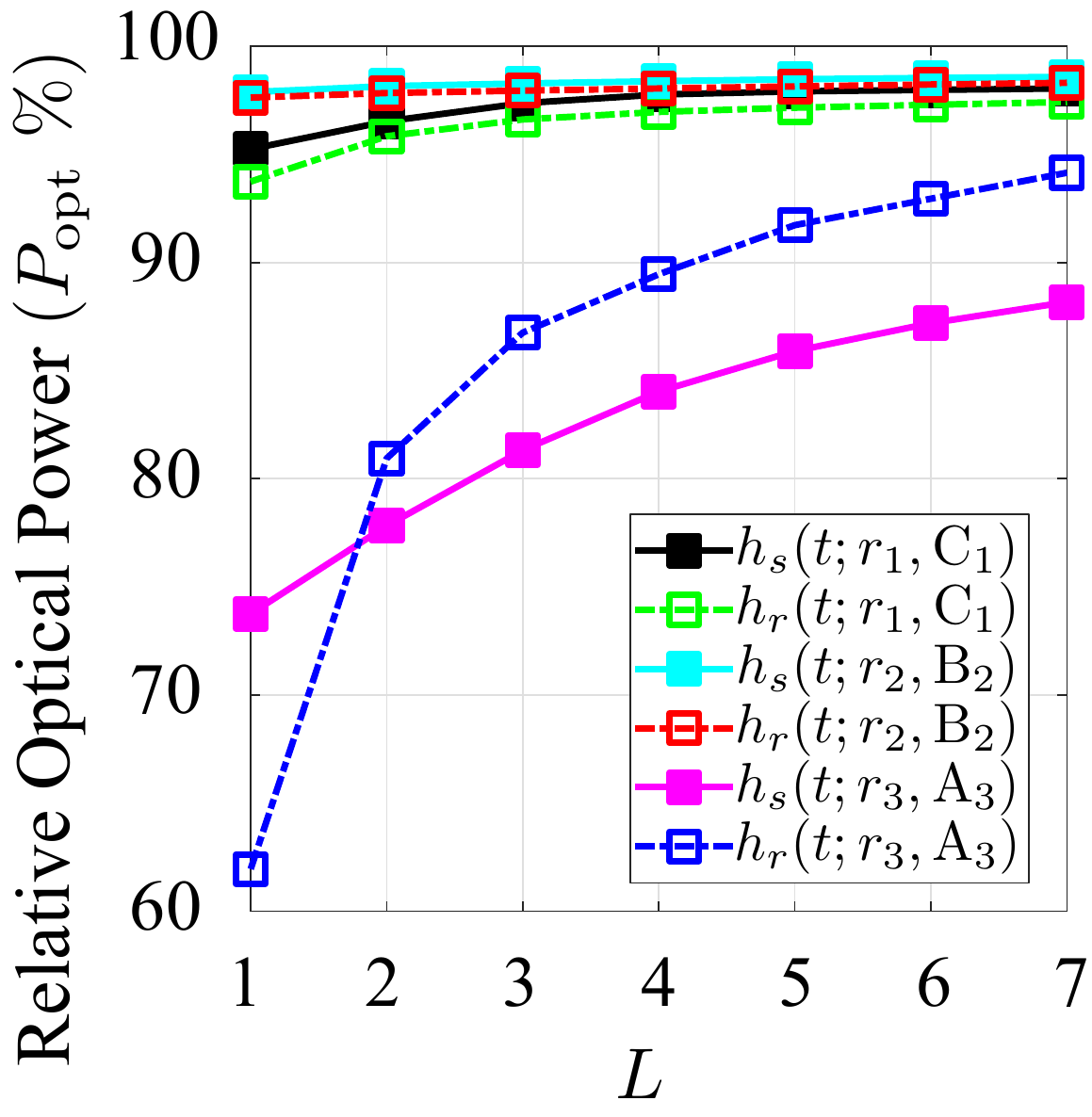}
		\caption{VL band}
		\label{fig:}
	\end{subfigure}
	\caption{Relative optical power, $P_{\text{opt}}~(\%)$, for the CIR when $1 \leq L \leq 7$ larger magnitude taps are chosen. The CIRs are given for both simplified, $h_s$, and realistic, $h_r$, cabin models for (a) IR band and (b) VL band sources.}
	\label{fig:perc_power}
\end{figure}
	
In practical wireless communication systems, the number of channel taps are limited in the time domain since they follow an exponential decay profile. To find the effective number of channel taps ($L$) for our simulations, the fractional optical power analysis depicted in Fig. \ref{fig:perc_power} is used. As illustrated in Fig. \ref{fig:perc_power}, the relative optical power, $P_\text{opt}\%$, represents the portion of optical power when first $L$ largest taps are picked from the \gls{CIR} as the effective channel. From the figure, it can be inferred that $L=7$ taps yield at least $80\%$ of the optical power of the initial \glspl{CIR} for all the \gls{UE} locations and both spectral regions. Consequently, the $7$ highest taps based \glspl{CIR} are obtained from \gls{MCRT} simulation results, are utilized in our \gls{BER} versus effective-\gls{SNR}-per-bit plots. The \gls{CP} length in the \gls{DCO-OFDM} system must be $N_\text{CP}\geq L$, to avoid \gls{ISI} as a rule of thumb. Hence, the \gls{CP} length of $N_\text{CP}=L$ is adopted in our simulations. By using \eqref{eq:clipping_noise}, we can obtain the frequency-domain signal at the \gls{RX} after the \gls{CP} addition and removal at the \gls{TX} and \gls{RX}, respectively as well as \gls{FFT} operation by
\begin{align}
Y[k] = H[k]\big( AX[k] + C[k] \big) + W[k], \quad \text{for }k\in\{1,2,\cdots,\frac{N}{2}-1\},
\end{align}
where the \gls{FFT} of $h[n]$, $x[n]$, $c[n]$ and noise term $w[n]$ are denoted by $H[k]$, $X[k]$, $C[k]$ and $W[k]$, respectively for the $k^\text{th}$ subcarrier.  It is important to note that the electrical domain frequency response characteristics of the front-end opto-electronic elements, $H_\text{elec}[k]$, could also be lumped into the channel model if necessary. Hence, the effective channel becomes, $H_\text{eff}[k]=H[k]H_\text{elec}[k]$. In this work, the electrical domain channel impairments, the analogue/digital and electrical/optical domain conversions are assumed to be ideal without loss of generality. Thus, the analytical average \gls{BER} expression for \gls{DCO-OFDM} under double-sided clipping and frequency selective channel effects could be calculated by using \cite{cy0201},
\begin{align}
\text{BER}_\text{theo} &= \frac{2}{N-2}\sum_{k=1}^{(N/2)-1}\frac{4M_{k,1}M_{k,2}-2\left( M_{k,1}+M_{k,2} \right)}{M_{k,1}M_{k,2}\log_2\left( M_{k,1}M_{k,2} \right)} \nonumber \\
&\times Q\left( \sqrt{\frac{6\gamma_k\log_2\left( M_{k,1}M_{k,2} \right)}{M_{k,1}^2 + M_{k,2}^2-2}} \right),
\label{eq:ABER}
\end{align}
\noindent where $M_k=M_{k,1}\times M_{k,2}$ denotes $M$ for $k^\text{th}$ subcarrier. The electrical domain effective-SNR-per-bit for the $k^\text{th}$ subcarrier after the \gls{ZF} channel equalization could also be calculated by
\begin{align}
\gamma_k = \frac{BA^2\text{E}\{ \lvert X[k] \rvert^2 \}}{R_k N\left( \sigma_c^2 + \sigma_w^2/\lvert H[k] \rvert^2 \right)},
\end{align}
\noindent where the \gls{OFDM} signal bandwidth is given by $B$. Also, the bit rate for the $k^\text{th}$ subcarrier could be calculated by
\begin{align}
R_k = B\log_{2}\left(M_k\right)\left( \frac{N-2}{N} \right)\left( \frac{N}{N+N_\text{CP}} \right) \quad \text{bits/sec.}
\end{align}
\noindent By using the above equation, the spectral efficiency of the $k^\text{th}$ subcarrier of \gls{DCO-OFDM} could also be calculated by $\eta_k = R_k/2B$ bits/sec/Hz.
\begin{figure}[!t]
	\centering
	\begin{subfigure}[t]{.78\columnwidth}
		\includegraphics[width=\columnwidth]{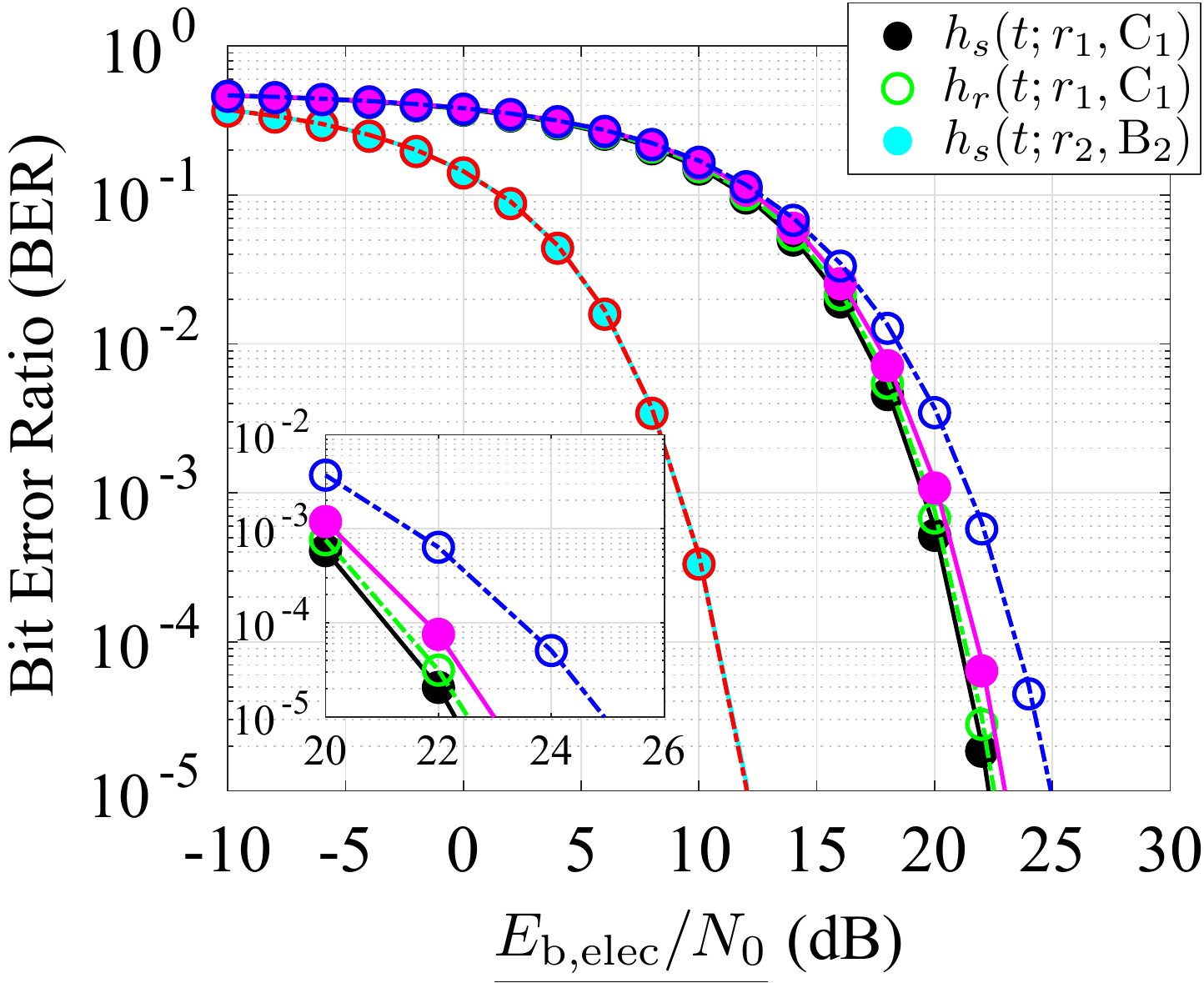}
		\caption{IR band, $M=4$ ($1~\text{bit/sec/Hz}$)}
		\label{fig:}
	\end{subfigure}\\
	\begin{subfigure}[t]{.78\columnwidth}
		\includegraphics[width=\columnwidth]{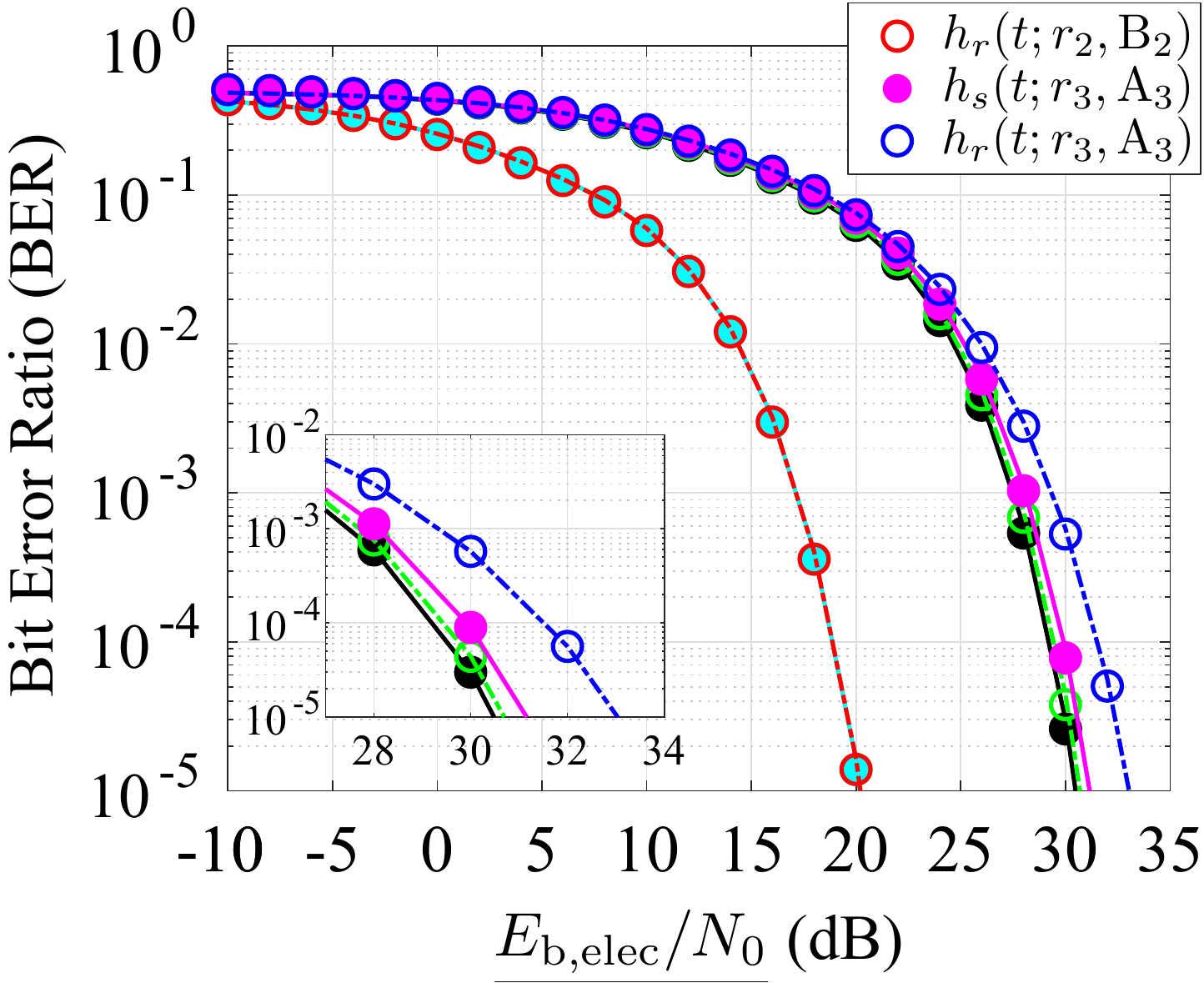}
		\caption{IR band, $M=64$ ($3~\text{bits/sec/Hz}$)}
		\label{fig:}
	\end{subfigure}
	\caption{BER vs. effective SNR-per-bit performance curves for a DCO-OFDM system under the MCRT channel dispersion and double-sided clipping impairments. The simulation results for realistic and simplified cabin models are given by dashed and solid lines, respectively. The theoretical results are presented by markers.}
	\label{fig:BER_SNR_IR}
\end{figure}
\begin{figure}[!t]
	\centering
	\begin{subfigure}[t]{.78\columnwidth}
		\includegraphics[width=\columnwidth]{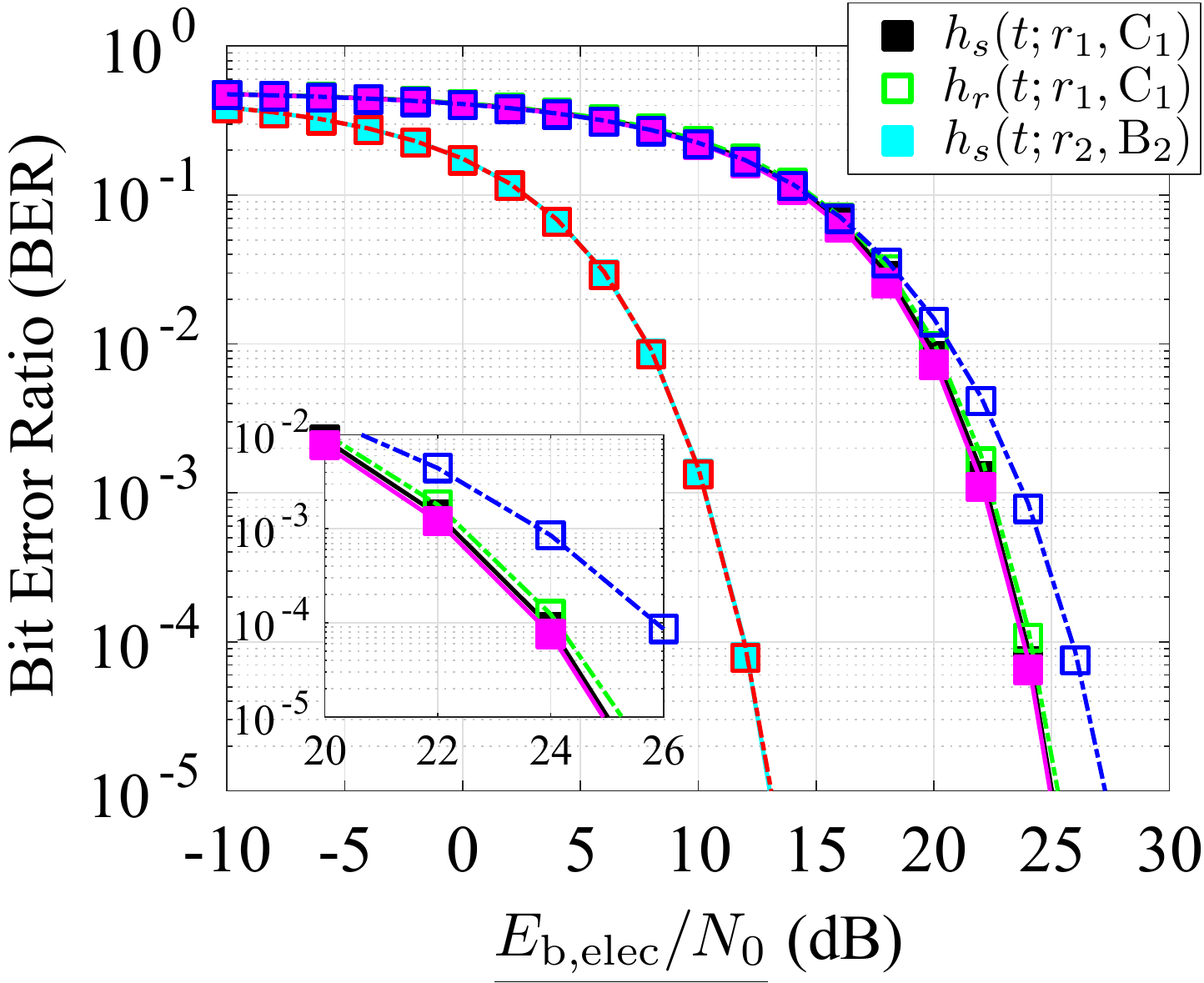}
		\caption{VL band, $M=4$ ($1~\text{bit/sec/Hz}$)}
		\label{fig:}
	\end{subfigure}\\
	\begin{subfigure}[t]{.78\columnwidth}
		\includegraphics[width=\columnwidth]{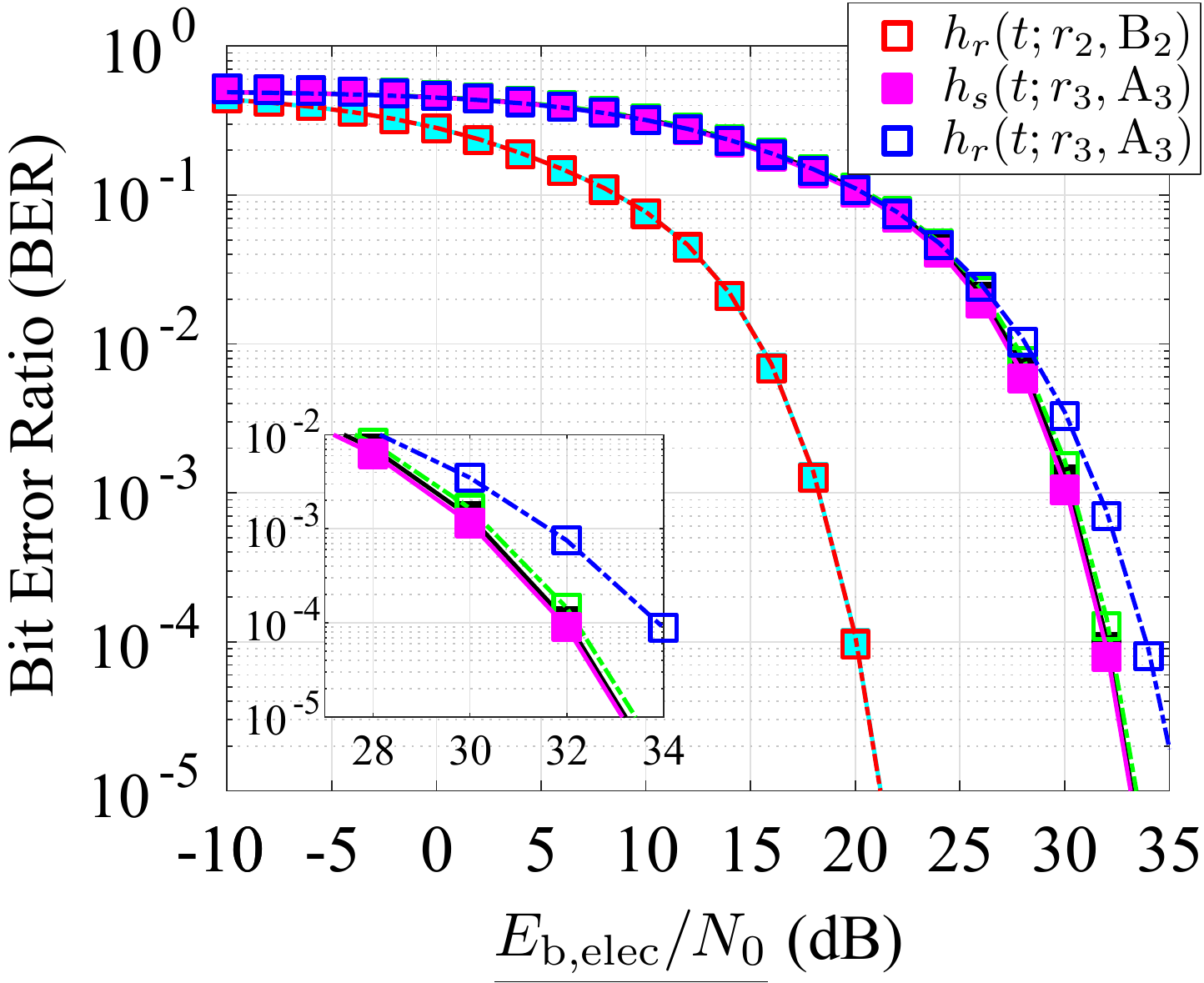}
		\caption{VL band, $M=64$ ($3~\text{bits/sec/Hz}$ )}
		\label{fig:}
	\end{subfigure}
	\caption{BER vs. effective SNR-per-bit performance curves for a DCO-OFDM system under the MCRT channel dispersion and double-sided clipping impairments. The simulation results for realistic and simplified cabin models are given by dashed and solid lines, respectively. The theoretical results are presented by markers.}
	\label{fig:BER_SNR_VL}
\end{figure}

In our \gls{BER} simulations, all the subcarriers are modulated by using the same normalized \gls{QAM} modulation which yields, $M_k=M,~\forall k$ and $\text{E}\{ \lvert X[k] \rvert^2 \}=1$. Since the magnitude of the channel taps, $\lvert H[k] \rvert$, are in the order of $10^{-5}$, in \gls{MCRT} results, the electrical \gls{PL} at the \gls{RX} becomes $-100$ dB. Therefore, the \gls{BER} plots are presented \gls{w.r.t.} the \emph{received} electrical-\gls{SNR}-per-bit value, which could be calculated by $\underline{E_\text{b,elec}/N_0}=E_\text{b,elec}/N_0-100$ \cite{phd_Fath}. The rest of \gls{DCO-OFDM} system parameters and their descriptions, which are used in the error performance simulations are given in Table \ref{table_berpar}.

The \gls{BER} vs. $\underline{E_\text{b,elec}/N_0}$ plots of \gls{DCO-OFDM} for average spectral efficiency of $1$ and $3$ bits/sec/Hz are given in Figs. \ref{fig:BER_SNR_IR} and \ref{fig:BER_SNR_VL}, respectively. The circle and square markers represent the computer simulation results for \gls{IR} and \gls{VL} bands, respectively, where the lines are the theoretical results obtained by using \eqref{eq:ABER}. Furthermore the different colours among each plot depicts the various locations of the \gls{UE} as explained at the beginning of the section. As illustrated in Fig. \ref{fig:BER_SNR_IR}, the error performance of seat B in the \gls{IR} band outperforms seats C and A result at least $10$ dB in both low and mid/high spectral efficiency regions. A similar trend can also be seen in \gls{VL} band, as depicted in Fig. \ref{fig:BER_SNR_VL}, where the error performance difference between seat B with seats C and A becomes at least $12$ dB for both spectral efficiency values. The main reason behind this error performance in point B2 is the very high received \gls{SNR} as well as a lack of multipath dispersion compared to other measurement points. For the \gls{UE} location A3, \gls{DCO-OFDM} error performance under both the \gls{IR} and \gls{VL} band channels with simplified cabin model outperformed the realistic cabin as much as $2$ dB in both spectral efficiency regions. This difference is also important as it shows the importance of cabin geometry on the practical system performance.
\section{Conclusion}
In this paper, a reading lights based broadband in-flight \gls{LiFi} system is investigated. Accordingly, an \gls{MCRT} based realistic \gls{DL} on-board \gls{LiFi} channel modelling technique for both \gls{IR} and \gls{VL} bands is presented. Accordingly, two narrow-body aircraft cabin models, one with accurate dimensions, surface geometry and curvatures, and another with planar surfaces are generated. Moreover, economy class cabin interior seating is also considered in our simulations. To model the source, receiver and surface coating optical characteristics, measurement based spatio-angular and spectral properties are used. Results show that the location of the \gls{UE} and cabin simplification have a significant impact on the channel parameters. The analytical expression and computer simulations based \gls{BER} curves obtained for \gls{DCO-OFDM} also confirmed the \gls{MCRT} results for a practical system performance. Consequently, favourable propagation characteristics and high link budget properties makes in-flight \gls{LiFi} a highly suitable candidate for on-board broadband \gls{5G} \gls{NR} applications.

\bibliographystyle{IEEEtran}
\bibliography{IEEEabrv,thesis}

%
\begin{IEEEbiography}[{\includegraphics[width=1in,height=1.25in,clip,keepaspectratio]{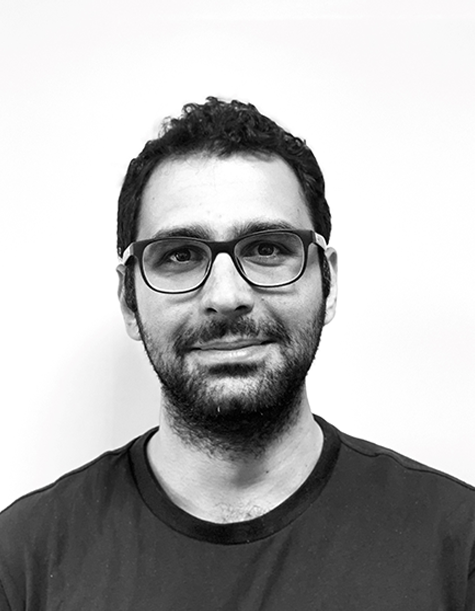}}]{Anil Yesilkaya}
	(Member, IEEE) received the B.Sc. (Hons.) and M.Sc. degrees in electronics engineering from Kadir Has University, Istanbul, Turkey, in 2014 and 2016, respectively. He received the PhD. degree in digital communications from the University of Edinburgh, Edinburgh, U.K., in 2021. He is currently working as a postdoctoral research associate in Horizon 2020 project 5G-CLARITY at the LiFi Research and Development Centre, University of Strathclyde. He was a recipient of the Best Paper Award from the IEEE International Conference on Communications (ICC) Optical Networks and Systems (ONS) Symposium in 2018.
\end{IEEEbiography}

\begin{IEEEbiography}[{\includegraphics[width=1in,height=1.25in,clip,keepaspectratio]{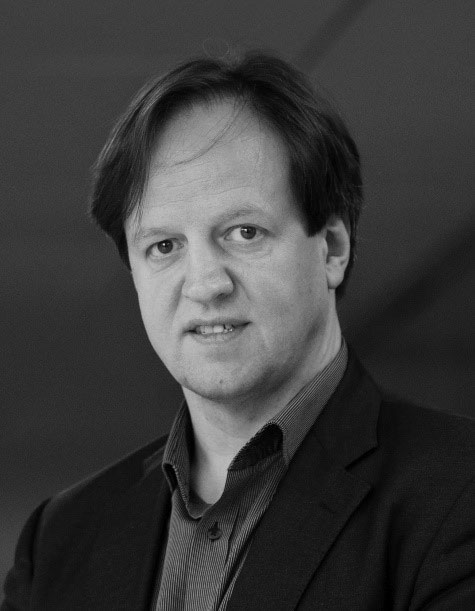}}]{Harald Haas}
	(Fellow, IEEE) received the Ph.D. degree from The University of Edinburgh in 2001. He is a Distinguished Professor of Mobile Communications at The University of Strathclyde/Glasgow, Visiting Professor at the University of Edinburgh and the Director of the LiFi Research and Development Centre. Prof Haas set up and co-founded pureLiFi. He currently is the Chief Scientific Officer. He has co-authored more than 600 conference and journal papers. He has been among the Clarivate/Web of Science highly cited researchers between 2017-2021. Haas’ main research interests are in optical wireless communications and spatial modulation which he first introduced in 2006. In 2016, he received the Outstanding Achievement Award from the International Solid State Lighting Alliance. He was the recipient of IEEE Vehicular Society James Evans Avant Garde Award in 2019. In 2017 he received a Royal Society Wolfson Research Merit Award. He was the recipient of the Enginuity The Connect Places Innovation Award in 2021. He is a Fellow of the IEEE, the Royal Academy of Engineering (RAEng), the Royal Society of Edinburgh (RSE) as well as the Institution of Engineering and Technology (IET).
\end{IEEEbiography} 








\end{document}